\documentclass[rmp,aps,reprint,amssymb,floatfix]{revtex4-1}

\usepackage{epsfig}
\usepackage{graphicx}
\usepackage[intlimits]{amsmath}
\usepackage[mathscr]{eucal}
\usepackage{bbm}
\usepackage{simplewick}
\usepackage{color}
\usepackage{enumerate}
\usepackage[position=top,caption=false]{subfig}
\begin{document}

\title{Light Hadron Masses from Lattice QCD}
\author{Zoltan Fodor}
\email[]{fodor@bodri.elte.hu}
\affiliation{Bergische Universit\"at Wuppertal, Gaussstr. 20, D-42119
Wuppertal, Germany.}
\author{Christian Hoelbling}
\email[]{hch@physik.uni-wuppertal.de}
\affiliation{Bergische Universit\"at Wuppertal, Gaussstr. 20, D-42119
Wuppertal, Germany.}

\begin{abstract}
This article reviews lattice QCD results for the light hadron
spectrum. We give an overview of different formulations of lattice
QCD, with discussions on the fermion doubling problem and improvement
programs. We summarize recent developments in algorithms and analysis
techniques, that render calculations with light, dynamical quarks
feasible on present day computer resources. Finally, we summarize
spectrum results for ground state hadrons and resonances using various
actions. 
\end{abstract}

\maketitle

\tableofcontents

\section{Introduction}

Already at the beginning of the $19^{\text{th}}$ century, it was
speculated by \cite{Prout:1815} that the hydrogen atom was the basic
building block for all other atoms. The mass of the proton, as mass of
the hydrogen atom, was known within a factor 2 accuracy already during
the late $19^{\text{th}}$ century \cite{Loschmidt:1865}. Later,
the development of mass spectrometry \cite{Goldstein:1886} allowed a
precision measurement of the $e/m$ ratio of the hydrogen nucleus
\cite{Wien:1902,Thomson:1907} and following the discovery of the
atomic nucleus by \cite{Rutherford:1911}, he could show that hydrogen
nuclei were present in other nuclei \cite{Rutherford:1919} and
coined for them the name protons.

The neutron was discovered 13 years later by \cite{Chadwick:1932}, who
also determined its mass with a 2 per mil accuracy. The first meson to
be discovered was the pion \cite{Powell:1947}, shortly followed by
the kaon \cite{Rochester:1947mi} and the $\Lambda$
\cite{Seriff:1950}, the first strange particles. While these
discoveries were made in cosmic ray experiments, the first resonance,
the $\Delta$ was discovered by \cite{Brueckner:1952zz} at a cyclotron
source. During the following years, these modern accelerators lead to
a proliferation of hadronic states and it became obvious that they
could not all be regarded as elementary.

This large number of hadronic states could first be successfully
described by their quark substructure \cite{GellMann:1961ky}, for
which finally quantum chromodynamics (QCD) was found as the dynamical
theory by \cite{Fritzsch:1973pi}. With the discovery of asymptotic
freedom \cite{Politzer:1973fx,Gross:1973id}, which built on earlier
work regarding the renormalizability of nonabelian gauge theories by
\cite{'tHooft:1972fi}, and the qualitative understanding of the
confinement phenomenon \cite{Wilson:1974sk} a coherent picture of
the strong interaction finally emerged. At energies that are large
compared to the typical QCD scale $\Lambda_{\text{QCD}}\sim
250\text{~MeV}$ the coupling is small and quarks and massless gluons
emerge as the fundamental degrees of freedom. At low energies however,
the spectrum of QCD consists of quark-gluon bound states that one
would like to identify with the experimentally observed
hadrons. Although this qualitative picture is quite compelling, it is
nevertheless very difficult to solve QCD in the low energy regime,
where it is a strongly coupled theory, and predict respectively postdict
the hadron spectrum from first principles.

In the present review, we summarize the current state of the art of
computing the light hadron spectrum, i.e. the spectrum of hadrons with
exclusively up, down and strange valence quarks, directly in QCD. In
sect.~\ref{sect:tech} we introduce the primary tool to study QCD in
the nonperturbative regime - lattice QCD. We review various possible
discretizations of continuum QCD in view of their usefulness for ab
initio calculations of light hadron masses with a small and controlled
total uncertainty. In sect.~\ref{sect:extr}, we review current methods
of extracting hadron masses in lattice QCD. We discuss efficient ways
of extracting ground state masses as well as current methods to
overcome the challenges in singlet and excited state spectroscopy. In
sect.~\ref{sect:physpred} we review methods for obtaining predictions
at the physical point (the point in parameter space at which the quark
masses have their physical values) in the infinite volume continuum
theory. Finally in sect.~\ref{sect:results} we summarize present and
notable past results and conclude with an overview of the present
understanding of the hadron spectrum from lattice QCD.  As a
convention when quoting lattice results the first error is statistical
and the second one (if given) systematic unless explicitly noted
otherwise.

\section{Lattice techniques}
\label{sect:tech}

Lattice field theory is in most cases the only known systematic way of
nonperturbatively computing Greens functions in quantum field
theories. It is especially useful in contexts where perturbative
treatment is usually inadequate, which is the case in low energy QCD.

Lattice gauge theory was introduced by \cite{Wilson:1974sk}\footnote{
  Independent developments of Smit and Polyakov were never published,
  see e.g.  \cite{Wilson:2004de}} and recent overviews include
\cite{Montvay:1994cy,Gupta:1997nd,DiPierro:2000nt,Smit:2002ug,Rothe:2005nw,DeGrand:2006zz,Gattringer:2010zz}\footnote{See
  also the classic introductory text by \cite{Creutz:1984mg}.}.

In general, a nonperturbative lattice calculation proceeds in 3
steps. First, one introduces a UV regulator into the theory by
means of a finite spacetime lattice. Then one computes Greens
functions in this discretized theory by means of stochastic
integration of the path integral and finally one removes the
regulator in order to obtain the continuum result. The last step
is possible in theories where the coupling does not diverge in
the UV regime. Due to asymptotic freedom, QCD does belong to this
class of theories and the cutoff can be removed.

In this section we will mainly focus on the first step of the above
procedure, the regularization of QCD on a spacetime lattice. This
regularization is not unique and the ambiguity is reflected in the
wide variety of lattice regularizations of QCD that are in use today
each carrying various advantages and disadvantages.  We start with a
brief introduction to the path integral formalism in
sect.~\ref{sect:basic} and the basics of the lattice discretization of
QCD in sect.~\ref{sect:lqcd}. We then introduce the basic concepts of
the stochastic evaluation of the discretized path integral in
sect.~\ref{sect:numerics} which are necessary to understand the
further developments of sect.~\ref{sect:doubl} and
sect.~\ref{sect:improve} where we discuss how to obtain efficient
lattice regularized theories that actually go over into QCD in upon
removal of the cutoff. Finally, in sect.~\ref{sect:aniso} we briefly
discuss anisotropic lattice regularizations of QCD that are relevant
for excited state spectroscopy.

The need for an efficient regularization arises particularly due to
the smallness of the light quark masses compared to the intrinsic QCD
scale $\Lambda_\text{QCD}$. On the one hand, the physical size of the
lattice needs to be much larger than the correlation length of the
system which in turn is given by the inverse of the mass of the
lightest particle in the spectrum, the pion. On the other hand, the
lattice cutoff needs to be much larger than $\Lambda_\text{QCD}$ in
order to not miss a substantial fraction of the nonperturbative
dynamics. These two requirements combined necessitate a large number
of lattice points if one would like to perform nonperturbative lattice
QCD calculations at physically light quark masses. In connection with
the fermion doubling problem the smallness of the light quark masses
causes yet further problems that are discussed in detail in
sect.~\ref{sect:doubl}.

Because of these effects, lattice QCD calculations until very recently
were restricted to quark masses larger (and in most cases
substantially so) than the physical ones. This in turn necessitated an
extrapolation in the light quark mass to the physical point in
addition to the already necessary continuum extrapolation. The fact
that physically light quark masses have been reached by reweighting
\cite{Aoki:2009ix} or directly, at large volumes and several different
values of the cutoff \cite{Durr:2010aw,Durr:2010vn} is to a large
extent due to recent advances in the construction of efficient lattice
regularizations that will be reviewed in this section.

\subsection{Basics of the Path Integral formalism}
\label{sect:basic}

We start by writing the partition function of a Euclidean quantum
field theory using the path integral formalism
\cite{Dirac:1933xn,Feynman:1948ur,Feynman:1948km,Feynman:1949zx,Feynman:1965xx}
as
\begin{equation}
\label{eq:pi}
\mathcal{Z}=\int{\mathcal D}\Phi e^{-S(\Phi)}
\end{equation}
with the action $S(\Phi)$ and $\Phi$ generically denoting all fields
of the theory. For bosonic fields one typically introduces
  periodic boundary conditions, while for fermion fields it is natural
  to introduce antiperiodic boundary conditions in time direction (see
  e.g. appendix A of \cite{Polchinski:1998rq}).  While this subtlety
  usually can be ignored, it does play some role when choosing the
  parity of interpolating operators as discussed in
  sect.~\ref{sect:oper}.  For a gauge theory with fermions one
specifically has
\begin{equation}
\label{eq:ctz}
\mathcal{Z}=\int{\mathcal D}A_\mu \int{\mathcal D}\psi{\mathcal D}\bar\psi
e^{-(\bar\psi M\psi+S_G)}
\end{equation}
with the Euclidean gauge action
\begin{equation}
S_G=\frac{1}{4}F_{\mu\nu}F_{\mu\nu}
\qquad
F_{\mu\nu}=\partial_\mu A_\nu-\partial_\nu A_\mu +ig[A_\mu,A_\nu]
\end{equation}
and the Euclidean Dirac operator in the case of one fermion flavor 
\begin{equation}
\label{eq:contfo}
M=\gamma_\mu D_\mu+\text{m}
\end{equation}
where the covariant derivative is given by
\begin{equation}
\label{eq:coder}
D_\mu=\partial_\mu+igA_\mu
\end{equation}
Note that in the massless case $M$ is antihermitian and -
depending on the gauge field configuration - may have exact zero
modes. These zero modes of the operator are related to the
topology of the underlying gauge field by the Atiyah-Singer index
theorem \cite{Atiyah:1967ih}

\begin{widetext}
 Using the rules of Grassmannian integration, one can generically rewrite
(\ref{eq:ctz}) as
\begin{equation}
\mathcal{Z}=\int{\mathcal D}A_\mu \int{\mathcal D}\psi{\mathcal D}\bar\psi
e^{-(\bar\psi M\psi+S_G)}
=\int{\mathcal D}A_\mu
\det(M)
e^{-S_G}
\end{equation}
Averages that correspond to expectation values of time ordered
operators in the operator formalism are generically obtained by
\begin{equation}
\label{eq:expo}
\langle O\rangle=
\frac{1}{\mathcal{Z}}
\int{\mathcal D}A_\mu \int{\mathcal D}\psi{\mathcal D}\bar\psi
O
e^{-(\bar\psi M\psi+S_G)}
\end{equation}
where $O$ denotes a generic observable composed of gauge and fermion
fields. The integration over the fermion fields can again be
explicitly performed resulting in the replacement of fermion bilinears
by propagators on a given gauge field background. For a single fermion
bilinear this explicitly reads
\begin{equation}
\label{eq:iprp}
\int{\mathcal D}\psi{\mathcal D}\bar\psi
\left(\psi^{s_f}_{c_f}(y)\bar\psi^{s_i}_{c_i}(x)\right)
e^{-\bar\psi M\psi}
=\det(M)
\left(M^{-1}\right)^{s_f s_i}_{c_f c_i}(y,x)
\end{equation}
while for a general expression it results in the usual combination of
all Wick contractions.
\end{widetext}

As one can see from (\ref{eq:ctz},\ref{eq:iprp}), virtual fermion
effects (``sea fermions'') are contained in the $\det(M)$ factor after
the fermion field integration. Ignoring this $\det(M)$ factor results
in an uncontrolled approximation to QCD, the quenched approximation
\cite{Marinari:1981qf}.

\subsection{QCD regularized on a lattice}
\label{sect:lqcd}

The path integral (\ref{eq:pi}) has to be performed over all field
configurations. In order to make it well-defined, we regulate it on a
finite spacetime lattice
\begin{equation}
x_\mu=n_\mu a^{(\mu)}
\quad\text{with}\quad
n_\mu\in\{0,\ldots,N_\mu-1\}
\end{equation}
where the $a^{(\mu)}$ are the lattice spacings in direction
$\mu$. Although more general topologies are possible in principle
(see e.g. \cite{Jersak:1996mj}), one usually imposes toroidal
boundary conditions $x_\mu+N_\mu a^{(\mu)}=x_\mu$. Here we will
also specialize to the common isotropic case in which the lattice
spacing in all directions are equal $a^{(\mu)}=a$. The
anisotropic case will be separately discussed in
section~\ref{sect:aniso}.
 
The fermion field $\psi$ is now a Grassmann vector, defined at
the discrete lattice points $x=n a$. We can write the naive
discretization of the free fermionic continuum action
as
\begin{equation}
\label{eq:fnf}
S_F(m)=a^4 \sum_{x=n a} \bar{\psi}(x)
\left(
\gamma_\mu {\mathcal D}_\mu+\text{m}
\right)
\psi(x)
\end{equation}
with
\begin{equation}
\label{eq:ddef}
{\mathcal D}_\mu=\frac{1}{2a}\left(V_\mu-V_\mu^\dag\right)
\end{equation}
and
\begin{equation}
\label{eq:freehop}
\left(V_\mu\right)_{xy}=\delta_{x+\hat{\mu},y}
\end{equation}
Note that this action was obtained by replacing the continuum
derivative operator $\partial_\mu$ with the simple lattice finite
difference operator ${\mathcal D}_\mu$. This choice is not unique and
this non-uniqueness can be exploited to construct efficient
fermion regularizations. Another feature of (\ref{eq:fnf}) is
that it does not describe a single fermion flavor even in the
continuum limit. The latter is known as the fermion doubling
problem \cite{Karsten:1980wd} and will be discussed in detail in
sect.~\ref{sect:doubl}. In sect.~\ref{sect:fimp} we will discuss
how the ambiguity in the fermion discretization can be utilized
to construct numerically efficient lattice fermion
regularizations.

The action (\ref{eq:fnf}) is invariant under a global symmetry
transformation
\begin{equation}
\psi(x)\rightarrow \Lambda\psi(x)
\quad
\bar\psi(x)\rightarrow \bar\psi(x)\Lambda^\dag
\end{equation}
with $\Lambda\in SU(3)$ for the case of QCD. This symmetry can be
promoted to a local one
\begin{equation}
\label{eq:loctr}
\psi(x)\rightarrow \Lambda(x)\psi(x)
\quad
\bar\psi(x)\rightarrow \bar\psi(x)\Lambda^\dag(x)
\end{equation}
by including a parallel transport $U_\mu(x)$ to the one-hop term
$V_\mu$ and thus replacing (\ref{eq:freehop}) with
\begin{equation}
\label{eq:hop}
\left(V_\mu\right)_{xy}=U_\mu(x)\delta_{x+\hat{\mu},y}
\end{equation}
The parallel transport $U_\mu(x)$ is the discretized version of
the path ordered product of continuum gauge fields $A_\mu(x)$
\begin{equation}
\label{eq:partrans}
U_\mu(x)={\mathscr P}e^{i g\int_x^{x+\hat{\mu}} dx^{\prime}_\mu A_\mu(x^{\prime})}
\end{equation}
with $g$ being the coupling constant and transforms as
\begin{equation}
U_\mu(x)\rightarrow\Lambda(x)U_\mu(x)\Lambda^\dag(x+\hat\mu)
\end{equation}
under gauge transformations. Note that one could in principle choose
different paths than the direct one in (\ref{eq:partrans}). As long as
the end points remain fixed, the action will be invariant under local
transformations (\ref{eq:loctr}). This non-uniqueness will play a role
when constructing efficient fermion discretizations in
sect.~\ref{sect:fimp}.

In order to construct a kinetic term for the gauge field we first note
that the trace over a closed loop of parallel transports is gauge
invariant. The simplest of these loops, the plaquette, is defined as
\begin{equation}
U_{\mu\nu}(x)=U_{\mu}(x)U_{\nu}(x+\hat\mu)U^\dag_{\mu}(x+\hat\nu)U^\dag_{\nu}(x)
\end{equation}
and has a naive continuum limit
\begin{equation}
U_{\mu\nu}\stackrel{a\rightarrow 0}{\longrightarrow}1+iga^2F_{\mu\nu}-\frac{1}{2}g^2a^4F^2_{\mu\nu}+O(a^6)
\end{equation}
The simplest discretization of the continuum
gauge action therefore reads \cite{Wilson:1974sk}
\begin{equation}
\label{eq:wg}
S_W=\beta\sum_{x,\mu>\nu}\left(1-\frac{1}{6}
\text{Tr}(U^\dag_{\mu\nu}(x)+U_{\mu\nu}(x))\right)
\end{equation}
with $\beta=6/g^2$ which has the continuum limit
\begin{equation}
\label{eq:cga}
S_W
\stackrel{a\rightarrow 0}{\longrightarrow}
\frac{1}{4}
\int d^4 x
\text{Tr}\left(
F_{\mu\nu}(x)F_{\mu\nu}(x)
\right)
+O(a^2)
\end{equation}
We see that (\ref{eq:wg}) which is known as the Wilson gauge action or
plaquette action has discretization errors of $O(a^2)$. Again, this
discretization is not unique and one can utilize this ambiguity to
find gauge actions with higher order discretization effects. This will
be discussed in detail in sect.~\ref{sect:gimp}. By combining
(\ref{eq:fnf}) with (\ref{eq:ddef},\ref{eq:hop}) and (\ref{eq:wg}) and
introducing the dimensionless quantities $\Psi=a^{3/2}\psi$,
$\bar\Psi=a^{3/2}\bar\psi$, $m=a\text{m}$ and $D_\mu=a{\mathcal D}_\mu$, we can write
the naive lattice QCD action as
\begin{equation}
\label{eq:naiv}
\begin{split}
S_n&=S_W+ \sum_{q=1}^{N_f} S_F(m_q)\\
&=\beta\sum_{x,\mu>\nu}\left(1-\frac{1}{6}
Tr(U^\dag_{\mu\nu}(x)+U_{\mu\nu}(x))\right)\\
&+
\sum_{q=1}^{N_f}
\sum_{x} \bar{\Psi}(x)
\left(
\gamma_\mu D_\mu+m_q
\right)
\Psi(x)
\end{split}
\end{equation}
where we have in addition taken the explicit sum over $N_f$ fermion
flavors $q$.

Before we go into the details of the fermion and gauge field
discretization, let us mention briefly how in principle the cutoff is
removed in lattice QCD.  As one can see from (\ref{eq:naiv}), the
lattice action exclusively consists of dimensionless quantities.  The
parameters of the action are the fermion masses $m_q$ and the coupling
$\beta$.  In order to remove the cutoff, i.e. to take the limit
$a\rightarrow 0$, one therefore has to tune these parameters such that
on the one hand the lattice spacing $a$ goes to zero while on the
other hand a certain set of dimensionful physical observables that are
used to define the physical content of the theory remain
constant. These trajectories in parameter space of $\beta$ and the
$m_q$ along which a set of physical observables remains constant as
the limit $a\rightarrow 0$ is taken are called lines of constant
physics.  A detailed discussion of how these can be defined is given in
sect.~\ref{sect:physpred}.

Along these lines of constant physics it is clear that correlation
lengths in physical units will go to a finite limit and therefore will
diverge in units of the lattice spacing $a$. In order to possess a
continuum limit it is therefore necessary for a lattice field theory
to exhibit a second order phase transition. Problems can arise if the
bare coupling constant diverges at a finite cutoff, the so called
Landau pole \cite{Landau:1955aa}. In that case the only line of
constant physics that does not show a divergence at finite cutoff is
the one with vanishing coupling, i.e. the trivial theory.  In order
for theories with a Landau pole problem to have a non vanishing
renormalized coupling (i.e. to be nontrivial) one must retain a finite
cutoff which prevents one from taking the continuum limit.  Such
theories can, however, still serve as effective theories. Consequences
for the lattice formulation of this class of theories are discussed
e.g. by
\cite{Gockeler:1997kt,Arnold:2002jk,Espriu:2003sa,Kogut:2005pm}.

\begin{figure}
\includegraphics[width=0.5\textwidth]{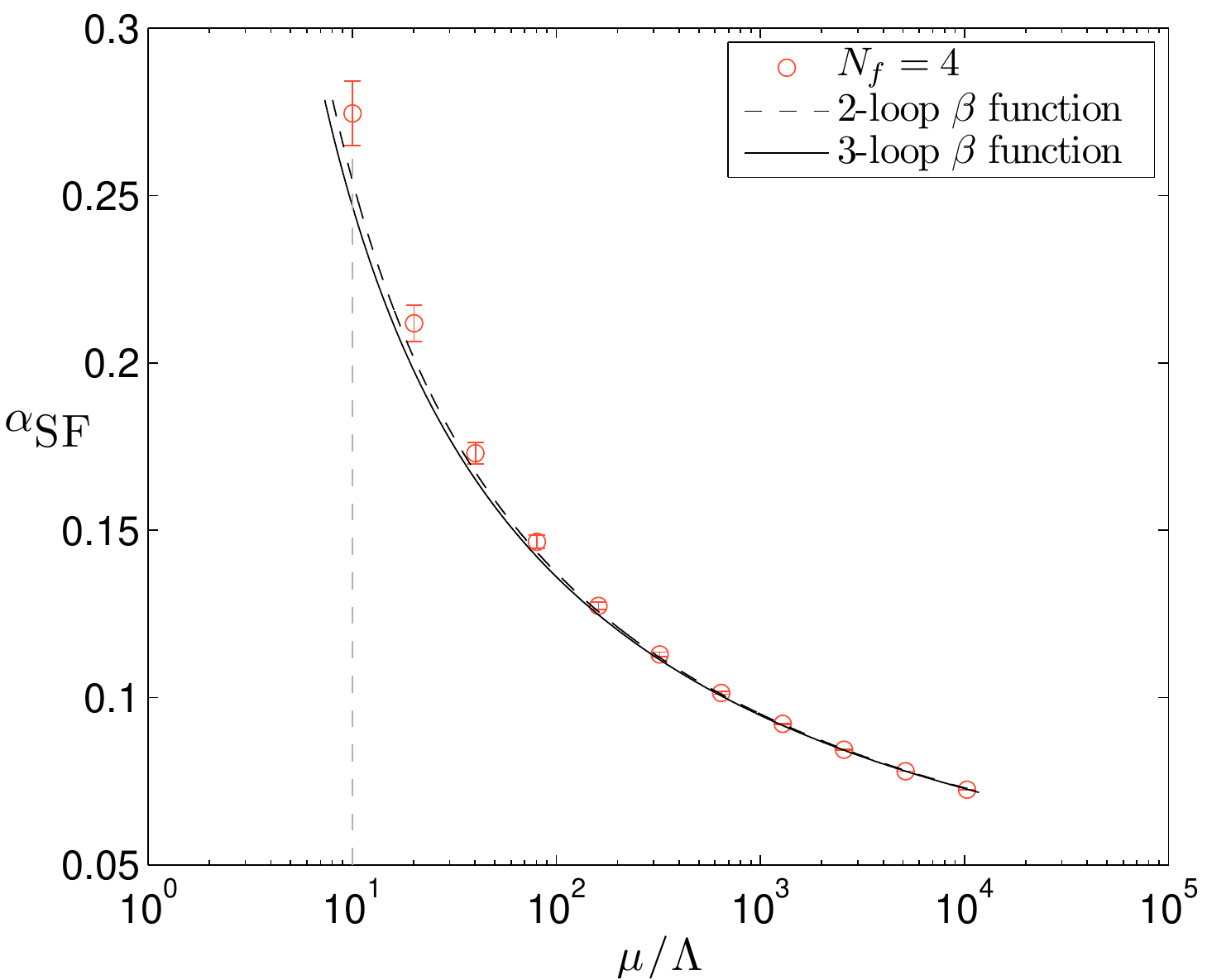}
\caption{
\label{fig:nf4run}
Nonperturbative running coupling constant of 4 flavor QCD in units of
the QCD scale $\Lambda$ from a lattice calculation of
\cite{Sommer:2010ui} compared to perturbative calculation. Plot
reproduced with friendly permission of Rainer Sommer.}
\end{figure}

Due to asymptotic freedom \cite{Politzer:1973fx,Gross:1973id}
however, no such problems are expected to arise in lattice QCD. The
perturbative expectation of the vanishing of the QCD coupling constant
at large scales has been confirmed by nonperturbative lattice
calculations in various settings
\cite{Bowler:1986rx,Luscher:1993gh,Gupta:1988pe,Bode:2001jv,DellaMorte:2002vm,Sommer:2010ui,Tekin:2010mm}. The
most recent result for $N_f=4$ QCD is plotted in
fig.~\ref{fig:nf4run}.

\subsection{Numerical evaluation of the Path Integral}
\label{sect:numerics}

Before we return to the task of constructing a lattice regularization
of QCD, we need to discuss some basics of the numerical evaluation of
the path integral (\ref{eq:expo}). In terms of dimensionless lattice
quantities, (\ref{eq:expo}) can be written as
\begin{equation}
\label{eq:expolat}
\langle O\rangle=
\frac{1}{\mathcal{Z}}
\int{\mathcal D}U_\mu \int{\mathcal D}\Psi{\mathcal D}\bar\Psi
O
e^{-(\bar\Psi M(U)\Psi+S_G(U))}
\end{equation}
with a generic gauge action $S_G(U)$ and fermion operator $M(U)$,
which in general depends on the gauge field $U$. We would now like to
perform the integration over both the fermion fields $\bar\Psi$,
$\Psi$ and the gauge filed $U$. Using the relation (\ref{eq:iprp}) and
its generalization for fermion multilinears, we can explicitly perform
the integration over fermion fields. With the understanding that we
have to replace fermion multilinears by the sum over all Wick
contractions, we may thus cast (\ref{eq:expolat}) into the form
\begin{equation}
\label{eq:expolint}
\langle O\rangle=
\frac{1}{\mathcal{Z}}
\int \prod_{x,\mu} d U_\mu(x) \det M(U)
O
e^{-S_G(U)}
\end{equation}
Generally and for QCD in particular, it is not possible to perform the
remaining integration over the gauge fields $U_\mu$ in closed
form. While both strong \cite{Wilson:1974sk} and weak (see
e.g.\cite{Capitani:2002mp}) coupling expansions are possible, neither
allows for a detailed quantitative understanding of the
nonperturbative dynamics of the system.

From a numerical perspective, (\ref{eq:expolint}) is a high dimensional
integral over  $4\times\prod_\mu N_\mu$ copies of the gauge group,
which in the case of QCD is $SU(3)$. The only category of numerical
methods that are suitable to perform such a high dimensional integral
are stochastic or Monte Carlo (MC) methods in which the space to be
integrated over is randomly sampled, i.e., where observables are
averaged over on randomly drawn gauge field configurations $U_\mu(x)$.
As Monte Carlo integration is stochastic in nature, there is always a
statistical error associated with it. This error has to be estimated
in a lattice calculation, which is typically achieved via the
jackknife or bootstrap methods (see e.g.\cite{citeulike:1386464}).

A straight Monte Carlo integration of (\ref{eq:expolint}) however
in which the gauge configurations $U_\mu$ are randomly produced
with equal weight is extremely inefficient. For interesting
parameter choices, all but a very small subset of relatively
smooth configurations are exponentially suppressed by the
exponent of the gauge action $S_G(U)$ and the fermion determinant
$\det M(U)$. In order to circumvent this problem, one can produce
gauge field configurations with a probability that is
proportional to $\det M(U)\times e^{-S_G(U)}$ and compute the
expectation value of an observable as an unweighted average over
these configurations. This technique, known as importance
sampling, requires an algorithm that produces gauge field
configurations with the proper weight. Typically this is achieved
via a Markov chain process, where a time series of gauge field
configurations is produced in which the $n^\text{th}$
configuration $U_\mu^{(n)}$ depends on the previous one
$U_\mu^{(n-1)}$.

A Markov chain is characterized by the transition probability
\begin{equation}
\label{eq:transm}
 p(m,n)=P(U^m|U^n)
\end{equation}
where the $U^i$ are all possible gauge configurations and the
conditional probability $P(U^m|U^n)$ is understood in the sense that
it denotes the probability of the system to go over from a
configuration $U^n$ to $U^m$ in one time step. The transition
probability $p$ acts on the space of all gauge configurations. It
fulfills the two basic relations
\begin{equation}
\label{eq:trprop}
 \forall m,n:p(m,n)\ge 0 \qquad \int dm p(m,n)=1
\end{equation}
where $\int dm$ denotes the integration over all possible gauge
field configurations. If in addition the transition probability
(\ref{eq:transm}) fulfills the detailed balance condition
\begin{equation}
\label{eq:detb}
\forall m,n: p(n,m)\rho_m=p(m,n)\rho_n
\end{equation}
with the desired equilibrium distribution $\rho_n=\det M(U^n)\times
e^{-S_G(U^n)}/\mathcal{Z}$, then one can easily show the following two
properties:
\begin{enumerate}
\item
  The transition probability maps the equilibrium distribution onto itself
\begin{equation}
\label{eq:eq1}
\rho_m=\int dn p(m,n)\rho_n
\end{equation}
\item Defining a distance $d(w,v)=\int dn |w_n-v_n|$ in the space
  of probability distributions, the application of the transition
  probability moves every probability distribution closer to the
  equilibrium distribution
\begin{equation}
\label{eq:eq2}
d(pw,\rho)\le d(w,\rho)
\end{equation}
\end{enumerate}
If in addition to (\ref{eq:eq1}) and (\ref{eq:eq2}) the system is
ergodic, i.e. if any configuration $U^i$ may be reached from any other
configuration $U^j$ with non-vanishing probability in a finite number
of time steps, then it is guaranteed that starting from an arbitrary
initial probability distribution we end up with the desired equilibrium
distribution $\rho$.

The time until the equilibrium distribution $\rho$ is reached (in the
sense that no statistically relevant drift towards the equilibrium
expectation value can be seen in any monitored observable) is usually
called thermalization phase and its shortness is an important quality
criterion of an algorithm. Once the system is thermalized, i.e. the
equilibrium distribution has been reached, it is advantageous if
consecutive configurations have as little correlation as possible. In
order to have a quantitative handle, it is customary to monitor the
autocorrelation time of certain observables within a Markov chain.

In the case of a pure gauge theory or the quenched approximation where
the fermion determinant factor $\det M(U)$ is missing and the weight
factor is proportional to the exponent of the gauge action
$e^{-S_G(U)}$, the update algorithms that produce the next element in
the Markov chain usually exploit the locality of the gauge action
$S_G(U)$. As pioneered by \cite{Creutz:1979zg,Creutz:1979kf}, one can
pick a certain gauge link $U_\mu(x)$ from the current gauge
configuration $U$ and produce a suggested new gauge configuration
$U^\prime$ by multiplying $U_\mu(x)$ with an element of the gauge
group. Since the gauge action $S_G(U)$ is a sum of local terms, the
change in the action $\delta S=S_G(U^\prime)-S_G(U)$ is readily
evaluated by recomputing those few terms that contain the flipped
gauge link. One can then perform a Metropolis \cite{Metropolis:1953am}
step, i.e. accept the gauge configuration $U^\prime$ as the next gauge
configuration in the Markov chain with probability $e^{-\delta S}$ if
the action has increased $\delta S>0$ or with probability $1$
otherwise. It is readily seen that this algorithm satisfies the
detailed balance condition (\ref{eq:detb}). Another frequently used
local update algorithms for pure gauge theories is the heatbath
\cite{Creutz:1980zw,Kennedy:1985nu}.  Supplemented by overrelaxation
steps
\cite{Adler:1981sn,Adler:1987ce,Brown:1987rra,Creutz:1987xi,Fodor:1994ih},
these algorithms are still the state of the art for pure gauge
theories.

Due to the nonlocal nature of the fermion determinant $\det M(U)$
an update in a theory with dynamical fermions is substantially
more complex and computationally demanding. For a lattice with
$N=\prod_\mu N_\mu$ sites, $M(U)$ is typically a
$(12\times N)^2$ matrix\footnote{Note that in the staggered
  fermion formulation the size of the matrix is reduced to
  $(3\times N)^2$. For a detailed discussion see
  sect.~\ref{sect:stag}.}  and therefore a direct computation of
$\det M(U)$ is prohibitively expensive for even moderately sized
lattices. Although alternative suggestions have been made
\cite{Fucito:1980fh,Kuti:1981dt,Berg:1981jy,Polonyi:1983tm,Scalapino:1981qs,Montvay:1984ui,Luscher:1993xx,Slavnov:1995ju},
one usually proceeds by introducing a bosonic (complex scalar)
pseudofermion field $\Phi$ \cite{Weingarten:1980hx}. The fermion
determinant may thus be written as
\begin{equation}
\label{eq:pseudof}
\det{M(U)}=
\int {\mathcal D}\Phi^\dag {\mathcal D}\Phi
e^{-\Phi^\dag M(U)^{-1}\Phi}
\end{equation}
The catch here is of course the appearance of the inverse fermion
matrix $M(U)^{-1}$ in (\ref{eq:pseudof}) which is again a nonlocal
object. In addition, the kernel operator $M(U)^{-1}$ has to exist
(i.e. the matrix $M(U)$ needs to be invertible) and be positive
definite hermitian in order to ensure the convergence of all gaussian
integrals over the pseudofermion field in (\ref{eq:pseudof}). From
(\ref{eq:naiv}) we see however, that the naive fermion operator is not
hermitian and neither will be the fermion operators we will construct
later on. As long as $\det M$ is real and positive definite however,
one may use the identity $\det M=\sqrt{\det\left(M^\dag M\right)}$ to
rewrite an arbitrary power of the fermion determinant as
\begin{equation}
\label{eq:pseudofreal}
\det{M(U)}^{2\alpha}=
\int {\mathcal D}\Phi^\dag {\mathcal D}\Phi
e^{-\Phi^\dag\left(M^\dag(U)M(U)\right)^{-\alpha}\Phi}
\end{equation}
The path integral can now be formulated in terms of bosonic variables
only with an additional term in the action
\begin{equation}
\label{eq:pseudofa}
S_F=\Phi^\dag\left(M^\dag(U)M(U)\right)^{-\alpha/2}\Phi
\end{equation}
Note that for actions where $M^\dag(U)M(U)$ does not couple even and
odd lattice sites one may choose to keep the pseudofermion fields
$\Phi_e$ on even lattice sites only and thereby obtain
\begin{equation}
\label{eq:pseudofeo}
\det{M(U)}^{\alpha}=
\int {\mathcal D}\Phi_e^\dag {\mathcal D}\Phi_e
e^{-\Phi_e^\dag\left(M^\dag(U)M(U)\right)^{-\alpha}\Phi_e}
\end{equation}

In order to efficiently integrate the system of pseudofermions and
gauge fields we follow
\cite{Callaway:1982eb,Callaway:1983ee,Polonyi:1983tm,Batrouni:1985jn,Duane:1985ym,Duane:1985hz,Duane:1986iw,Duane:1987de}
and reinterpret the total action of the system 
\begin{equation}
\label{eq:stot}
S=S_G+\Phi^\dag\left(M^\dag(U)M(U)\right)^{-\alpha/2}\Phi
\end{equation}
as the potential part of a fictitious Hamiltonian
\begin{equation}
\label{eq:hmch}
{\mathcal H}=\frac{1}{2}\pi^2+S(\phi)
\end{equation}
with conjugate momenta $\pi$ where $\phi$ collectively denotes all
pseudofermion and gauge fields. One can then proceed to choose some
initial momenta and integrate the canonical equations of motion
\begin{equation}
\label{eq:eqm}
\dot{\phi}=\pi
\qquad
\dot{\pi}=-\frac{\partial S}{\partial\phi}
\end{equation}
numerically in a fictitious time $\tau$ along a ``classical''
path. The classical partition function corresponding to the set of all
such classical trajectories is given by
\begin{equation}
\label{eq:classz}
Z=\int {\mathcal D} \pi {\mathcal D} \phi e^{-H}=
\int {\mathcal D} \pi e^{-\frac{1}{2}\pi^2} \int {\mathcal D} \phi e^{-S}
\end{equation}
As the gaussian integration over the momenta $\pi$ only
gives an irrelevant prefactor, (\ref{eq:classz}) does reproduce
the correct probability distribution in the original
theory. Assuming ergodicity, one can obtain the correct
distribution of classical paths (\ref{eq:classz}) by periodically
refreshing the momenta $\pi$ with a random value from a gaussian
distribution. The expectation value of an observable can thus be
obtained by averaging it along all classical trajectories in the
update chain. The inexact nature of the numerical integration
introduces a systematic error, which however can be corrected by
a final Monte Carlo accept/reject step of the complete trajectory
(see \cite{Duane:1987de}). This algorithm is known as hybrid
Monte Carlo (HMC).

The Hamiltonian in (\ref{eq:hmch}) is readily constructed in the case
where $\alpha$ in (\ref{eq:pseudofreal}) or (\ref{eq:pseudofeo}) is a
positive integer. For a general fractional power $\alpha$ one can
resort to a polynomial \cite{deForcrand:1996ck,Frezzotti:1997ym} or
rational \cite{Clark:2003na,Clark:2004cp} approximation of the
desired fractional power of $M^\dag(U)M(U)$. These versions of the HMC
algorithm are known as polynomial HMC (PHMC) and rational HMC (RHMC)
respectively.

In order to integrate the equations of motion (\ref{eq:eqm})
numerically, one has to compute the derivative of the action
(\ref{eq:stot}) with respect to the gauge field
\begin{equation}
\label{eq:force}
\frac{\partial S}{\partial U_\mu(x)}=
\frac{\partial S_G}{\partial U_\mu(x)}+
\Phi^\dag\frac{\partial\left(M^\dag(U)M(U)\right)^{-\alpha}}{\partial U_\mu(x)}\Phi
\end{equation}
This derivative is commonly known as the force term. The computationally
expensive part of (\ref{eq:force}) is the second term, the fermionic
force term. In the special case $\alpha=1$ and using the shorthand
notation ${\mathcal M}(U)=M^\dag(U)M(U)$ it may be written as
\begin{equation}
\label{eq:fforce}
\Phi^\dag\frac{{\partial\mathcal M}(U)^{-1}}{\partial U_\mu(x)}\Phi
=
-\Phi^\dag{\mathcal M}(U)^{-1}\frac{\partial{\mathcal M}(U)}{\partial U_\mu(x)}{\mathcal M}(U)^{-1}\Phi
\end{equation}
We see that a single inversion of ${\mathcal M}(U)$ on a pseudofermion
vector is required to compute (\ref{eq:fforce}). For general
fractional exponents $\alpha$ one may introduce a rational
approximation
\begin{equation}
r(\mathcal M(U))\simeq{\mathcal M}(U)^{-\alpha}
\end{equation}
with
\begin{equation}
r(x)=\sum_i\frac{\alpha_i}{x+\beta_i}
\end{equation}
In this case, the fermion
force may be written as
\begin{widetext}
\begin{equation}
\Phi^\dag r(\mathcal M(U))\Phi\\
=
-\sum_i\Phi^\dag\alpha_i\left({\mathcal M}(U)+\beta_i\right)^{-1}\frac{\partial{\mathcal M}(U)}{\partial U_\mu(x)}\left({\mathcal M}(U)+\beta_i\right)^{-1}\Phi
\end{equation}
\end{widetext}
which can be computed using one single multilinear matrix inversion
\cite{deForcrand:1995bs,Glassner:1996gz,Frommer:1995ik,Jegerlehner:1996pm} on a single vector.

An alternative integration scheme, the hybrid molecular dynamics (HMD)
R-algorithm, was proposed by \cite{Gottlieb:1987mq}. Although it has
seen considerable use in the past, it has largely been replaced by the
RHMC algorithm. It is a pure molecular dynamics algorithm that, in
contrast to pseudofermion algorithms, does not allow for a final MC
step to correct for finite step size errors accumulated along the
integration trajectory. Due to this feature, detailed balance is
fulfilled by the R-algorithm only in the limit of a vanishing step size
in contrast to HMC-type algorithms that are exact even at finite step
size.

An efficient HMC algorithm has to simultaneously satisfy two criteria:
On the one hand, the acceptance rate should be high (one typically
aims for $\sim 80-90\%$) and on the other hand the autocorrelation time
should be small. The autocorrelation between successive configurations
can be decreased by a longer integration trajectory separating them.
This however leads to larger numerical integration errors and
consequently to a lower acceptance rate. A trivial remedy consists of
decreasing the time step in the numerical integration, which however
is computationally expensive because the fermion force has to be
computed more often. It is therefore advantageous to use higher order
integration schemes
\cite{2002PhRvE..66b6701O,2002PhRvE..65e6706O,2003CoPhC.151..272O,Takaishi:2005tz}
that allow a larger time step in the numerical integration while
keeping the acceptance rate high. 

Another method for speeding up HMC type algorithms consists of
introducing different time steps for pseudofermions and gauge fields
\cite{Sexton:1992nu}. Splitting off the UV modes of the spectrum by
mass preconditioning \cite{Hasenbusch:2001ne,Hasenbusch:2002ai} or via
domain decomposition
\cite{Luscher:2003vf,Luscher:2003qa,Luscher:2005rx} and integrating IR
and UV part with different time steps leads to a substantial
additional speedup. This speedup is especially large if combined with
the suppression of UV modes and other improvements of the fermion
regularization that will be discussed in sect.~\ref{sect:fimp}.

As noted in sect.~\ref{sect:basic}, ignoring the effects of the
fermion determinant results in the quenched
approximation. Since it bypasses the most computationally demanding
part of the ensemble generation, it was extensively used in the early
years of lattice QCD and is still useful for certain conceptual
studies. Although it is an uncontrolled approximation, it may be
justified by noting that it becomes exact in the large $N_c$
limit. Furthermore, by choosing the proper scale setting observable
(see sect.~\ref{sect:physpred}) a large part of the dynamical fermion
corrections might cancel and effectively be absorbed into a
redefinition of the coupling constant.\footnote{Another attempt to
  justify the use of the quenched approximation was made by
  \cite{Anthony:1982fe,Duffy:1983hf}. They suggested to extrapolate to
  a positive number of quark flavors by computing observables in the
  quenched approximation and at an effective negative number of quark
  flavors.}

\subsection{The fermion doubling problem and its solutions}
\label{sect:doubl}

We now return to the free, naive fermion action (\ref{eq:naiv})
\begin{equation}
\label{eq:naivfa}
\bar\Psi\left(\gamma_\mu D_\mu+m\right)\Psi
\end{equation}
The fermion operator reads
\begin{equation}
\label{eq:naivopf}
M=\gamma_\mu D_\mu+m
\end{equation}
which in Fourier space becomes
\begin{equation}
\label{eq:naivpop}
M(p)=\frac{i}{a}\sum_\mu \gamma_\mu \sin(ap_\mu)+m
\end{equation}
The momentum space propagator is consequently given by
\begin{equation}
\label{eq:naivprop}
D(p)=M^{-1}(p)=
\frac{-\frac{i}{a}\sum_\mu \gamma_\mu \sin(ap_\mu)+m}
{\left(\frac{1}{a}\sum_\mu \sin(ap_\mu)\right)^2+m^2}
\end{equation}
which in addition to the physical pole at $p^2=-m^2$ has $15$
additional poles located at the edges of the Brillouin zone. The poles
are located at $(p-\Pi)^2=-m^2$, where $\Pi$ is any of the 16
four-momenta
\begin{equation}
\label{eq:brim}
\Pi=(p_0,p_1,p_2,p_3) \qquad \text{with } p_\mu\in\{0,\pi/a\}
\end{equation}
This rather fundamental obstacle of putting fermion fields on the
lattice is known as the doubling problem. Physically, we can trace
this problem back to the well known axial anomaly of a continuum
theory. In the massless limit, a classical fermionic theory is
invariant under the chiral transformation
\begin{equation}
\label{eq:chiral}
\begin{split}
\Psi(x) & \rightarrow \Psi^{\prime}(x)=e^{i\phi\gamma_5} \Psi(x) \\
\bar{\Psi}(x) & \rightarrow \bar{\Psi}^{\prime}(x)=\bar{\Psi}(x) e^{i\phi\gamma_5}
\end{split}
\end{equation}
As demonstrated by \cite{Adler:1969gk,Bell:1969ts}, the conservation
of the corresponding Noether current, the axial vector current, is
destroyed by quantum fluctuations. In a lattice regulated theory
however the existence of a classical symmetry implies a conserved
current. The anomaly of the physical fermion axial vector current is
canceled by the anomaly of unphysical doublers as demonstrated by
\cite{Karsten:1980wd}\footnote{See also
  \cite{Chodos:1977dh,Kerler:1981tb}}.

It was later shown by
\cite{Nielsen:1980rz,Nielsen:1981xu,Nielsen:1981hk}, that no lattice
fermion regularization exists that fulfills all of the following
conditions at the same time
\begin{enumerate}[i]
\item\label{nu:nodub}
Absence of doubler fermions
\item\label{nu:chis}
Continuum chiral symmetry in the massless case
\item\label{nu:loc}
Locality in a sense that $M(x,y)\rightarrow 0$ vanishes exponentially as $x-y\rightarrow\infty$
\item\label{nu:ctl}
Correct continuum limit
\end{enumerate}

\begin{figure}
\subfloat[Fermionic case]{
\label{fig:fermkern}
\includegraphics[width=0.5\textwidth]{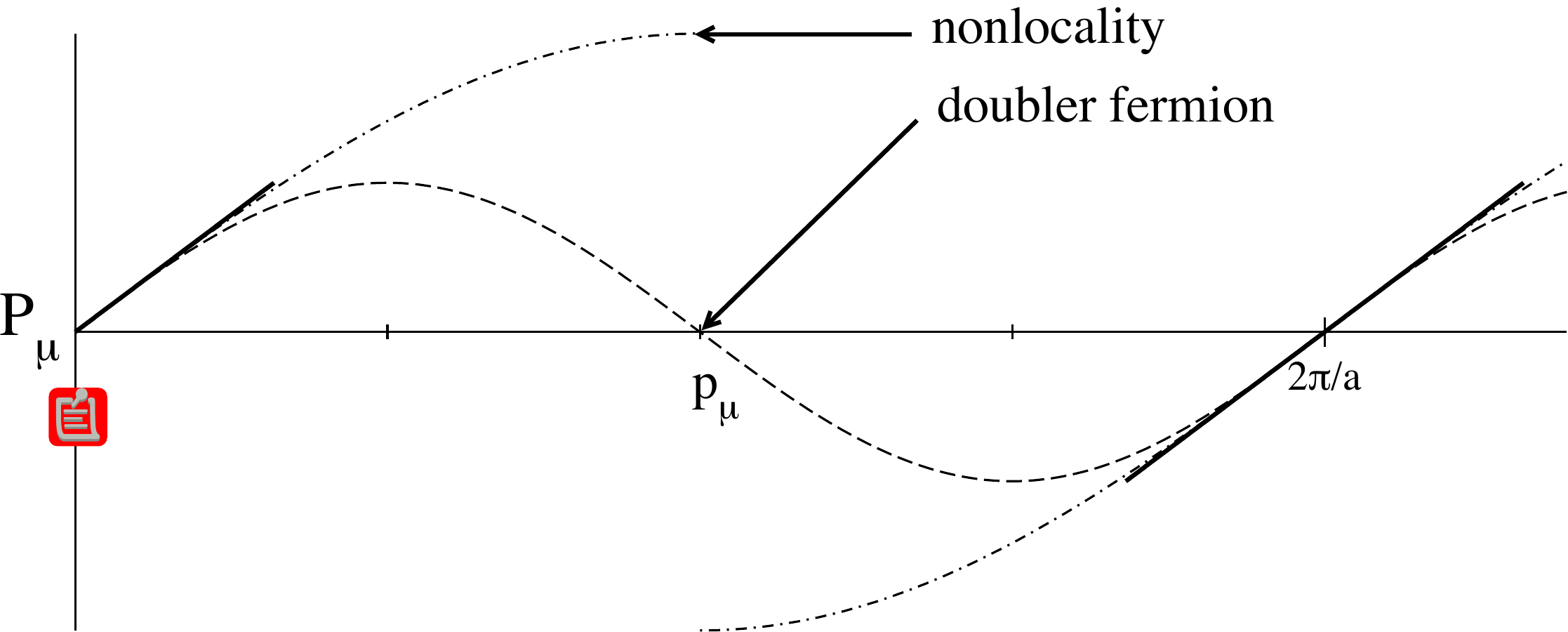}
}
\\
\subfloat[Bosonic case]{
\label{fig:boskern}
\includegraphics[width=0.5\textwidth]{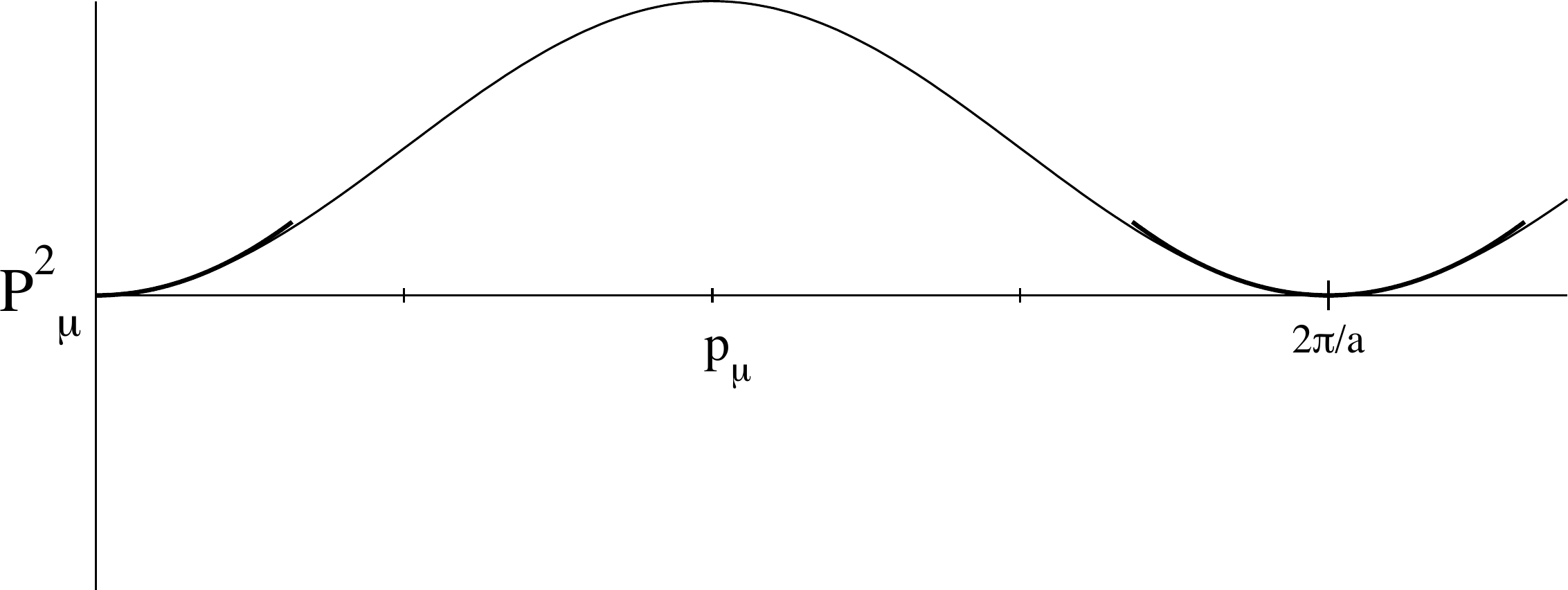}
}
\caption{ \label{fig:nini} { In the fermionic case
    \protect\subref{fig:fermkern}, periodicity of the lattice momentum $P_\mu$
    requires a second root (doubler) or jump within the Brillouin zone
    (nonlocality). In the bosonic case \protect\subref{fig:boskern}, only the
    squared lattice momentum is required to be periodic which can be
    fulfilled without a jump or an additional root.  }
}
\end{figure}

This result can be understood by noting that a general lattice fermion
operator $M$ which anticommutes with $\gamma_5$ in the $m=0$ case can
be written as
\begin{equation}
M(p)=m+i\sum_\mu\gamma_\mu P_\mu(a p)+\sum_\mu \gamma_\mu\gamma_5 R_\mu(a p)
\end{equation}
The requirement that it reproduces the correct continuum theory
implies, that for small $a$ the lattice momentum $P_\mu$ goes over
into the continuum momentum $p_\mu$ while $R_\mu\rightarrow
0$. Additionally, $P_\mu$ is periodic in every direction with period
$2\pi/a$. As shown in fig.~\ref{fig:nini}, these restrictions on
$P_\mu$ imply that either it has a second root in the first Brillouin
zone, which gives an additional pole in the propagator, i.e. a doubler
fermion, or that it has at least one discontinuity, which makes the
fermion operator $M(x,y)$ nonlocal. Note that in comparison the
discretized version of the continuum action of a scalar field
\begin{equation}
\label{eq:scct}
S_\phi=\frac{1}{2}\phi^\dag\left(\partial_\mu\partial_\mu-m^2\right)\phi
\end{equation}
in momentum space reads
\begin{equation}
\label{eq:scft}
S_\phi=-\frac{1}{2}\phi^\dag\left(m^2+\sum_\mu\left(P^2\right)_\mu\right)\phi
\end{equation}
which only depends on the discretized momenta squares
\begin{equation}
\left(P^2\right)_\mu=\frac{2}{a^2}\left(1-\cos ap_\mu\right)
\end{equation}
These are naturally periodic with a period of $2\pi/a$ as displayed in
fig.~\ref{fig:boskern}.

One therefore has to give up on any one of the above requirements for
lattice regularizations of fermions. Obviously, one can not give up
the requirement (\ref{nu:ctl}) of a correct continuum limit. Giving up
the locality requirement (\ref{nu:loc}) on the other hand has been
suggested, among others, by \cite{Drell:1976mj} who proposed
\begin{equation}
P_\mu=p_\mu
\end{equation}
by \cite{Rebbi:1986ra}, who suggested
\begin{equation}
P_\mu=\sin{a p_\mu}\frac{2\sum_\nu\sin{\frac{a p_\nu}{2}}}{a\sum_\nu\sin{a p_\nu}}
\end{equation}
and by \cite{Gross:1987by} whose construction involves non-symmetric
difference operators and contains nonrenormalizable terms in the
continuum limit. However, all of these approaches turned out to be
problematic and have been abandoned.

Among the remaining two options, we will first discuss latice fermion
regularizations that give up on the requirement (\ref{nu:nodub}) and
therefore describe more than a single flavor in the continuum
limit. Among these,
\cite{Karsten:1981gd,Wilczek:1987kw,Creutz:2007af,Borici:2007kz} have
suggested different implementations of minimally doubled fermions,
i.e. lattice fermions which have one single doubler only. As this
single doubler has to be placed somewhere within the Brillouin zone,
all of these formulations share the characteristic that in Fourier
space there is a distinguished direction, namely the direction from
the physical particles pole to the pole of the single doubler fermion.
Therefore, a number of discrete lattice symmetries are broken
\cite{Bedaque:2008xs} resulting in a more complicated renormalization
pattern and, generically, in a fine-tuning of the parameters of the
action \cite{Capitani:2010nn}. Currently fundamental properties of
minimally doubled fermions are still being clarified and applications
to hadron spectroscopy or other phenomenologically relevant
computations are not available in the literature yet.

\subsubsection{Staggered fermions}
\label{sect:stag}

A less minimal but more symmetric way of putting doublers on the lattice
is given by the staggered fermion formulation that was developed in a
series of papers by
\cite{Kogut:1974ag,Banks:1975gq,Susskind:1976jm}. Staggered fermions are
obtained from the naive fermion action (\ref{eq:naivfa}) by noting a
fourfold exact degeneracy (in the interacting theory) that can be
exposed by a spin-diagonalization
\begin{equation}
\label{eq:spindiag}
\Psi(x)=\Gamma(x)\chi(x)
\quad
\bar\Psi(x)=\bar\chi(x)\Gamma^\dag(x)
\end{equation}
with
\begin{equation}
\label{eq:stagclif}
\Gamma(x)=\prod_\mu\gamma_\mu^{\frac{x_\mu}{a}}
\end{equation}
In terms of $\bar\chi$ and $\chi$, (\ref{eq:naivfa}) can be written as
\begin{equation}
\label{eq:stagfa}
\bar\chi\left(\eta_\mu D_\mu+m\right)\chi
\qquad
\eta_\mu(x)=(-1)^{\sum_{\nu>\mu}x_\nu}
\end{equation}
where $\eta_\mu(x)$ is a pure phase factor making explicit the
decoupling of the 4 spin components of $\chi$. Defining $\chi$ on a
single component only, we have reduced the fermion content of the
theory by a factor of 4, from 16 four-component spinors to 16 single
component modes that are still symmetrically distributed over the
Brillouin zone at the momenta $\Pi$ given in (\ref{eq:brim}).

The staggered fermion operator
\begin{equation}
\label{eq:stagop}
M=\eta_\mu D_\mu+m
\end{equation}
is antihermitian in the massless case, i.e. its eigenvalues are
restricted to the imaginary axis for $m=0$. Therefore any finite
mass $m>0$ provides an IR cutoff and the operator is invertible. In
addition, the massless staggered fermion operator does preserve a
remnant of the chiral symmetry of the naive fermion operator (\ref{eq:chiral})
\begin{equation}
\label{eq:stageps}
\begin{split}
\chi(x) & \rightarrow \chi^{\prime}(x)=e^{i\phi\epsilon} \chi(x) \\
\bar{\chi}(x) & \rightarrow \bar{\chi}^{\prime}(x)=\bar{\chi}(x) e^{i\phi\epsilon}
\end{split}
\end{equation}
where $\epsilon=(-1)^{\sum_\mu x_\mu}$. For the staggered fermion
operator (\ref{eq:stagop}) this implies that 
\begin{equation}
\label{eq:stageh}
\epsilon M=M^\dag\epsilon
\end{equation}
is hermitian. Therefore, the eigenvalues of
$M$ come in complex conjugate pairs and $\det M$ is real and
positive. Furthermore, we see that
\begin{equation}
M^\dag M=\eta_\mu D_\mu^\dag \eta_\nu D_\nu+|m|^2
\end{equation}
does not couple odd and even sites so that the pseudofermion
representation (\ref{eq:pseudofeo}) may be used for the fermion
determinant.

As demonstrated by
\cite{Sharatchandra:1981si,vandenDoel:1983mf,Golterman:1984cy,KlubergStern:1983dg,Gliozzi:1982ib,Daniel:1986zm},
these 16 spinor components may again be interpreted as 4 fermion
flavors (also referred to as tastes in the literature), each one
described by a 4-component spinor. 
Following \cite{Golterman:1985dz},
we split the lattice coordinate $x=y+h$ into a piece $y$ that
describes the origin of the the elementary $2^4$ hypercube that x is
located in and an offset $h$ which describes the location of $x$
within this hypercube. We construct
\begin{equation}
\label{eq:staggam}
X(y)=\frac{1}{4}\sum_{h}\Gamma(h)U(y+h,y)\chi(y+h)
\end{equation}
where $h_\mu\in\{0,a\}$ and $U(y+h,y)$ is any parallel transport from $y+h$ to
$y$. Interpreting $X(y)$ as a 16 component vector, we can define an
arbitrary fermion bilinear with spin structure $\gamma_s$ and flavor
structure $\gamma_f$ as
\begin{equation}
\bar X \left(\gamma_s\otimes\xi_f\right)X=
\text{Tr}\left(\bar X \gamma_s X \gamma_f^\dag \right)
\end{equation}
and the staggered fermion action (\ref{eq:stagfa}) can be written as
\begin{equation}
\label{eq:stflop}
\bar X
\left(
(\mathbbm{1}\otimes\mathbbm{1})m
+
(\gamma_\mu\otimes\mathbbm{1})
{\hat D}_\mu
+
(\gamma_5\otimes\xi_\mu\xi_5)
{\hat C}_\mu
\right)
X
\end{equation}
with the first and second derivative operators on the coarse lattice
\begin{equation}
\label{eq:thop}
{\hat D}_\mu=\frac{1}{4}\left({\hat V}_\mu-{\hat V}_\mu^\dag\right)
\quad
{\hat C}_\mu=\frac{1}{4}\left({\hat V}_\mu-2\mathbbm{1}+{\hat V}_\mu^\dag\right)
\end{equation}
where
\begin{equation}
\left({\hat V}_\mu\right)_{xy}=U_\mu(x+\hat{\mu})U_\mu(x)\delta_{x+2\hat{\mu},y}
\end{equation}
Note that the third term in (\ref{eq:stflop}) that implies a mixing of
the four remaining flavors (tastes) is an artefact of the taste
assignment (\ref{eq:staggam}) which does not respect the full set of
symmetries of the staggered action. In fact, for the noninteracting
case \cite{Adams:2004mf} has found a taste assignment that is diagonal
in the tastes and local.

A massless four flavor continuum theory does have a classical $U(4)$
chiral symmetry which gets reduced to $SU(4)$ by the anomaly. This
implies a 15-plet of massless pseudoscalar Goldstone particles (pions)
and an additional massive one ($\eta^\prime$). As we have seen in
(\ref{eq:stageps}), staggered fermions retain a $U(1)$ subgroup of this
symmetry. In the spin-flavor basis, the remnant staggered chiral
symmetry reads
\begin{equation}
\label{eq:stagchir}
\begin{split}
X(y) & \rightarrow X^{\prime}(y)=e^{i\phi(\gamma_5\otimes\xi_5)} X(y) \\
\bar{X}(y) & \rightarrow \bar{X}^{\prime}(y)=\bar{X}(y) e^{i\phi(\gamma_5\otimes\xi_5)}
\end{split}
\end{equation}
This symmetry is spontaneously broken implying a single Goldstone
particle, the pseudoscalar in taste space, that is exactly massless.

In order to obtain one resp. two flavors in the functional integral
(\ref{eq:expolint}), it is customary to take the quartic resp. square
root of the 4-flavor staggered functional determinant $\det M$. This
procedure, commonly referred to as rooting, was introduced in
\cite{Marinari:1981qf} in the context of the Schwinger model. On a
technical level this is realized in a pseudofermion based algorithm by
a fractional power $\alpha=1/4$ resp. $\alpha=1/2$ in
(\ref{eq:pseudofeo}).

On a more fundamental level, the validity of rooted staggered fermions
relies upon the assumption that there exists a local lattice fermion
operator that squared or to the fourth power has the same functional
determinant $\det M$ as the 4-flavor staggered operator. In fact,
\cite{Adams:2004mf} demonstrated that in the free
case such an operator can be found. In the interacting case however,
the situation is more complex. As \cite{Durr:2004ta} have
demonstrated, rooted staggered fermions are in the wrong universality
class in the strictly massless case $m=0$ implying that the chiral
$m\rightarrow 0$ limit does not commute with removing the cutoff
$a\rightarrow 0$, a result already anticipated by
\cite{Smit:1986fn}. This observation reflects the fact that the
staggered chiral symmetry (\ref{eq:stageps},\ref{eq:stagchir}) is not
anomalously broken.  In fact, the staggered chiral symmetry is
retained even in the single flavor rooted theory implying an exactly
massless $\eta^\prime$ for $m=0$ and as demonstrated by
\cite{Bernard:2006zw,Bernard:2006vv,Bernard:2007qf,Prelovsek:2005rf}
nonunitarity of the rooted theory at finite cutoff.  On the other
hand, there is some numerical indication that out of the chiral limit
rooted staggered fermions do indeed have correct continuum behavior
for many observables
\cite{Durr:2004ta,Durr:2004as,Durr:2003xs,Davies:2003ik,Davies:2004hc,Aubin:2004fs,Follana:2007uv,Bazavov:2010ru,Bazavov:2009bb}. More
formally, \cite{Bernard:2006ee} have shown by a symmetry argument that
rooted staggered fermions can not be described by a local
operator. However, analytical calculations
\cite{Bernard:2006ee,Giedt:2006ib,Bernard:2007ma,Shamir:2006nj}
indicate that the nonlocal terms vanish in the continuum limit and
consequently that rooted staggered fermions do have the correct
continuum limit as long as the proper order of limits is observed.
Implications of the delicate nature of the staggered fermion continuum
limit were also extensively discussed in the literature
\cite{Bernard:2004ab,Durr:2006ze,Creutz:2007yg,Bernard:2006vv,Hasenfratz:2006nw,Creutz:2007pr}.
For recent reviews see e.g. \cite{Durr:2005ax,Sharpe:2006re}.

The representation of the spin-taste structure by different points
within an elementary hypercube (\ref{eq:stagclif},\ref{eq:staggam})
and the taste breaking term in the action (\ref{eq:stflop}) imply some
additional complications for extracting hadron masses with staggered
fermions that will further be discussed in sect.~\ref{sect:extr}.

\subsubsection{Wilson fermions}

We now turn our attention to lattice fermion formulations that fully
lift the naive flavor degeneracy and are able to naturally describe a
single flavor theory in the continuum limit. It was first realized by
\cite{Wilson:1975id} that the fermion doubling problem can be solved
by adding a laplacian term to the naive fermion operator
(\ref{eq:naivfa})
\begin{equation}
\label{eq:wilsonac}
S_W=\bar\Psi\left(\gamma_\mu D_\mu+m+\frac{r}{2}\Box\right)\Psi
\end{equation}
where
\begin{equation}
\label{eq:box}
\Box=\sum_\mu C_\mu
\qquad
C_\mu=V_\mu-2+V_\mu^\dag
\end{equation}
with the parallel transport $V_\mu$ defined in (\ref{eq:hop}) and the
Wilson parameter $r$ that is usually set to $1$. The additional term
in the action, the so-called Wilson term, may be interpreted as a
momentum dependent mass term.  Note that in contrast to naive and
staggered fermions, the Wilson fermion operator is generally not
normal due to the additional laplacian term, although it still is in
the free case. The free Wilson operator
\begin{equation}
\label{eq:wilsonop}
M_W=\gamma_\mu D_\mu+m+\frac{r}{2}\Box
\end{equation}
in momentum space reads
\begin{widetext}
\begin{equation}
\label{eq:wilsonpop}
M_W(p)=\frac{i}{a}\sum_\mu \gamma_\mu \sin(ap_\mu)+m-\frac{r}{a}\sum_\mu\left(
\cos(ap_\mu)-1
\right)
\end{equation}
\end{widetext}
Comparing (\ref{eq:wilsonpop}) to the naive operator (\ref{eq:naivpop})
we see that the additional term
$\frac{r}{a}\sum_\mu\left(\cos(ap_\mu)-1\right)$ does vanish as $O(a)$
for any fixed {\it physical momentum} $p$. On the other hand, for
a fixed {\it lattice momentum} $ap$ the additional term gives
a contribution that is divergent as $O(1/a)$ except for $p=0$. In
particular, all doubler modes with $n$ momentum components $\pi/a$
do receive an additional mass of $2rn/a$ thus effectively removing them
from the spectrum in the continuum limit.

Since the additional laplacian term in (\ref{eq:wilsonac}) does not
anticommute with $\gamma_5$, the exact chiral symmetry of naive fermions
(\ref{eq:chiral}) is broken as required by the Nielsen-Ninomiya theorem
\cite{Nielsen:1980rz,Nielsen:1981xu,Nielsen:1981hk}. In fact, the
laplacian term commutes with $\gamma_5$ due to its trivial spin
structure. Consequently, the Wilson operator obeys the relation
\begin{equation}
\label{eq:g5hermw}
M_W^\dag=\gamma_5 M_W\gamma_5
\end{equation}
which is known as $\gamma_5$-hermiticity. It implies that the operator
\begin{equation}
\label{eq:hwil}
\gamma_5 M_W=\left(\gamma_5 M_W\right)^\dag
\end{equation}
is hermitian and the eigenvalues of $M_W$ are either real or come in
complex conjugate pairs. Consequently, $\det(M_W)$ is real. In order
for the pseudofermion representation (\ref{eq:pseudofreal}) to be well
defined, $\det(M_W)$ needs to be positive definite in addition, which
is guaranteed if $m>0$. However, due to the breaking of chiral
symmetry the fermion mass is not protected against additive
renormalization. The bare fermion mass receives corrections that are
divergent in the continuum limit and needs to be renormalized. Due to
this additive renormalization, the bare fermion mass corresponding to
a physically interesting renormalized mass often turns out to be
negative. While pairs of complex conjugate eigenvalues still give a
positive contribution to $\det(M_W)$ in this case, the real eigenmodes
only do so if the number of negative ones is even. Furthermore, even
eigenmodes that are positive but very small pose serious problems for
matrix inverters. In the case of an exact zero eigenmode, it is not
possible to define the fermion determinant in terms of the
pseudofermion fields according to (\ref{eq:pseudofreal}). Hitting such
a configuration within numerical precision will result in a failure of
the matrix inversion to properly converge \cite{Bardeen:1997gv}. These
configurations are known as ``exceptional'' and the appearance of even
a single exceptional configuration in a Markov chain indicates that
one is not able to properly sample a relevant region in configuration
space. Ensembles exhibiting an exceptional configuration therefore
have to be discarded.

In practice, exceptional configurations therefore set a lower limit to
the masses one can reach with Wilson-type fermions. One has to make
sure that all eigenmodes of the fermion matrix are sufficiently
separated from zero. While these restrictions were initially very
severe, they do not present a substantial obstacle for current state
of the art lattice calculations. The use of large physical volumes,
small lattice spacings, improved gauge actions (see
sect.~\ref{sect:gimp}) and smeared link fermion actions (see
sect.~\ref{sect:fimp}) all reduce the probability of exceptional
configurations appearing in a simulation.

\begin{figure}
\includegraphics[width=0.5\textwidth]{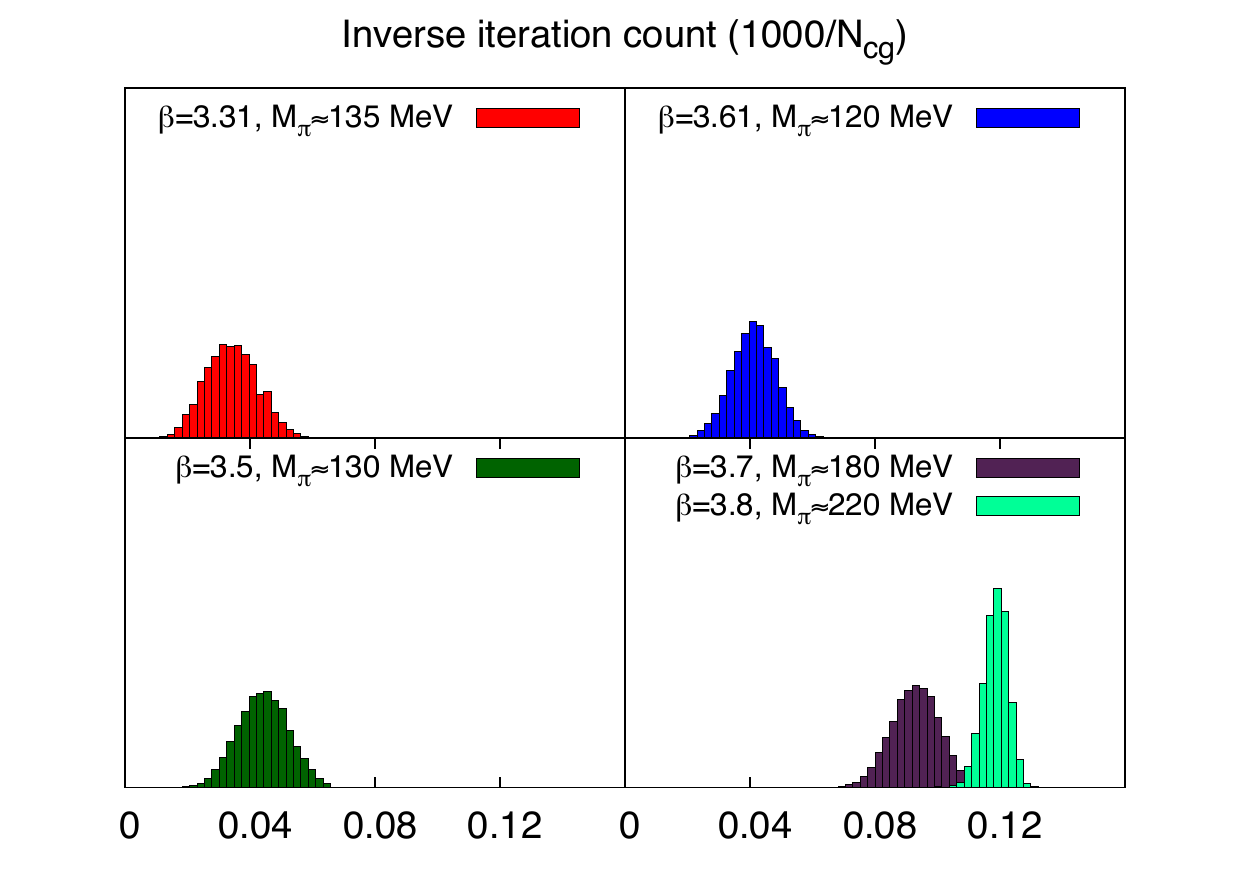}
\caption{
\label{fig:iic}
Inverse iteration count of the fermion matrix inverter for Wilson-type
fermions and a number of different ensembles with bare couplings
$\beta$ and approximate pion masses $M_\pi$. The lower tail of the
distributions shows a clear separation from $0$ indicating the absence
of exceptional configurations in the ensembles. From
\cite{Durr:2010aw} }
\end{figure}

Due to the strong correlation between the
condition number of the fermion matrix and the iteration count of the
inverter and because the relative fluctuations of the largest
eigenvalue are small, one can use the distribution of the inverse
iteration count instead of the distribution of the lowest eigenmode. A
tail of this distribution that extends towards the origin is a clear
and direct indication of problems with exceptional configurations
while a clear separation from $0$ demonstrates the absence of
exceptional configurations and positivity of the fermion
determinant. Such a distribution is plotted in fig.~\ref{fig:iic} for
a recent study with light Wilson-type fermions \cite{Durr:2010aw}.

The dominance of low modes in the computational cost of inverting a
Wilson-type Dirac operator has lead to efforts of preconditioning the
inversion by removing or effectively projecting out a relatively small
number of low modes. These techniques are generally known as deflation
methods
\cite{deForcrand:1995bs,Neff:2001zr,Giusti:2002sm,Giusti:2004yp,DeGrand:2004qw,Stathopoulos:2007zi,Luscher:2007se,Darnell:2007dr}. They
can lead to a huge decrease in the cost of computing propagators on
gauge configurations, especially in circumstances where one needs to
compute many propagators on the same gauge configuration. In this
case, one can perform a rather expensive eigenmode projection step
since it only has to be performed once for each gauge configuration.
For a further discussion see also sect.~\ref{sect:oper}

On a more fundamental level, real modes of the Wilson operator are
related to the topological charge of the gauge configuration by the
index theorem \cite{Atiyah:1967ih}. In the continuum limit, real modes
become exactly degenerate zero modes that are tied to the gauge field
topology
  \cite{Smit:1986fn,Setoodeh:1988ds,Vink:1988xt,Bardeen:1997gv,Hasenfratz:1998ri,Gattringer:1997ci}.

Recently it has been proposed to construct a Wilson-like operator that
instead of lifting the flavor degeneracy of the naive operator
(\ref{eq:naivopf}) does lift the taste degeneracy of the staggered
operator (\ref{eq:stagop})
\cite{Adams:2010gx,deForcrand:2011ak,Hoelbling:2010jw} (for earlier
work in this direction see also
\cite{Golterman:1984cy,Golterman:1985dz,Becher:1982ud,Mitra:1982my,Golterman:1984ds,Gockeler:1984rq,Mitra:1983bi}). Conceptual
aspects of this formulation are still being studied and no application
to hadron spectroscopy or other phenomenologically relevant
computations are available in the literature yet.

\subsubsection{Twisted mass fermions}
\label{sect:twist}

Twisted mass fermions \cite{Frezzotti:2000nk} are a variant of the
Wilson fermion formulation that has recently gained attention. The
basic idea is to perform a chiral rotation, that is not affected by an
anomaly, on the mass term. Because the transformation has to be
anomaly free, the number of flavors to be chirally rotated has to be
even. In the simplest case of two flavors the mass term reads
\begin{equation}
\label{eq:stmt}
\overline{m} e^{i\alpha\gamma_5\tau_3}=m+i\mu\gamma_5\tau_3
\end{equation}
where
\begin{equation}
\tan(\alpha)=\frac{\mu}{m}
\qquad
\overline{m}=m^2+\mu^2
\end{equation}
and $\tau_3$ is the diagonal Pauli matrix in flavor space. Due to the
opposite twist angles between the two flavors, the anomaly cancels
and the chiral rotation of the mass term (\ref{eq:stmt}) may be
absorbed into a chiral rotation of the fermion fields
\begin{equation}
\label{eq:tmchs}
\bar\Psi\rightarrow \bar\Psi e^{i\alpha/2\gamma_5\tau_3}
\qquad
\Psi\rightarrow e^{i\alpha/2\gamma_5\tau_3}\Psi
\end{equation}
provided that the massless part of the fermion operator is invariant
under the chiral transformation (\ref{eq:tmchs}). For Wilson fermions
however, chiral symmetry is explicitly broken. Replacing the standard
mass term in (\ref{eq:wilsonop}) with a twisted mass term of the form
(\ref{eq:stmt}) therefore results in a different theory where the two
flavor fermion matrix is given by
\begin{equation}
\label{eq:tmop}
M_{tm}=\gamma_\mu D_\mu+\frac{r}{2}\Box+m+i\mu\gamma_5\tau_3
\end{equation}
This represents the twisted mass fermion matrix in the so-called
twisted basis. The basis is called twisted because (\ref{eq:tmop})
does describe the physically uninteresting case of a complex mass
term.  In order to obtain physically interesting predictions for a
theory with a real mass term from (\ref{eq:tmop}), the chiral rotation
in the mass term has to be supplemented by an equivalent
transformation of the fermion fields (\ref{eq:tmchs}). In the new
basis, (\ref{eq:tmop}) describes the physically interesting case of
real mass fermions. This rotated basis of the fermion fields is
therefore usually referred to as the physical basis.

Wilson fermions with a twisted mass term do not obey standard time
reversal and parity transformation symmetries but modified versions
thereof. However, standard $CPT$ symmetry is fulfilled and the
behavior with regard to chiral and flavor symmetry is the same as for
standard Wilson fermions, i.e. chiral symmetry is broken while flavor
symmetry is exactly conserved in the twisted basis. In the physical
basis (\ref{eq:tmchs}) however a subset of the flavor and axial
symmetries get transformed into each other. Consequently, flavor
symmetry is broken while part of the chiral symmetry is restored at
maximal twist $\alpha=\pi/2$. This implies that at maximal twist on
the one hand there are isospin breaking cutoff effects
\cite{Scorzato:2004da,Bar:2010jk} while on the other hand the cutoff terms are
generally of $O(a^2)$ \cite{Frezzotti:2003ni,Aoki:2004ta}. Note, that
the bare mass is not protected against additive renormalization and
therefore the mixing angle $\alpha$ gets renormalized. In order to
achieve maximal renormalized twist, the bare mass needs to be
tuned. This tuning is routinely done as part of any twisted mass
calculation (see e.g. \cite{Baron:2011sf}).

Introducing a pair of non-degenerate quarks is usually done with the
help of an additional mass term that carries a nondiagonal flavor
structure $\tau_1$ resulting in a non-degenerate 2-flavor fermion
operator of the form
\begin{equation}
\label{eq:tmndop}
\hat{M}_{tm}=\gamma_\mu D_\mu+\frac{r}{2}\Box+m+i\mu\gamma_5\tau_3+\epsilon \tau_1
\end{equation}
\cite{Frezzotti:2003xj,Chiarappa:2006ae}.  An alternative suggestion
has been proposed by \cite{Pena:2004gb}.

From (\ref{eq:tmop}) it follows, that the two flavor operator has a
determinant which is bounded from below by $\mu^2$
\cite{Frezzotti:2000nk} and a spectral gap of size $\mu$ around the
real axis \cite{Gattringer:2005vp}. Furthermore, the twisted mass
fermion matrix fulfills a generalized form of the
$\gamma_5$-hermiticity  condition of the Wilson operator
(\ref{eq:g5hermw}) 
\begin{equation}
\label{eq:g5hermtm}
M_{tm} ^\dag=\gamma_5\tau_1 M_{tm} \tau_1 \gamma_5
\end{equation}
that similarly implies the appearance of eigenmodes in complex
conjugate pairs. In contrast to Wilson fermions however there are no
real eigenmodes due to the spectral gap \cite{Gattringer:2005vp} and
therefore no exceptional configurations. Note that (\ref{eq:g5hermtm})
also holds for the non-degenerate $\hat{M}_{tm}$ from
(\ref{eq:tmndop}).

Numerical evidence has been found with twisted mass fermions for a
line of first order phase transition in the bare quark mass that
extends into the twisted mass direction
\cite{Farchioni:2004us,Farchioni:2004fs}. This observation can be
understood in terms of the phase structure of lattice QCD with Wilson
fermions as proposed by \cite{Sharpe:1998xm} from analysis of the
effective chiral potential. It represents one of two possibilities of
finding a minimum - the other one being the appearance of an
unphysical phase where parity and flavor are spontaneously broken
\cite{Aoki:1983qi,Aoki:1996af} . Evidence for the Aoki phase was found
at coarser lattice spacings \cite{Ilgenfritz:2003gw,Sternbeck:2003gy}.

In general, it is mandatory for all simulations with Wilson-type
fermions to avoid being too close to the line of first order phase
transition. The situation is particularly challenging to twisted mass
fermions at maximal twist however since the line of first order phase
transition opens up in the twisted mass direction at the critical bare
mass. The minimum pion mass one can reach with twisted mass fermions
at a given lattice spacing is estimated by \cite{Shindler:2007vp} to
be $\sim 300\text{~MeV}$ at $a\sim0.07-0.1\text{~fm}$ which is roughly
consistent with recent numerical results \cite{Baron:2010bv}.

For further details about twisted mass fermions we refer the
interested reader to a  recent review \cite{Shindler:2007vp}.

\subsubsection{Chirally symmetric fermions}
\label{sect:chsyf}

Although the Nielsen-Ninomiya theorem does not allow one to retain the
continuum form of chiral symmetry for local, doubler free fermions, an
exact symmetry may be found at finite lattice spacing that goes over
into the continuum chiral symmetry upon removal of the cutoff
\cite{Ginsparg:1981bj,Luscher:1998pqa,Hasenfratz:1998ri,Narayanan:1994gw}.
The continuum form of chiral symmetry implies that the massless
fermion operator $M$ does anticommute with $\gamma_5$. One can
generalize that relation by introducing a modified
\begin{equation}
\label{eq:g5hat}
\hat\gamma_5=\gamma_5\left(
1-2aRM
\right)
\end{equation}
and demanding that
\begin{equation}
\label{eq:gwr}
\gamma_5 M+M\hat\gamma_5=0
\end{equation}
The condition (\ref{eq:gwr}) is known as the Ginsparg-Wilson relation
\cite{Ginsparg:1981bj} and the operator $R$ in (\ref{eq:gwr}) has to
be local and is known as the Ginsparg-Wilson kernel.  From
(\ref{eq:g5hat}) one can see that the action
\begin{equation}
\label{eq:gwac}
S_\text{GW}=\bar\Psi M \Psi
\end{equation}
is invariant under the chiral symmetry
\begin{equation}
\label{eq:gwsym}
\begin{split}
\bar\Psi&\rightarrow\bar\Psi \left(1+i\gamma_5\right)\\
\Psi&\rightarrow \left(1+i\hat\gamma_5\right) \Psi 
\end{split}
\end{equation}
The continuum form of chiral symmetry can be regained in observables
by constructing them with the chirally rotated fermion field
\begin{equation}
\label{eq:chirot}
\hat\Psi=\tilde 1\Psi
\qquad
\tilde 1=1-aRM
\end{equation}
instead of the bare $\Psi$. As the full chiral symmetry is preserved
by Ginsparg-Wilson fermions, all consequences of this symmetry such as
the appearance of exact zero modes, an exactly conserved axial current
\cite{Kikukawa:1998py} and the anomalous breaking of the flavor
singlet part of the symmetry are also retained. It has been proven by
\cite{Horvath:1998cm,Bietenholz:1999dg} that chirally symmetric
lattice fermion operators can not be ultralocal, i.e. they can not be
realized by couplings to a finite number of nearest neighbors. It is
therefore necessary to prove the locality of chiral fermion actions in
the sense that the coupling decreases exponentially with distance with
an exponent that is on the order of the cutoff and not a physical
mass.

Different fermion operators fulfilling the Ginsparg-Wilson relation
have been suggested. The overlap operator
\cite{Narayanan:1992wx,Narayanan:1993sk,Narayanan:1993ss,Narayanan:1994gw,Neuberger:1997fp,Neuberger:1998wv}
is an explicit construction that corresponds to the unitary part of a Wilson operator
at negative bare mass $-\rho$. In the massless case, the fermion matrix is
given by
\begin{equation}
\label{eq:ovm0}
M_o=\rho\left(1+\frac{M_W(-\rho)}{\sqrt{M^\dag_W(-\rho) M_W(-\rho)}}\right)
\end{equation}
and a real mass term may be added as
\begin{equation}
\label{eq:ovm}
M_o(m)=M_o+\tilde 1 m
\end{equation}
The overlap operator (\ref{eq:ovm0}) obeys the Ginsparg-Wilson
relation with an ultralocal $R=\frac{1}{2\rho}$. Locality of the
overlap operator has been established numerically
\cite{Hernandez:1998et} (see also \cite{Golterman:2003qe}). Note that
by construction the overlap operator is normal.

Overlap fermions are numerically extremely demanding. In contrast to
all previously discussed fermion discretizations, the fermion operator
$M_o$ is not sparse. The multiplication of $M_o$ on a vector has to
proceed via approximating the inverse matrix square-root in
(\ref{eq:ovm0}). While it is possible to do this with polynomial
\cite{Giusti:2002sm} or rational
\cite{Neuberger:1998my,Edwards:1998yw,vandenEshof:2002ms}
approximations, it typically requires at least $O(100)$ applications
of the kernel Wilson operator to perform one matrix-vector
multiplication with $M_o$.

Due to the exact chiral symmetry, overlap fermions are free of
exceptional configurations at finite mass. This leads however to a
nontrivial 
technical problem for dynamical overlap fermions that was first
observed by \cite{Fodor:2003bh}. The nonanalyticity
of (\ref{eq:ovm0}) implies a divergence of the fermionic force term
(\ref{eq:fforce}) at certain points in configuration
space. Specifically, such a divergence occurs at topological sector
boundaries where the number of zero modes changes. These points have
to be treated separately in the HMC integration
\cite{Fodor:2003bh,Cundy:2005pi,Egri:2005kp} specifically at fine
lattices, as the simulation can get stuck in one topological
sector.\footnote{Note that while the nonanalyticity problem is
  particularly apparent for overlap quarks its origin is
  physical. Upon removing the cutoff, every fermion formulation should
ultimately develop nonanalyticities at the topological sector
boundaries.}
 Alternatively, one can artificially constrain the simulation
to a single topological sector via the addition of extra Wilson
fermions with large negative mass
\cite{Izubuchi:2002pq,Vranas:2006zk,Fukaya:2006vs} and treat this
constraint as an additional finite volume effect.

Historically, overlap fermions were first formulated in five
dimensions based on the realization that one may have chiral domain
wall defects in a $2n+1$ dimensional vector-like gauge theory
\cite{Callan:1984sa,Frolov:1992ck,Kaplan:1992bt}. This five
dimensional form of chiral fermions was further developed by
\cite{Shamir:1993zy} and is known as domain wall fermions.  Domain
wall fermions do have an exact chiral symmetry only in the limit that
the fifth dimension is large. On the lattice, this can of course not
be realized and there is a remnant breaking of chiral symmetry
\cite{Blum:1996jf} (for a recent update on the size of this effect see
e.g. \cite{Aoki:2010dy}).

On a technical level, domain wall fermions are realized by
5-dimensional Wilson fermions at a negative bare mass M. The gauge
field is 4-dimensional only and identical for each slice in the fifth
dimension. Along the fifth dimension $s$, gauge links are set to
$U_5(x,s)=\mathbbm{1}$ for $s\neq 0$ except at the defect location
$s=0$, where they are set to $U_5(x,0)=\mathbbm{1}m$. According to
\cite{Shamir:1993zy}, a left resp. a right handed chiral mode will
form at the positive resp. negative side of the defect in the limit of
an infinite fifth dimension and these modes couple with the mass term
$m$. A remnant coupling of the chiral modes through the bulk will
appear for a finite fifth dimension that will be suppressed
exponentially in the size of the fifth dimension $N_5$.

Due to the residual chiral symmetry breaking, domain wall fermions do
suffer a small additive mass renormalization known as residual mass
$m_\text{res}$. Like in the case of Wilson fermions it is therefore
necessary in principle to take negative bare mass values for reaching
arbitrarily small but positive renormalized quark masses. One could
therefore encounter exceptional configurations, but due to the
smallness of $m_\text{res}$ this is not a problem in current
simulations. The extent of the fifth dimension is typically around
$N_5=16$ in present day calculations rendering domain wall fermions
numerically more expensive than Wilson fermions by about an order of
magnitude.

Another variant of chirally symmetric lattice fermions operators is
known as perfect action or fixed point fermions
\cite{Hasenfratz:1993sp,DeGrand:1995ji,Bietenholz:1995cy}. Perfect
actions are obtained by following the renormalization group flow of a
blocking transformation to the renormalized trajectory that ends in a
fixed point. Therefore, their form is not explicitly given but needs
to be determined by following the renormalization group flow. Up to
truncation errors, the action so obtained is classically perfect in
the sense that it has no remaining cutoff effects in the classical
theory (see sect.~\ref{sect:fimp} for a more detailed discussion).

From a numerical perspective, fixed point actions are expensive to
simulate. In principle, fixed point fermion operators are not sparse
matrices and neither can they be explicitly constructed out of a
sparse matrix as is the case for overlap fermions. Consequently, one
needs to truncate the operator to a finite range and the chiral
symmetry is only approximate. The resulting additive mass
renormalization is small however and no problems with exceptional
configurations have been seen \cite{Gattringer:2003qx}. The same paper
also reports that the numerical cost is increases between one and two
orders of magnitude compared to Wilson fermions.  For a review of
truncated perfect action fermios see \cite{Bietenholz:2006ni}.

Yet another variant of approximately chiral lattice fermions is
obtained by inserting a truncated expansion of a general fermion
operator into the Ginsparg-Wilson relation (\ref{eq:gwr}) and
explicitly solving for the expansion coefficients
\cite{Gattringer:2000js}.  Numerical properties and cost of this
variant of approximately chiral fermions are roughly comparable to
those of the truncated perfect action as demonstrated in
\cite{Gattringer:2003qx}.

\subsection{Constructing efficient regularizations}
\label{sect:improve}

As mentioned in sect.~\ref{sect:lqcd}, lattice discretizations of
continuum actions are not unique. The essential step in
discretizing a continuum action is the replacement of derivative
terms by lattice finite difference operators. Disregarding
quantum effects, it is easy to see how the discretization of
derivative operators can be systematically improved by adding
finite difference operators with increasing distances. For the
simple case of one-dimensional symmetric difference operators
\begin{equation}
\begin{split}
\Delta_1
f(x)&=\frac{f(x+a)-f(x-a)}{2a}\\
&=f^\prime(x)+\frac{a^2}{6}f^{\prime\prime\prime}(x)+O(a^4)\\
\Delta_2
f(x)&=\frac{f(x+2a)-f(x-2a)}{4a}\\
&=f^\prime(x)+\frac{4a^2}{6}f^{\prime\prime\prime}(x)+O(a^4)
\end{split}
\end{equation}
we find discretization errors of $O(a^2)$. In the linear combination
\begin{equation}
\Delta^i=\frac{4 \Delta_1-\Delta_2}{3}
\end{equation}
however the $O(a^2)$ terms
cancel and therefore 
\begin{equation}
\Delta^i f(x)=f^\prime(x)+O(a^4)
\end{equation}
Generally speaking, one can systematically improve discretized
continuum operators by taking liner combinations of lattice operators
and imposing conditions on the coefficients such that the continuum
limit is correct and leading order discretization effects are
cancelled.

Finding the proper coefficients in the linear combination is of course
not always as trivial as in the example above. Specifically, one needs
to keep in mind that when computing observables in a quantum theory,
the classically computed coefficients can receive radiative
corrections. One can then look at a certain set of observables and try
to cancel higher order effects in them on a quantum level. Such a
strategy has first been suggested by
\cite{Symanzik:1983dc,Symanzik:1983gh} who realized that a lattice
Lagrangian in general is equivalent order by order in $a$ and $g^2$ to
a continuum local effective Lagrangian.

Other strategies of finding improved discretizations are based on the
mean field approximation or the renormalization group as will be
detailed below. One common feature that all of the methods share is
that they use the freedom in defining a lattice regularization of a
continuum operator in order to suppress unphysical UV fluctuations in
the lattice action. There is no a priori guide for determining which
specific improvement out of the rather large number of possibilities
is optimal. It is therefore essential to consider the potential
benefits of a specific improvement in relation to its computational
cost. In the end, the optimal action will be the one that provides the
smallest error (including all systematics) on the physical observables
(i.e. the hadron masses) for a given amount of available computer
time. To find a good improvement strategy one therefore needs to find
the right balance of different improvements such that the overall
error is minimized.

\subsubsection{Gauge field improvement} 
\label{sect:gimp}

\begin{figure}
\includegraphics[width=0.5\textwidth]{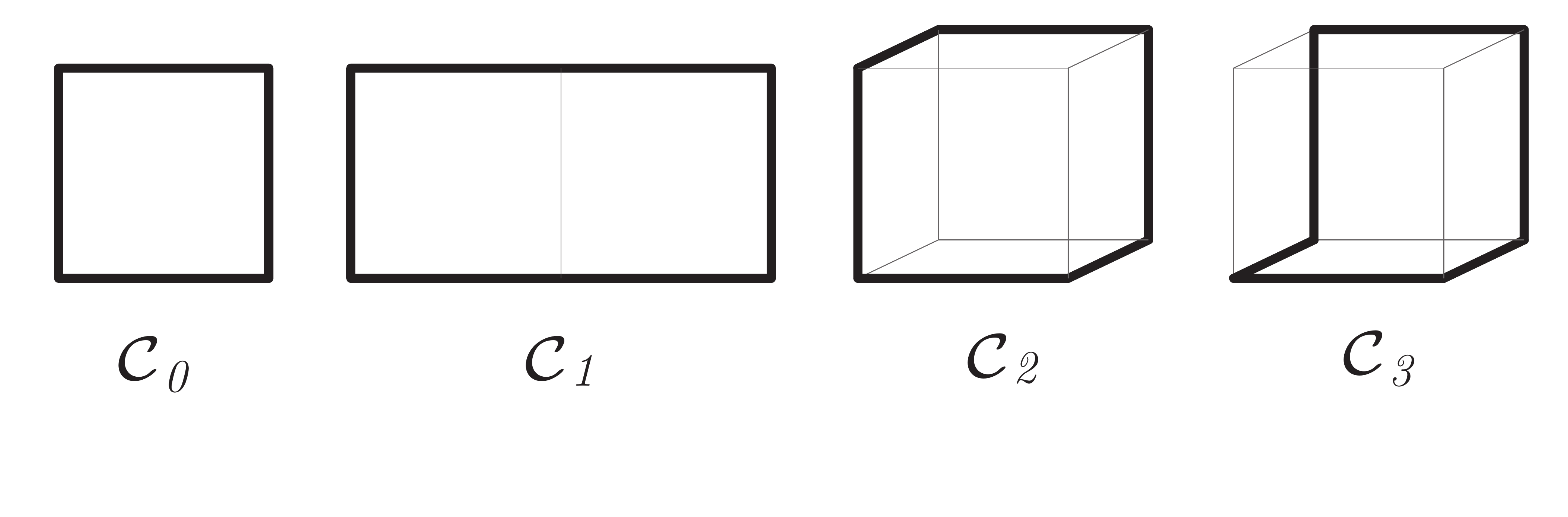}
\caption{
\label{fig:plaq6}
The four possible forms of closed gauge loops on a hypercubic lattice
with 6 or less gauge links corresponding to the members of the sets
$\mathcal{C}_0,\ldots,\mathcal{C}_3$ as described in the text.
}
\end{figure}

The simple Wilson gauge action (\ref{eq:wg}) only contains the
elementary plaquette and has discretization errors of $O(a^2)$. As
demonstrated by \cite{Weisz:1982zw,Luscher:1984xn,Luscher:1984xne},
one can improve the scaling by taking proper linear combinations of
the elementary Wilson plaquette and more extended gauge loops. The
coefficients must of course be chosen such that in the continuum limit
one still obtains the correct continuum gauge action
(\ref{eq:cga}). In addition however, one can demand that all
corrections that are of $O(a^2)$ vanish classically. The desired
continuum action has the form
\begin{equation}
\mathcal{O}_0=
\text{Tr}\left(
F_{\mu\nu}F_{\mu\nu}
\right)
\end{equation}
All together,
there are three different terms that represent possible $O(a^2)$
corrections to this form
\begin{equation}
\label{eq:o6terms}
\begin{split}
\mathcal{O}_1&=
\text{Tr}\left(
D_\mu F_{\mu\nu} D_\mu F_{\mu\nu}
\right)\\
\mathcal{O}_2&=
\text{Tr}\left(
D_\sigma F_{\mu\nu} D_\sigma F_{\mu\nu}
\right)\\
\mathcal{O}_3&=
\text{Tr}\left(
D_\mu F_{\mu\nu} D_\sigma F_{\sigma\nu}
\right)
\end{split}
\end{equation}
where the covariant derivative $D_\mu$ is given by
(\ref{eq:coder}). In order to reach classical improvement, we have to
demand that these additional terms vanish.

On the lattice, a gauge action can generically be written as a sum of
terms of the form
\begin{equation}
\label{eq:gg}
S_i=\beta\sum_{\mathcal{P}\in \mathcal{C}_i}\left(1-\frac{1}{3}
\text{ReTr}(U(\mathcal{P})\right)
\end{equation}
where the $U(\mathcal{P})$ are path ordered products of gauge links
along a closed path $\mathcal{P}$. For the simple Wilson gauge action
(\ref{eq:wg}), which we will refer to as $S_0$ in this context, the
set of closed gauge loops $\mathcal{C}_0$ consists of all elementary
plaquettes. Going one order higher, we see that there are also three
possible forms of 6-link loops that are the minimal extensions of the
elementary plaquette. The first one of them is the planar $2\times 1$
loop while the other ones extend into the elementary hypercube (see
fig.~\ref{fig:plaq6}). Not distinguishing between loops that differ
only by rotation, we label the sets of loops of these different forms
$\mathcal{C}_1,\mathcal{C}_2,\mathcal{C}_3$.  The expansion of the
corresponding gauge actions (\ref{eq:gg}) in terms of
continuum operators (\ref{eq:o6terms}) up to next-to-leading order reads
\cite{Luscher:1984xn,Luscher:1984xne}
\begin{equation}
\label{eq:o6paths}
\begin{split}
S_0&=-\frac{1}{4}\mathcal{O}_0+\frac{1}{24}\mathcal{O}_1+\ldots\\
S_1&=-2 \mathcal{O}_0+\frac{5}{6}\mathcal{O}_1+\ldots\\
S_2&=-2 \mathcal{O}_0-\frac{1}{6}\mathcal{O}_1+\frac{1}{6}\mathcal{O}_2+\frac{1}{6}\mathcal{O}_3+\ldots\\
S_3&=-4 \mathcal{O}_0+\frac{1}{6}\mathcal{O}_1+\frac{1}{2}\mathcal{O}_3+\ldots
\end{split}
\end{equation}
where volume sums are implied. For a  general linear combination
\begin{equation}
\label{eq:genga}
S=\sum_{i=0}^4 c_i S_i
\end{equation}
the correct continuum limit is therefore imposed by the condition
\begin{equation}
\label{eq:coco}
c_0+8c_1+8c_2+16c_3=1
\end{equation}
The cancellation of all the higher order operators (\ref{eq:o6terms})
may be achieved by imposing
\begin{equation}
\label{eq:tllwc}
c_0+20c_1=0
\qquad
c_2=c_3=0
\end{equation}
which leads to a choice of coefficients $c_0=5/3$ and $c_1=-1/12$. The
resulting action is known as tree level L\"uscher-Weisz action. It
realizes the tree level (i.e. classical) $O(a^2)$ improvement of the
Wilson plaquette gauge action (\ref{eq:wg}). Even before the work of
L\"uscher and Weisz these coefficients have been found by
\cite{Curci:1983an} with a different matching procedure.

As mentioned above, in a quantum theory radiative corrections can in
general reintroduce $O(a^2)$ terms into observables. Generically,
these corrections are proportional to $g^2$ such that the tree level
L\"uscher-Weisz action is correct up to $O(g^2a^2)$ terms as opposed
to $O(a^4)$ in the classical theory. In order to correct for these
radiative effects, one may look at different on-shell observables.
Looking at the scattering amplitudes of massive gluons one can
construct an action with $O(g^4a^2)$ scaling by modifying $c_0$, $c_1$
and $c_2$ with $O(g^2)$ terms \cite{Luscher:1985zq}. The resulting
coefficients read $c_0=5/3+0.237 g^2$, $c_1=-1/12-0.02521g^2$ and
$c_2=-0.00441g^2$. Note that the one loop coefficients are explicitly
scale dependent via $g^2$.

It is of course possible to go beyond perturbative Symanzik
improvement and concentrate on specific discretization terms that are
found to be important in nonperturbative calculations. One case of
particular relevance was found by
\cite{Parisi:1980pe,Lepage:1992xa}. As the gauge field is represented
on the lattice in exponentiated form
\begin{equation}
\label{eq:gfe}
\begin{split}
U_\mu(x)&=e^{igaA_\mu(x)}\\
&=1+igaA_\mu(x)-\frac{g^2a^2}{2}A_\mu^2(x)+\ldots
\end{split}
\end{equation}
the theory contains vertices with an arbitrary number of
gluons that give rise to tadpole diagrams. Although these
are formally suppressed by powers of $a$, UV divergences in the
tadpole loops apear and these terms are in fact scaling as powers of
$g^2$ instead.

This problem may be adressed in mean field theory by redefining the
relation between the lattice $U_\mu$ and the continuum
$A_\mu$. Formally writing the gauge field $U_\mu$ as a product of an
IR and an UV part
\begin{equation}
\label{eq:uvir}
U_\mu=U_\mu^\text{(UV)}U_\mu^\text{(IR)}
\end{equation}
and replacing the UV part by its mean field value $u_0$, one can now
identify the physically relevant IR part $U_\mu^\text{(IR)}$ of the
lattice gauge field with the continuum field obtaining a relation
\begin{equation}
\label{eq:tadi}
U_\mu=u_0 e^{igaA_\mu(x)}
\end{equation}
between the lattice gauge field $U_\mu$ and the continuum $A_\mu$. One
can therefore implement tadpole improvement by replacing all gauge
links $U_\mu$ in lattice operators by $U_\mu/u_0$.  Because UV
fluctuations are dominant, one can obtain a good estimate of $u_0$ by
simply taking the expectation value of the trace of a gauge link in a
fixed gauge or, alternatively, defining
\begin{equation}
\label{eq:tadpl}
u_0=\left\langle\frac{1}{3}\text{ReTr}(U_{\mu\nu})\right\rangle^{1/4}
\end{equation}
In both cases $u_0$ can be either determined in perturbation theory or
by measuring it directly as part of a nonperturbative calculation. In
the latter case, care has to be taken to determine $u_0$
self-consistently because it is both a parameter in the action and an
observable.

There is a third, independent strategy of improving the gauge action
that is based on the renormalization group. As already mentioned in
sect.~\ref{sect:chsyf}, one may try to follow the renormalization
group flow of a blocking transformation in the space of all possible
actions towards the renormalized trajectory. The renormalized
trajectory is the trajectory of the renormalization group flow that
starts from a fixed point at the critical surface where the
correlation length diverges. Any point along the renormalized
trajectory therefore corresponds to the action at a certain finite
correlation length that has vanishing irrelevant operator
contributions and therefore reproduces continuum physics without
cutoff effects. Actions along the renormalized trajectory are thus
called perfect actions.

The exact position of the renormalized trajectory is elusive since the
renormalization flow is not known analytically. Strategies of finding
approximations are based on the fact that the renormalized trajectory
is attractive under blocking transformations as they reduce irrelevant
operators. One should also keep in mind that a perfect action
generally lives in an infinite dimensional space of couplings that has
to be truncated for practical purposes and that a thus truncated
perfect action is not guaranteed to exhibit the smallest scaling
violations possible among actions living in that subspace.

Studying the repeated application of a blocking transformation in a
truncated subspace of gauge loops, \cite{Iwasaki:1983ck} suggested an
action of the L\"uscher-Weisz form (\ref{eq:genga}) but with a set of
coefficients $c_0=3.648$, $c_1=-0.331$, and $c_2=c_3=0$. Also within
the same truncation scheme, the doubly blocked Wilson (DBW2) action
\cite{Takaishi:1996xj,deForcrand:1999bi} has been obtained by double
blocking from Wilson configurations. It has the coefficients
$c_0=12.2704$, $c_1=-1.4088$, and $c_2=c_3=0$. Note that these
coefficients do not have an explicit scale dependence and therefore
can only cancel quantum effects at the scale where they are computed.

A different strategy has been followed by
\cite{Hasenfratz:1993sp,DeGrand:1995ji} who obtained a classically
perfect action by a saddle point integration around $g=0$ followed by
a truncation to a rather large set of couplings.

\subsubsection{Fermion field improvement}
\label{sect:fimp}

In the case of lattice gauge actions, the Wilson plaquette action
(\ref{eq:wg}) provided a unique starting point for all improvement
efforts. In the case of fermion actions the range of unimproved
actions is quite diverse and so are their major discretization
effects. There are some common improvements that positively affect all
fermion actions, but the improvement strategies are sufficiently
distinct for different discretizations so we will discuss them
separately. We start by first discussing the improvement of Wilson
fermions.

The Wilson fermion action (\ref{eq:wilsonac}) has leading
discretization effects of $O(a)$. These can be cancelled classically
by adding a two-hop term to the action \cite{Hamber:1983qa}
\begin{equation}
\label{eq:hwac}
S_\text{HW}=\bar\Psi\left(\gamma_\mu (2D_\mu-{\hat D}_\mu)+
\frac{r}{2}(2\Box-{\hat\Box})
+m\right)\Psi
\end{equation}
with $\hat\Box=\sum_\mu\hat{C}_\mu$ and the covariant two-hop operators
from (\ref{eq:thop}). One could in principle compute the coefficients
of the one-hop and two-hop terms in perturbation theory or
nonperturbatively to achieve further improvement. This was not further
pursued in practice however since \cite{Sheikholeslami:1985ij}
discovered that one can remove $O(a)$ discretization terms with a more
local operator. This additional term is the discretized magnetic
moment operator and the corresponding action reads
\begin{equation}
\label{eq:swac}
S_\text{SW}=S_W
-\frac{rc_\text{SW}}{2}\sum_{\mu<\nu}
\bar\Psi
\sigma_{\mu\nu}F_{\mu\nu}
\Psi
\end{equation}
where the field strength $F_{\mu\nu}(x)$ is usually obtained by taking
an average of the imaginary parts of all the plaquette around
the point $x$. Due to the arrangement of the four plaquettes involved
in the average, the additional term and the resulting action are
commonly referred to as the clover term and the clover action. Tree
level improvement is achieved by setting the coefficient
$c_\text{SW}=1$. The resulting action has discretization effects of
$O(g^2a)$ and $O(a^2)$. Numerically, the $O(g^2a)$ and the $O(a^2)$
corrections are competing and it is not possible to say a priori which
of these effects are dominant for a specific observable at a specific
lattice spacing. Further improvement is possible by either
perturbation theory \cite{Wohlert:1987rf,Luscher:1996vw}, mean field
tadpole improvement or nonperturbative methods \cite{Luscher:1996ug}.
It turns out that the clover action (\ref{eq:swac}) is equivalent to
the Hamber-Wu action (\ref{eq:hwac}) up to $O(a^2)$ terms in terms of
rotated fermion fields \cite{Heatlie:1990kg,Martinelli:1990ny}.

Although suggestions exist in the literature for improving the Wilson
operator further by adding more extended terms
\cite{Alford:1996nx,DeGrand:1998pr,Durr:2010ch}, it is not much
pursued due to the computational overhead.

In contrast to Wilson fermions the staggered action has leading
discretization effects of $O(a^2)$. These can be eliminated on a
classical level by replacing the one-hop derivative of the staggered
action (\ref{eq:stagfa}) with a suitable combination of one-hop and
three-hop derivative operators such that $O(a^2)$ terms cancel
\cite{Naik:1986bn}. The three-hop term is used rather than the two-hop
term in order not to interfere with the staggered flavor structure.

The main concern for staggered fermions however is not the Symanzik
improvement but rather the minimization of taste breaking
effects. Since the different fermion components sit at the edges of
the Brillouin zone, they interact via the exchange of hard gluons with
momenta on the order of the cutoff scale. Suppressing these
interactions, i.e. reducing the unphysical UV fluctuations, is
therefore especially important for staggered fermions.

\begin{figure}
\includegraphics[width=0.5\textwidth]{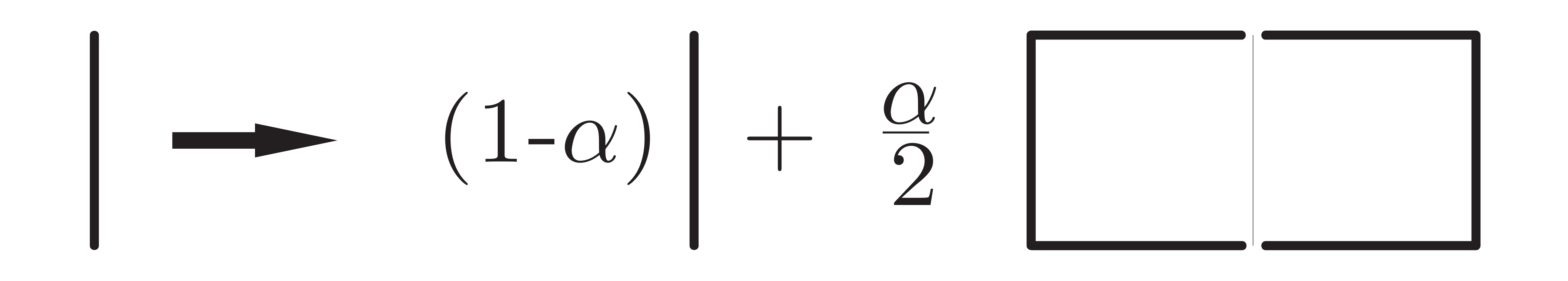}
\caption{
\label{fig:apesmear}
The principle of APE smearing displayed in the two dimensional
case. The ``thin'' gauge link is replaced by a weighted average over
the gauge link and the staples, which is then usually backprojected onto
the gauge group.
}
\end{figure}

The primary method in use today for reducing unphysical UV noise is
link smearing (also known as UV filtering or fattening). Since link
smearing is used for Wilson-type and chirally symmetric fermion
actions as well, we will discuss it in a more general context.
Although on a technical level it can be implemented by modifying the
gauge fields only, it is important to remember that it strictly is a
modification of the fermion action only. The original suggestion, put
forward by the APE collaboration \cite{Albanese:1987ds}, is commonly
known as APE smearing. The basic idea is that one can use the freedom
in defining the parallel transport in the covariant one-hop term
(\ref{eq:hop}) of any fermion operator to suppress UV fluctuations. It
is not necessary that one takes the same gauge link than used in the
gauge action but instead a linear combination of various paths that
have the correct starting and end points. In the case of APE smearing
these paths are, in addition to the original gauge link, all three
link connections of the same two points (see fig.~\ref{fig:apesmear})
usually referred to as staples.\footnote{Note that in the literature
  the sum over all three-link connections is also sometimes referred
  to as the staple.}

One can define an APE smeared gauge link $U^\text{(APE)}_\mu(x)$ from
the original gauge links $U_\mu(x)$ in d dimensions via
\begin{equation}
\label{eq:apesmear}
U^\text{(APE)}_\mu(x)=
(1-\alpha) U_\mu(x)+
\frac{\alpha}{2(d-1)}
\Omega_\mu(x)
\end{equation}
where we used the staple sum
\begin{equation}
\label{eq:stapsum}
\Omega_\mu(x)=
\sum_{\pm\nu\neq\mu}
U_\nu(x)
U_\mu(x+\hat\nu)
U^\dag_\nu(x+\hat\mu)
\end{equation}
with the identity $U_{-\mu}(x)=U^\dag_\mu(x-\hat\mu)$. The smearing
parameter $\alpha$ determines the relative weight of the staple vs.
the original link and is typically set to a value $\alpha\sim 0.6$.
The resulting gauge links $U^\text{(APE)}_\mu(x)$ are no more an
element of the gauge group and it is therefore customary to
backproject them onto the gauge group via
\begin{equation}
\label{eq:backproj}
 U^\prime=\frac{U^\text{(APE)}}{\sqrt{{U^\text{(APE)}}^\dag U^\text{(APE)}}}
\qquad
\hat{U}=\frac{U^\prime}{\left(\det{U^\prime}\right)^{1/3}}
\end{equation}
Where $U^\prime$ is unitary and $\hat{U}$ also has unit
determinant.\footnote{For an alternative suggestion on doing the
  backprojection see \cite{Durr:2010ch}.}

As the backprojection (\ref{eq:backproj}) is not analytic, the
fermionic force term (\ref{eq:fforce}) in the pseudofermion field
integration may exhibit singularities. Although there are suggestions
to remedy this situation by leaving out the second step of the
backprojection (\ref{eq:backproj}) and use $U^\prime$ only
\cite{Hasenfratz:2007rf}, it is customary in dynamical simulations to
use the analytic link smearing suggested by
\cite{Morningstar:2003gk}. They define the so-called stout link
as\footnote{We use the term stout link as it is commonly done in the
  literature. Note however that in the original paper the term stout
  link was used in a slightly different way.}
\begin{equation}
\label{eq:stout}
V_\mu(x)=e^{\rho S_\mu(x)}U_\mu(x)
\end{equation}
where
\begin{equation}
S_\mu(x)=\frac{1}{2}\left(A_\mu(x)-\frac{1}{N}\text{Tr}A_\mu(x)\right)
\end{equation}
with
\begin{equation}
A_\mu(x)=\Omega_\mu(x)U^\dag_\mu(x)-U_\mu(x)\Omega^\dag_\mu(x)
\end{equation}
The parameter $\rho$ is a smearing parameter that, similarly to
$\alpha$ in the case of APE smearing, determines the relative weights
of the original link and the staple. For small smearing parameters,
APE and stout link smearing are equivalent if one sets
$\alpha=2(d-1)\rho$ \cite{Capitani:2006ni}. With a proper matching of
the smearing parameters one can find a close correspondence even if
their values are large \cite{Hasenfratz:2007rf}.

\begin{figure}
\includegraphics[width=0.5\textwidth]{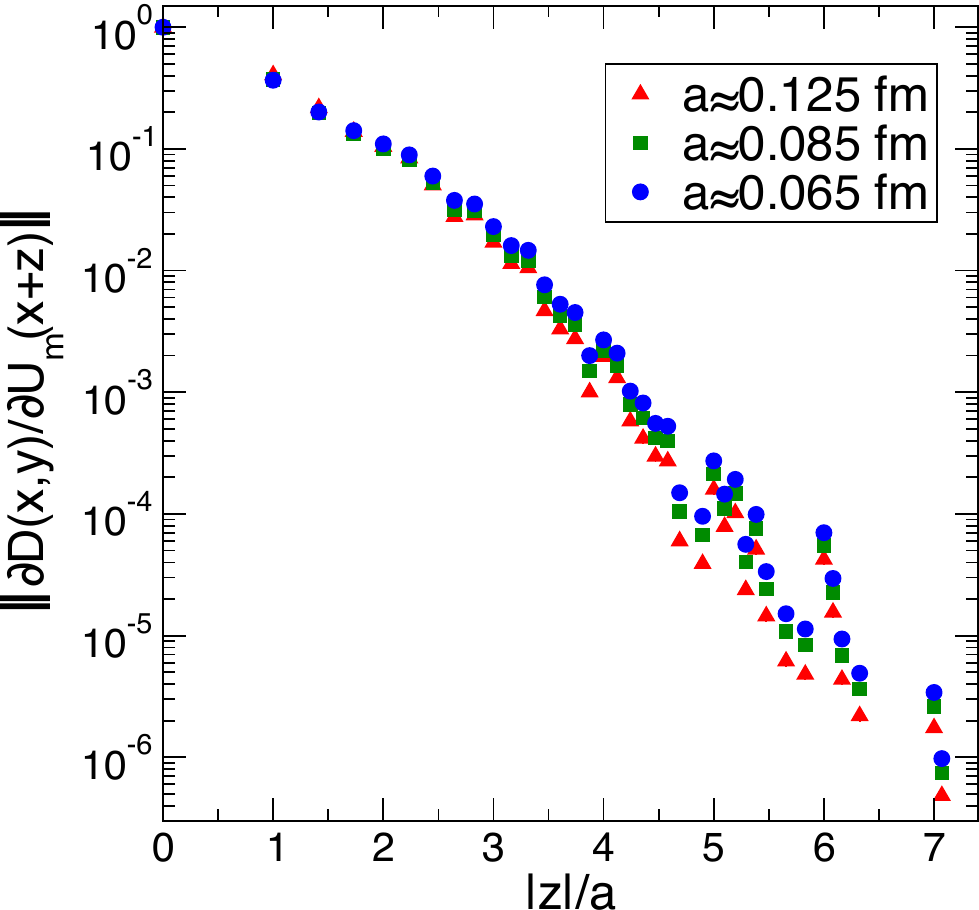}
\caption{
\label{fig:smearloc}
Locality of a fermion operator coupling to 6-times stout smeared gauge
field from \cite{Durr:2008zz}. The stout smearing parameter is set to
$\rho=0.11$. At small distances the coupling decreases exponentially
with an effective mass of $\sim 2.2 a^{-1}$ that is proportional to
the lattice cutoff. At euclidean distances larger than $\sqrt{50}a$
couplings are zero due to the ultralocality of the smearing procedure.}
\end{figure}

Both APE and stout smearing, as in general all link smearing
techniques, generate a smeared gauge field from the original one, wich
is usually called thin link. It is therefore straightforward to apply
the smearing prescription repeatedly on the already smeared links and
use this multiply smeared gauge link field for constructing the
fermionic operator. As long as the smearing parameter and the number
of smearing steps is held constant, it amounts to an ultralocal
redefinition of the fermion operator and does not affect the continuum
limit. In fact, the locality range of an ultralocal fermion operator
itself is not at all affected by smearing the gauge links. Gauge link
smearing does not introduce any new couplings into the fermion
operator. What is affected by gauge link smearing is the fermion to
gauge field coupling which becomes more extended. For smeared gauge
links the fermion matrix elements are affected by changes of the
original gauge field at further distances. If one keeps the number of
smearing steps constant when going to the continuum limit however,
this redefinition is still ultralocal. In addition if the smearing
parameter is not excessive, one expects an exponential decrease of the
gauge field to fermion coupling within the ultralocality range with an
exponent that is proportional to the cutoff. This has been numerically
demonstrated for a 6-times stout link smeared tree level improved
Wilson operator in \cite{Durr:2008zz} (see fig.~\ref{fig:smearloc}).

A variant of APE link smearing where one tries to maximize the
smearing while only taking into account links that have a distance of
at most one single lattice unit to the original link is known as
hypercubic (HYP) smearing \cite{Hasenfratz:2001hp}. One step of HYP
smearing combines three steps of APE smearing where in the first two
APE steps all dimensions that would give distance two contributions in
any one direction are disregarded in forming the staple sum. The
analytic version of HYP smearing along the lines of
\cite{Morningstar:2003gk} is known as hypercubic exponential (HEX)
smearing \cite{Capitani:2006ni}.

In the context of Wilson-type fermions smearing has been found to
drastically reduce the additive mass renormalization, especially in
combination with $O(a)$ improvement
\cite{Capitani:2006ni}. Furthermore, renormalization constants and the
value of the improvement coefficient $c_\text{SW}$ from
(\ref{eq:swac}) are much closer to their tree level values and the
normality of the operator is improved
\cite{Durr:2005an,Hoffmann:2007nm,Horsley:2008ap,Durr:2008rw}. Most
importantly, gauge link smearing reduces the fluctuations in the real
modes that lead to exceptional configurations
\cite{DeGrand:1998mn,Stephenson:1999ns}. Thus it is possible with
smeared link Wilson-type fermions to reach physical quark masses
\cite{Durr:2010aw}, which has proven to be difficult without link
smearing. These are clear indications that gauge link smearing
efficiently suppresses UV fluctuations and ameliorates the chiral
symmetry breaking that is inherent in Wilson type fermions.

One can combine nonperturbative improvement of the Wilson action with
smearing as is done e.g. in the SLINC (stout link improved
nonperturbative clover) action \cite{Horsley:2008ap}.

It is also possible to use different gauge link definitions for
different parts of the fermion operator. An operator of this form, the
so-called fat link irrelevant clover (FLIC) fermions suggested by
\cite{Zanotti:2001yb} use an $O(a)$ improved Wilson operator with
smeared links in the continuum irrelevant terms and original thin
links for the rest.

The improvements of Wilson type fermions carry over to their use as
kernel operators for the overlap construction. Overlap operators with
smeared Wilson kernels are generally cheaper computationally, have
renormalization constants closer to their tree level values and
require less fine tuning of the negative mass parameter $\rho$
\cite{Kovacs:2002nz,DeGrand:2002vu,Durr:2005an,Durr:2005ik}.

In the context of staggered fermions the main advantage of smeared
links is the suppression of taste violation
\cite{Blum:1996uf,Lepage:1997id,Lepage:1998vj,Lagae:1998pe,Orginos:1998ue,Orginos:1999cr}\footnote{For
  a study of reduced taste violation with an improved, unsmeared
  action see \cite{Bernard:1997mz}.}.  In the context of staggered
fermions, the staples used in APE smearing are referred to as fat3
staples. Adding to these staples ones that extend in a second lattice
direction and consist of five links one arrives at the so-called fat5
links and finally adding staples that extend in all three other
lattice directions and consist of seven links each one obtains smeared
gauge links referred to as fat7. Finally, staples that consist of five
links and extend two lattice spacings in one direction are referred to
as the Lepage term \cite{Lepage:1998vj}.  Adding a Naik term to the
staggered action, replacing the gauge links with fat7 gauge links plus
a Lepage term and tadpole improving the action one obtains the
so-called ``asqtad'' action that is frequently used in current
staggered fermion calculations and has scaling corrections of
$O(g^2a^2)$. More recently, two steps of fat7 link smearing, the first
one without the second one with Lepage term, with a gauge group
projection after the first smearing step were suggested by
\cite{Follana:2006rc}. Staggered fermions with this variant of link
smearing are known as ``highly improved staggered quarks'' or
HISQ.

Besides these smearing terms that were specifically designed to reduce
taste splitting, simple stout link smearing has also been applied to
the staggered fermion operator. Depending on the quark mass, the level
of taste splitting has found to be generally comparable to that of the
HISQ action or slightly less than for the specially designed asqtad
action \cite{Aoki:2009sc,Borsanyi:2010bp}.

\subsection{Anisotropic discretizations}
\label{sect:aniso}

As we will discuss in sect.~\ref{sect:extr}, information about excited
states can be extracted from correlation functions at short euclidean
distances. It is therefore desirable for excited hadron spectroscopy
to have a fine resolution in time direction while keeping the
resolution in space direction coarser in order to keep the overall
computational effort small.

Choosing an anisotropic discretization explicitly breaks the
hypercubic lattice remnant of the Lorentz symmetry down to the
subgroup of spatial cubic rotations. Consequently spatial and temporal
lattice extents receive different normalization and the renormalized
anisotropy generally differs from the input bare one.

Generalizing the simple case of the Wilson plaquette action
(\ref{eq:wg}) to include an anisotropy one obtains
\begin{widetext}
 \begin{equation}
\label{eq:wganiso}
S_{AW}=\beta\xi_0\sum_{x,i>j}\left(1-\frac{1}{3}
\text{ReTr}(U_{ij}(x))\right)\\
+
\frac{\beta}{\xi_0}\sum_{x,i}\left(1-\frac{1}{3}
\text{ReTr}(U_{i0}(x))\right)
\end{equation}
where $\xi_0=a^s_0/a^t_0$ is the bare anisotropy factor for the gauge
action, i.e. the ratio of bare spatial to temporal lattice
spacings. Similarly one can define an anisotropic Wilson fermion
operator as a generalization of the isotropic case (\ref{eq:wilsonop})
\begin{equation}
\label{eq:wopaniso}
\begin{split}
M_\text{aniso}=m+\nu_s\left(\gamma_i D_i+\frac{r}{2}\sum_i C_i\right)+
\nu_t\left(\gamma_0 D_0+\frac{r}{2} C_0\right)
\end{split}
\end{equation}
\end{widetext}
with $C_\mu$ defined in (\ref{eq:box}) and $\nu_{s/t}$ the speed of
light in spatial or temporal direction. In order to obtain the same
renormalized aspect ratio for both fermion and gauge actions one needs
to tune $\nu_s$ and $\nu_t$. An additional tuning of the gauge action
anisotropy is only required if one wants to tune to a specific
renormalized aspect ratio.

The improvement programme for gauge and fermion fields as outlined in
sect.~\ref{sect:improve} can be carried over to anisotropic lattices
if one keeps in mind the separation between spatial and temporal split
and ultimate tuning of both gauge and fermion field
anisotropies. Since spatial and temporal directions are in any case
treated differently for anisotropic lattices, one may even use
different improvements on spatial and temporal field components that
are specifically suited for either coarse or fine lattices. For
further details on implementation and parameter tuning of anisotropic
scalar, gauge and fermion actions see
e.g. \cite{Burgers:1987mb,Karsch:1989fu,Csikor:1998ui,Engels:1999tk,Alford:2000an,Klassen:1998ua,Chen:2000ej,Harada:2001ei,Umeda:2003pj,Morrin:2006tf,Edwards:2008ja}

\section{Extraction of hadron masses}
\label{sect:extr}

In sect.~\ref{sect:tech} we set up the framework for regularizing QCD
on a discrete spacetime lattice. In this section we discuss how to
extract the observables of interest, the hadron masses and energy
levels, from lattice QCD. The emphasis in this section is on the
technical details of extracting hadron masses in a nonperturbative
lattice QCD calculation. The question of how these measured hadron
masses can be turned into physical predictions is discussed in
sect.~\ref{sect:physpred}.

We start by discussing the basic concept of extracting energy levels
in lattice QCD in sect.~\ref{sect:lvl} with an emphasis on the
extraction of the ground state. The efficiency of this extraction of
energy levels depends on the choice of source and sink operators,
which is reviewed in sect.~\ref{sect:oper}. Finally, we discuss the
particular challenges involved in extracting excited states in
sect.~\ref{sect:exst}.

\subsection{Extraction of energy levels in lattice QCD}
\label{sect:lvl}

The principle of extracting energy levels of hadrons from lattice QCD
is relatively straightforward. Given a specific fermion matrix $M(U)$
on a gauge field background $U$, the Feynman propagator $S_U(x,y)$ of
the fermion field on this given gauge field background is
\begin{equation}
\label{eq:qprop}
S_U(y,x)=\left(M_U^{-1}\right)_{y,x}
\end{equation}
where we have suppressed additional color and spinor indices that both
$M$ and $S$ carry. These quark propagators on a fixed gauge field
background are the basic building blocks from which hadronic
observables may be built. Note that for every action that fulfills the
$\gamma_5$-hermiticity  condition (\ref{eq:g5hermw}) one can write
\begin{equation}
\label{eq:qpropdag}
S_U(x,y)=\gamma_5  S^\dag_U(y,x) \gamma_5
\end{equation}
to exchange source and sink points of the propagator. For staggered
fermions an equivalent relation is provided by the
$\epsilon$-hermiticity (\ref{eq:stageh}) that implies
\begin{equation}
\label{eq:qpropdags}
S_U(x,y)=\epsilon S^\dag_U(y,x)\epsilon
\end{equation}
Having constructed a hadronic observable, one can simply average it
over the configurations that were produced using an importance
sampling technique (see sect.~\ref{sect:numerics}) to obtain the path
integral expectation value (\ref{eq:expolint}) up to a statistical
precision which is limited by the size of the ensemble of
configurations.

Let us assume that we are interested in the mass of a certain hadronic
state $|h\rangle$ that we do not know how to construct explicitly. We
choose two (not necessarily different) interpolating operators
$\mathcal{O}_{i/f}$ that have a non-vanishing overlap with $|h\rangle$
\begin{equation}
\label{eq:opov}
\langle 0|\mathcal{O}_{i/f}|h\rangle\neq 0
\end{equation}
and compute the expectation value of the correlation function
\begin{equation}
\label{eq:hadprop}
\begin{split}
G(t,0)&=\langle 0|\mathcal{O}_f(t)\mathcal{O}^\dag_i(0)|0\rangle\\
&=\langle 0|e^{\mathcal{H}t}\mathcal{O}_f(0)e^{-\mathcal{H}t}\mathcal{O}^\dag_i(0)|0\rangle
\end{split}
\end{equation}
between times $0$ and $t$. Inserting a complete set of eigenstates of
the Hamiltonian $\mathcal{H}$ in the standard fashion we find
\begin{equation}
\label{eq:partun}
G(t,0)=\sum_n
\frac{\langle 0|\mathcal{O}_f|n\rangle\langle
  n|\mathcal{O}^\dag_i|0\rangle}
{2E_n}
 e^{-E_nt}
\end{equation}
where $E_n$ is the energy of the $n^\text{th}$ eigenstate of
$\mathcal{H}$ above the vacuum energy.  If the state $|h\rangle$ we
are interested in happens to be the lowest energy state with the
quantum numbers of $\mathcal{O}_{i/f}$, one simply needs to go to
asymptotic times for all other states to die out in the correlation
function (\ref{eq:opov})
\begin{equation}
\label{eq:asyt}
G(t,0)
\stackrel{t\rightarrow\infty}{\longrightarrow}
\frac{\langle 0|\mathcal{O}_f|h\rangle\langle
  h|\mathcal{O}^\dag_i|0\rangle}
{2M_h}
 e^{-M_ht}
\end{equation}
and to extract the mass $M_h$ and the product of matrix elements
$\langle 0|\mathcal{O}_f|h\rangle\langle
h|\mathcal{O}^\dag_i|0\rangle$ from it.

It is of course not possible to go to asymptotic times on a finite
lattice. However, since the higher energy states are dying off
exponentially in euclidean time with an exponent that is their energy
difference to the ground state, it is often possible to reach a
distance that is effectively asymptotic for limited lattice time
extents. The time interval that starts from where one can not see the
effect of higher energy states and ends at a possible loss of signal
is called the plateau region. For ground state spectroscopy it is
desirable to extend this plateau region in order to get a
statistically clean signal. This is often achieved by choosing the
operators $\mathcal{O}_{i/f}$ in such a way that they have a large
overlap with the ground state and small overlap with all other
states. This will further be detailed in sect.~\ref{sect:oper}.

Another common strategy to extend the plateau range is to take either
$\mathcal{O}_i$ or $\mathcal{O}_f$ to be a sum of local
operators $\mathcal{O}_l$ over an entire time slice
\begin{equation}
\mathcal{O}_{i/f}(t)=\sum_{\vec{x}}\mathcal{O}_l(t,\vec{x})
\end{equation}
With this choice either the initial or the final state is projected to
zero spatial momentum and consequently all higher momentum excitation
that may appear in the sum over states (\ref{eq:partun}) are cancelled.

Since our lattices have a torus topology with a period $T$ in time
direction, there is a backward contribution that is dominant for
$T-t>t$. It has a similar form than (\ref{eq:asyt}) with the
difference, that the complete set of states has now been inserted on
the other side
\begin{equation}
\label{eq:back}
G(t,0)=
\stackrel{T-t\rightarrow\infty}{\longrightarrow}
b\frac{\langle 0|\mathcal{O}^\dag_i|\bar{h}\rangle \langle \bar{h}|\mathcal{O}_f|0\rangle}
{2M_{\bar{h}}}
e^{-M_{\bar{h}}(T-t)}
\end{equation}
Note the appearance of the ground state $\bar{h}$. It coincides with
$h$ except for cases where the lowest state that couples to both
$\mathcal{O}_i$ and $\mathcal{O}_f$ is different from the lowest state
that couples to both $\mathcal{O}^\dag_i$ and $\mathcal{O}^\dag_f$.
The factor $b=\pm 1$ has been inserted to account for a sign flip that
occurs for interpolating operators with an odd number of quark fields
when the timeslice is crossed that incorporates antiperiodic boundary
conditions in time direction. Without loss of generality, we assume
here that this time slice is traversed in the backward contribution.

In addition one also has in principle contributions from propagators
that wrap onc or more times around the lattice in time
direction. These contributions are tiny however - each additional
wrapping gives a suppression factor $e^{-TM_h}$
or $e^{-TM_{\bar{h}}}$ - and the resulting geometric series can be
summed up. Putting all this together, we find that in the plateau
range the correlation function (\ref{eq:hadprop}) is given by
\begin{equation}
\label{eq:gcf}
\begin{split}
 G(t,0)=&
\mathcal{A}_f\frac{ e^{-M_ht}}
{2M_h\left(1-be^{-TM_h}\right)}\\
+&
\mathcal{A}_b\frac{e^{-M_{\bar{h}}(T-t)}}
{2M_{\bar{h}}\left(1-be^{-TM_{\bar{h}}}\right)}
\end{split}
\end{equation}
with the matrix elements
\begin{equation}
\label{eq:hamp}
\begin{split}
\mathcal{A}_f&=\langle 0|\mathcal{O}_f|h\rangle\langle
  h|\mathcal{O}^\dag_i|0\rangle\\
\mathcal{A}_b&=b\langle 0|\mathcal{O}^\dag_i|\bar{h}\rangle \langle
\bar{h}|\mathcal{O}_f|0\rangle
\end{split}
\end{equation}

One undesirable feature of (\ref{eq:gcf}) is the exponential decay of
the signal with euclidean time. For large time separations $t$ or
$T-t$, the signal exponentially vanishes.

\begin{figure}
\includegraphics[width=0.5\textwidth]{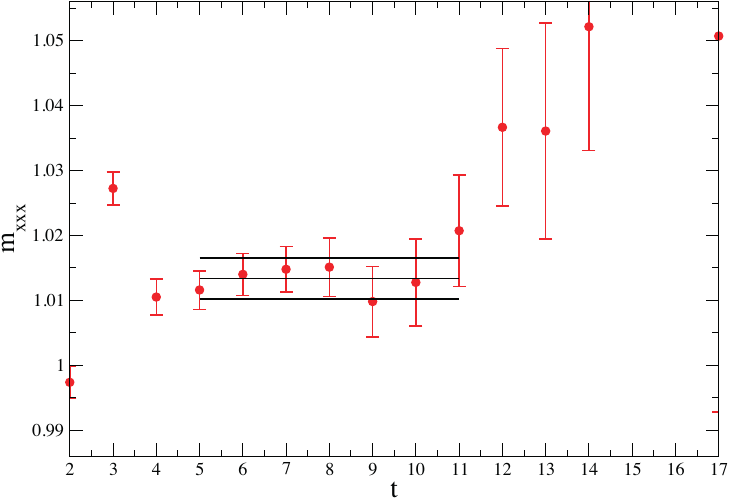}
\caption{
\label{fig:masplat}
Plot of the effective mass $M_\text{eff}(t+a/2)$ vs. $t/a$
(\ref{eq:meff}) of a decuplet baryon from a recent lattice calculation
\cite{Aoki:2010dy}. One can see clearly the onset of the plateau and
the eventual loss of signal. The plateau region is indicated together
with the value of the mass obtained from a fit to the correlation
function. Plot reproduced with friendly permission of the RBC-UKQCD
collaboration.}
\end{figure}

In order to check for the existence and extent of the plateau region,
one can define an effective mass
\begin{equation}
\label{eq:meff}
M_\text{eff}(t+a/2)=\ln\frac{G(t+a,0)}{G(t,0)}
\end{equation}
which will be $M_h$ or $-M_{h^\prime}$ in the region where either the
first or the second exponential dominate in
(\ref{eq:gcf}).\footnote{In the case where $G(t,0)$ is either
  symmetric or antisymmetric one can modify (\ref{eq:meff}) such that
  it gives $M_\text{eff}=M_h$ throughout the entire plateau
  region. See \cite{Fleming:2009wb} for a more thorough discussion of
  effective masses.}  As one can see in fig.~\ref{fig:masplat}, the
effective mass plot is very useful for identifying the plateau region
of a correlation function. One should however keep in mind that the
time $t$ for which the asymptotic regime (\ref{eq:asyt}) is reached
may vary widely. It is possible that the coupling of an operator to
the lowest energy state of the same quantum numbers is nonzero but so
tiny that the ground state is not reachable.

We now turn to the explicit form of the interpolating operators
$\mathcal{O}_{i/f}$. For Wilson fermions the simplest form they can
assume is that of a local operator with the correct quantum
number. For pseudoscalar mesons composed of two different quark
flavors one can e.g. take
\begin{equation}
\label{eq:simpp}
\mathcal{P}_l(x)=\bar\Psi_1(x)\gamma_5\Psi_2(x)
\end{equation}
or
\begin{equation}
\label{eq:simpa0}
{\mathcal{A}_0}_l(x)=\bar\Psi_1(x)\gamma_0\gamma_5\Psi_2(x)
\end{equation}

\begin{figure}
\subfloat[Flavor non-singlet]{
\label{fig:nons}
\includegraphics[width=0.5\textwidth]{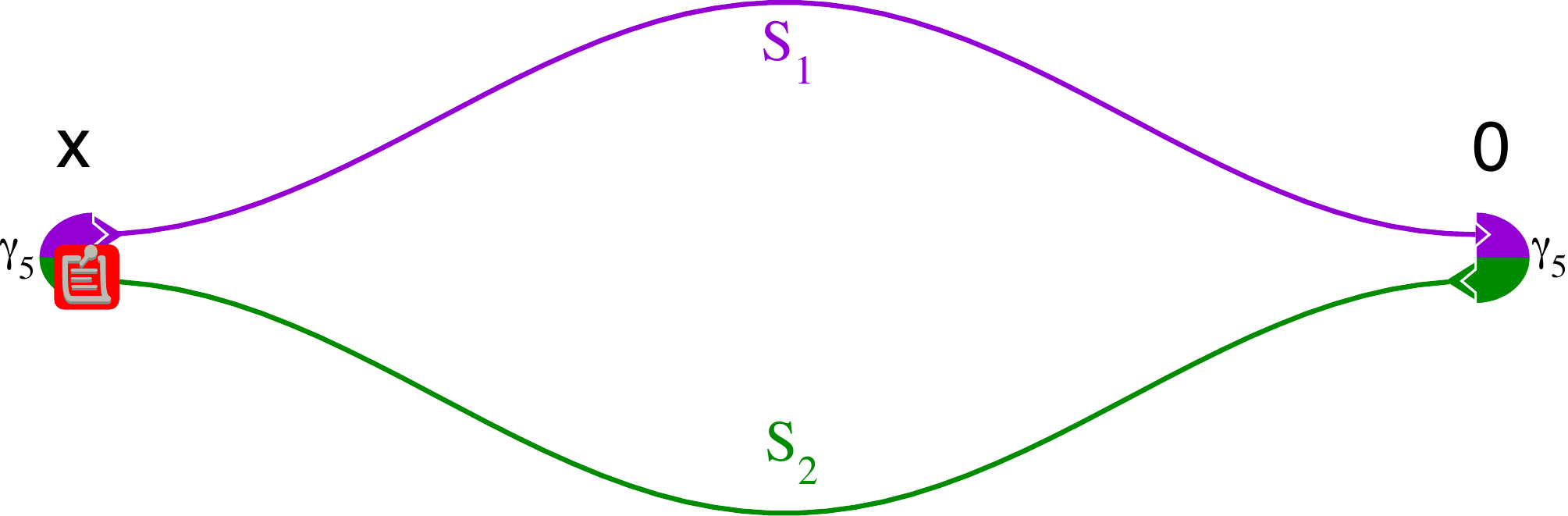}
}
\\
\subfloat[Flavor singlet connected]{
\label{fig:singc}
\includegraphics[width=0.5\textwidth]{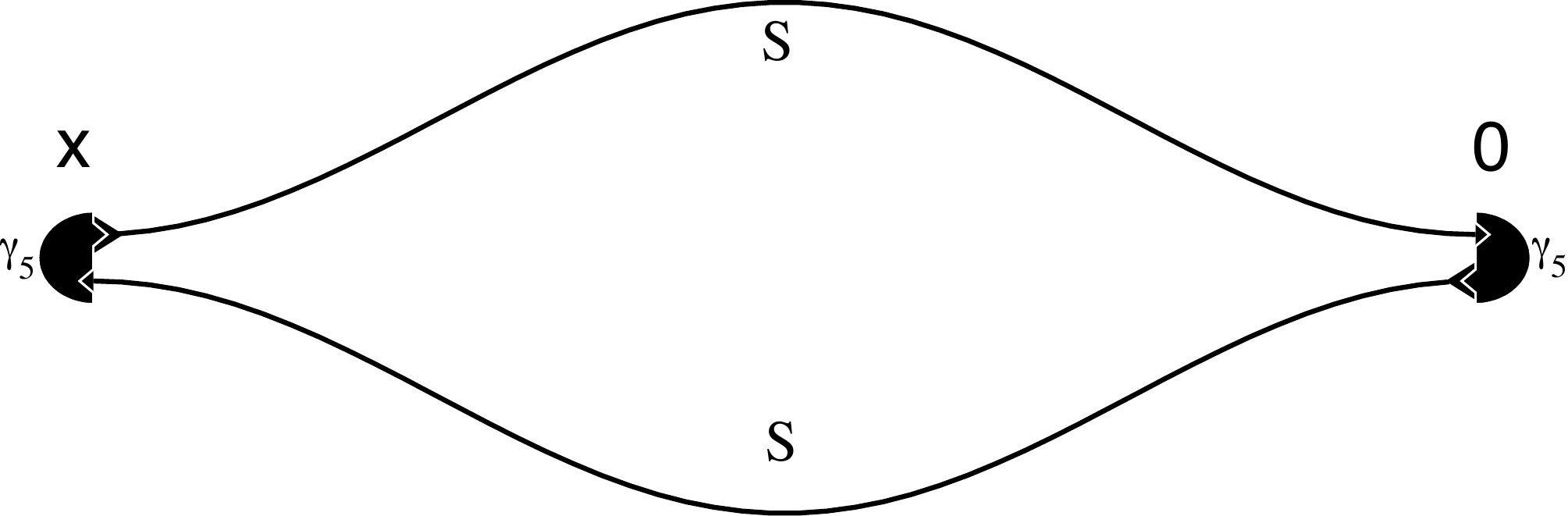}
}
\\
\subfloat[Flavor singlet disconnected]{
\label{fig:singd}
\includegraphics[width=0.5\textwidth]{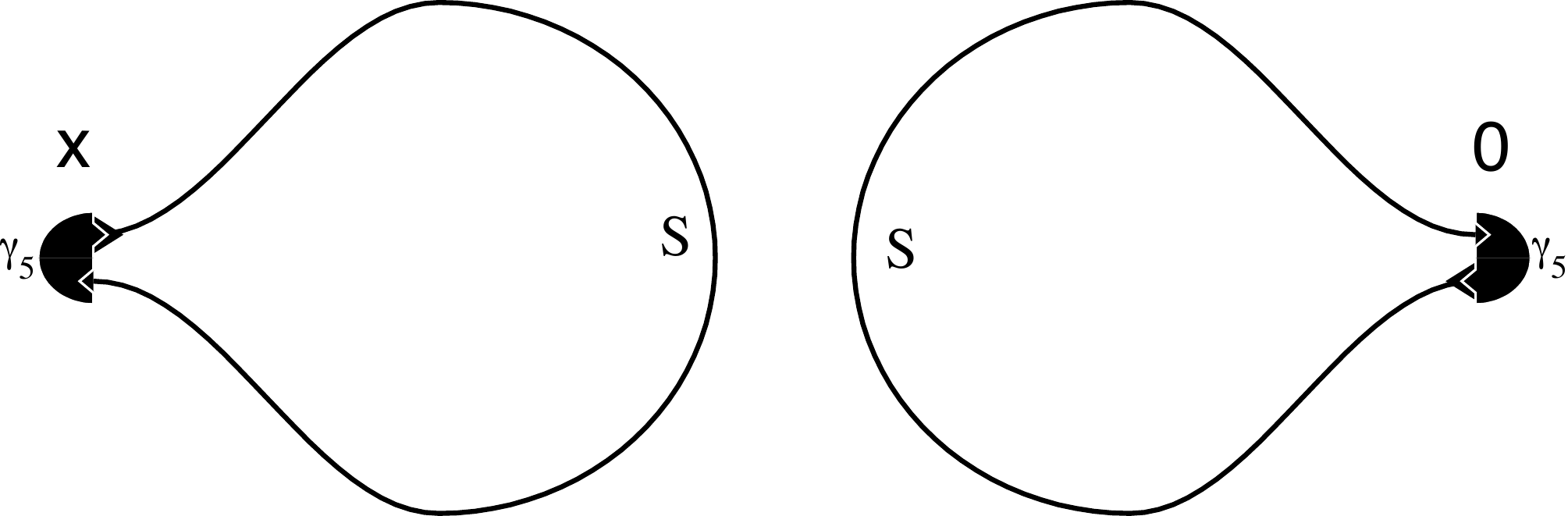}
}
\caption{
\label{fig:ppprop}
Graphical representation of the contraction for a flavor
non-singlet \protect\subref{fig:nons} mesonic Greens function (\ref{eq:pp},\ref{eq:ppfin}) and
of the connected \protect\subref{fig:singc} and disconnected
\protect\subref{fig:singd} contributions to a
flavor singlet mesonic Greens function (\ref{eq:ppsing}).
}
\end{figure}

In order to demonstrate how to construct a proper lattice observable,
we take as an example $\mathcal{O}_f(t)=\mathcal{P}_l(t,\vec{x})$
and $\mathcal{O}_i(0)=\mathcal{P}_l(0,\vec{0})$
and plug these operators into (\ref{eq:hadprop}). Using $x=(t,\vec{x})$, the resulting Greens
function is
\begin{equation}
\label{eq:pp}
\begin{split}
G_{PP}(x,0)&=\langle 0|\mathcal{P}_l(x)\mathcal{P}^\dag_l(0)|0\rangle\\
&=\langle 0|
\bar\Psi_1(x)\gamma_5\Psi_2(x)
\bar\Psi_2(0)\gamma_5\Psi_1(0)
|0\rangle\\
&=
\langle 0|
\contraction{}{\bar\Psi}{_1(x)\gamma_5\Psi_2(x)\bar\Psi_2(0)\gamma_5}{\Psi}
\contraction{\bar\Psi_1(x)\gamma_5}{\Psi}{_2(x)}{\bar\Psi}
\bar\Psi_1(x)\gamma_5\Psi_2(x)
\bar\Psi_2(0)\gamma_5\Psi_1(0)
|0\rangle
\\
&=-\left\langle\text{Tr}\left(
S_1(0,x)\gamma_5
S_2(x,0)\gamma_5
\right)\right\rangle
\end{split}
\end{equation}
where in the last line the expectation value $\langle\cdots\rangle$
denotes the (properly weighted) average over all gauge configurations
and $S_i$ denote the Feynman propagator of the quark
flavor $i$ on a given gauge field background. A graphical
representation of this greens function in terms of quark propagators
is displayed in fig.~\ref{fig:nons}. Using the
$\gamma_5$-hermiticity relation (\ref{eq:qpropdag}), (\ref{eq:pp}) can
be cast in the form
\begin{equation}
\label{eq:ppfin}
G_{PP}(x,0)=-\left\langle\text{Tr}\left(
S^{\dag}_1(x,0)
S_2(x,0)
\right)\right\rangle
\end{equation}
which can be obtained in practice by computing the inverse of the
fermion matrix on just one source point $0$. Note that this is true
for an arbitrary sink point $x$ so that in particular one does not
need to perform more inversions when either
summing over all sink
points $\vec{x}$ in a given time slice or computing the Greens
function from $0$ to a different sink point in an arbitrary time slice.

In case of flavor singlet interpolating operators
\begin{equation}
\mathcal{P}(x)=\bar\Psi(x)\gamma_5\Psi(x)
\end{equation}
the Greens function contains one more Wick contraction
\begin{equation}
\label{eq:ppsing}
\begin{split}
G^{\text{(s)}}_{PP}(x,0) =&\langle
0|\mathcal{P}(x)\mathcal{P}^\dag(0)|0\rangle\\
=&\langle 0|
\bar\Psi(x)\gamma_5\Psi(x)
\bar\Psi(0)\gamma_5\Psi(0)
|0\rangle\\
=&
\langle 0|
\contraction{}{\bar\Psi}{(x)\gamma_5\Psi(x)\bar\Psi(0)\gamma_5}{\Psi}
\contraction{\bar\Psi(x)\gamma_5}{\Psi}{(x)}{\bar\Psi}
\bar\Psi(x)\gamma_5\Psi(x)
\bar\Psi(0)\gamma_5\Psi(0) 
|0\rangle\\
+&
\langle 0|
\contraction{}{\bar\Psi}{(x)\gamma_5}{\Psi}
\bar\Psi(x)\gamma_5\Psi(x)
\contraction{}{\bar\Psi}{(0)\gamma_5}{\Psi}
\bar\Psi(0)\gamma_5\Psi(0)
|0\rangle\\
=&-\left\langle\text{Tr}\left(
S^{\dag}(x,0)S(x,0)
\right)\right\rangle\\
&+\left\langle\text{Tr}\left(
S(0,0)\gamma_5
\right)
\text{Tr}\left(
S(x,x)\gamma_5
\right)\right\rangle
\end{split}
\end{equation}
that leads to a quark line disconnected contribution to the Greens
function that is displayed in fig.~\ref{fig:singd} in addition to the 
connected piece (fig.~\ref{fig:singc}) that is also present
in the flavor non-singlet case (\ref{eq:ppfin}). Since the source and
sink points coincide in the disconnected piece, the
$\gamma_5$-hermiticity relation (\ref{eq:qpropdag}) does not provide
any further simplification as in the case of the connected piece.

For twisted mass, chiral and clover fermions the construction of
operators has to be done in the properly transformed basis (see
(\ref{eq:tmchs},\ref{eq:chirot}) for the case of twisted mass and
chiral fermions respectively) but is identical to the Wilson case
otherwise.

For staggered fermions the construction of interpolating operators is
complicated by the presence of four interacting tastes for each
fermion flavor. Roughly speaking, when using the spin-flavor basis
introduced in sect.~\ref{sect:stag} as a guide, one can construct
mesonic interpolating operators along the same lines as for Wilson
type fermions (the rigorous construction can be found in
\cite{Golterman:1984cy,KlubergStern:1983dg,Kilcup:1986dg}). One should
note however that due to the distribution of spin degrees of freedom
within an elementary hypercube all operators that have a taste
structure different from their spin structure are necessarily not
localized to a single lattice point. One relevant example of an
operator that is local and also leads to a correlation function that
is positive on every gauge configuration is
\begin{equation}
\label{eq:simpps}
\mathcal{P}^{5}_l(y)=\bar{X}_1(y)(\gamma_5\otimes\xi_5)X_2(y)
\end{equation}
for which the goldstone pion is the ground state.

The simplest baryon operators for Wilson quarks that have the nucleon
as a ground state are the local
\begin{equation}
\label{eq:nuop1}
\mathcal{N}^\text{(1)}_l=
\epsilon_{abc}\left(u^T_aC\gamma_5 d_b\right)u_c
\end{equation}
and
\begin{equation}
\label{eq:nuop2}
\mathcal{N}^\text{(2)}_l=
\epsilon_{abc}\left(u^T_aC d_b\right)\gamma_5 u_c
\end{equation}
where $a,b,c$ are color indices and the charge conjugation operator
$C=\gamma_0\gamma_2$ s. We have suppressed explicit spin indices and
coordinates. The difference between (\ref{eq:nuop1}) and
(\ref{eq:nuop2}) is that in the first case one starts off with a
pseudoscalar ``diquark'' (the bracketed expression) whereas in the
second case the diquark is scalar and the $\gamma_5$ only enters when
combining the diquark with the remaining $u$. This difference in spin
structure leads to a very different nonrelativistic limit of the two
operators - $O(1)$ for $\mathcal{N}^\text{(1)}$ vs. $O(p^2/E^2)$ for
$\mathcal{N}^\text{(2)}$ - which in turn implies that the relative
overlap of $\mathcal{N}^\text{(1)}$ with the ground state as compared
to the excited states is much larger than that of
$\mathcal{N}^\text{(2)}$ \cite{Bowler:1984dh,Leinweber:1994nm}. For
decuplet baryons, one can construct an interpolating operator by
replacing the pseudoscalar diquark in (\ref{eq:nuop1}) with a vector
one \cite{Chung:1981cc,Leinweber:1992hy}. An interpolating operator
coupling to the $\Delta^{++}$ can for example be obtained by
\begin{equation}
\label{eq:delop}
\mathcal{D}_\mu=
\epsilon_{abc}\left(u^T_aC\gamma_\mu u_b\right)u_c
\end{equation}
Greens functions $G_{\mu\nu}$ that arise from using (\ref{eq:delop})
for both source and sink are however not pure spin $3/2$
\cite{Leinweber:1992hy}. A spin projection to a pure spin $3/2$ state
may be performed using the projection operator
\cite{VanNieuwenhuizen:1981ae,Benmerrouche:1989uc}
\begin{equation}
\label{eq:spproj}
P^{3/2}_{\mu\nu}=\delta_{\mu\nu}-\frac{1}{3}\gamma_\mu\gamma_\nu-
\frac{1}{3p^2}\left(\gamma_\tau p_\tau\gamma_\mu p_\mu+p_\mu\gamma_\nu\gamma_\tau p_\tau\right)
\end{equation}
Numerical evidence suggests that even the unprojected correlator
couples almost exclusively to the spin $3/2$ state \cite{Leinweber:1992hy}.

For staggered fermions, zero momentum baryon operators have been
constructed by \cite{Golterman:1984dn}. This construction does not
rely on the spin-flavor interpretation of baryons but instead is based
on analyzing the discrete spacetime symmetry group of staggered
fermions. One interesting feature of this construction is that there
exists an operator that couples to the $\Delta$ but does not couple to
the nucleon while on the other hand every operator that couples to the
nucleon also couples to the $\Delta$. Since the mass difference
$\delta M$ of these two states is rather small, the $\Delta$ dies off
slowly in euclidean time $t$ as $\propto e^{-\delta M t}$ and it is
challenging to find a plateau region.  Effects of different quark
flavors have recently been added to this formulation by
\cite{Bailey:2006zn}.

An alternative approach to extracting hadron masses via measuring the
free energy of a system with a finite baryon density has been
suggested by \cite{Fodor:2007ga}. This method however requires the
introduction of a chemical potential and its practical use is severely
limited due to the sign problem.

\subsection{The role of operators}
\label{sect:oper}

As we have seen in sect.~\ref{sect:lvl}, the choice of operators
determines the relative strength of the coupling to different states
with the same quantum numbers and therefore potentially has a huge
influence on the quality of signal. We first consider the case of the
simple method of improving a local operator by summing it over either
source or sink timeslice to obtain a projection to zero spatial
momentum.

When writing the hadron correlation function (\ref{eq:hadprop}) in
terms of Wick contractions of the quark fields, it is often possible
to use the relations (\ref{eq:qpropdag}) or (\ref{eq:qpropdags}) in
order to obtain an expression where the quark field sources are
restricted to timeslice $0$ (an explicit example for the flavor
non-singlet pseudoscalar propagator was given in (\ref{eq:ppfin})). In
these cases, if one only needs to do the momentum projection at the
sink timeslice $t$, it is a trivial summation. Performing the momentum
zero projection on the source side in principle requires computing the
inverse of the fermion matrix for every point in the source timeslice
$t=0$. Since this would involve a prohibitively large number of
fermion matrix inversions, a range of methods has been developed to
deal with this problem and related ones where one needs information of
the fermion propagator from every source point to every sink point on
the lattice
\cite{Bitar:1988bb,Bernardson:1993yg,Kuramashi:1993ka,Dong:1993pk,deDivitiis:1996qx,Eicker:1996gk,Michael:1998sg,McNeile:2000xx,Wilcox:1999ab,Neff:2001zr,Duncan:2001ta,DeGrand:2002gm,Bali:2005fu,Foley:2005ac,Boucaud:2008xu,Bali:2009hu}. These
methods, commonly known as all-to-all techniques, are based on either
stochastic estimates or eigenmode approximations of the full
propagator or a combination of both.

The basic idea behind a stochastic estimate of the all-to-all
propagator is to compute the inverse of the fermion matrix on a number
of source vectors $\xi^{(i)}$
\begin{equation}
\label{eq:noisprop}
\sigma_U^{(i)}=M_U^{-1}\xi^{(i)}
\end{equation}
Provided that the $\xi^{(i)}$ fulfill the condition
\begin{equation}
\label{eq:noisid}
\sum_i{\xi^{(i)}}^\dag(y)\xi^{(i)}(x)=\delta_{x,y}\mathbbm{1}
\end{equation}
one can obtain the propagator between any two lattice points on a
fixed gauge background $U$ as
\begin{equation}
\label{eq:funprop}
S_U(y,x)=\sum_i
{\xi^{(i)}}^\dag(y)\sigma_U^{(i)}(x)
\end{equation}
It is often useful to restrict the vectors $\xi^{(i)}$ to a subspace
of the lattice, i.e. to modify the relation
(\ref{eq:noisid}) so that it is zero outside a certain subspace. If
one wants to obtain full all-to-all propagators it is then of course
necessary to have more than one set of source vectors $\xi^{(i)}$ so
that all sets together cover the entire space desired. This method is
sometimes referred to as partitioning or dilution and commonly used
subspaces include individual spin components and individual time
slices.

We are interested in the case where the source vectors $\xi^{(i)}$ are
random vectors and the relation (\ref{eq:noisid}) is fulfilled
stochastically. Examples of sets of stochastic source vectors that
fulfill (\ref{eq:noisid}) are $Z_2$ noise where every component of
$\xi^{(i)}$ is randomly chosen to be either of $\pm 1$ or $U(1)$ noise
where every element is a complex random number with unit modulus. The
question of how many source vectors are necessary to get an optimal
signal for a specific observable at minimum computational cost has to
be answered numerically. It is important to note however that since
one is usually not interested in the propagator on a specific gauge
configuration $U$ and the path integral sum does commute with the sum
over all source vectors $\xi^{(i)}$, the optimal number of sources per
configuration might even turn out to be one.

Eigenmode approximations generally assume that the inverse of the
fermion operator is well approximated by restricting it to a subspace that
is spanned by a number of its lowest eigenmodes. These eigenmodes are
then computed and an approximation of the full matrix inverse can be
found. This truncation represents an uncontrolled approximation, but
it may easily be supplemented with a stochastic estimate of the effect
of the higher eigenmodes. One only needs to project out the components
of the original source vector that lie in the space of low eigenmodes,
treat them exactly and use stochastic estimators on the orthogonal
compliment.

Until now we have only discussed (sums of) local operators,
i.e. operators where all the quarks originate from a single lattice
point. Hadrons however are extended objects with quark distributions
that have finite widths and different shapes. One strategy to improve
the overlap between an interpolating field operator and a specific
hadronic state therefore consists of trying to model its quark
distribution. One possibility that is often used is to replace the
quark point-sources $\Psi(t,\vec{x})$ by a sum over quark sources at
neighboring sites. In its simplest form, one can take
\begin{equation}
\label{eq:qsmear}
\Psi^\prime(t,\vec{y})=\sum_{\vec{x}}G(\vec{y},\vec{x})\Psi(t,\vec{x})
\end{equation}
where $G(\vec{y},\vec{x})$ is the smearing kernel that determines the
relative weight of the source points. A computationally convenient
restriction of the smearing kernel is the factorization ansatz
\cite{Bacilieri:1988fh,DeGrand:1990dz}
\begin{equation}
\label{eq:gaussource}
G(\vec{y},\vec{x})=g(\vec{x})g(\vec{y})
\end{equation}
where effectively the quark fields are independently smeared. Typical
choices for the form of $g(\vec{x})$ include a wall
\cite{Bitar:1990cb}, a hard sphere or box \cite{Bacilieri:1988fh}, a
gaussian \cite{DeGrand:1998pr,DeGrand:1998jq} or a radial exponential
\cite{Ali_Khan:2001tx} of different size.  Note that in general
$\Psi^\prime(t,\vec{y})$ in (\ref{eq:qsmear}) is not gauge invariant
and one therefore needs to work in a fixed gauge on the source
timeslice.  A more sophisticated method of smearing the quark source
that is gauge invariant is known as Wuppertal smearing (or gauge
invariant gaussian smearing) \cite{Gusken:1989qx} where a covariant
laplacian is added to the unit operator and repeatedly applied $N$
times
\begin{equation}
\label{eq:wuppsmear}
G=\left(\mathbbm{1}+\sum_{i=1}^3\alpha\left(V_i+V_i^\dag\right)\right)^N
\end{equation}
The one-hop term $V_i$ is given by (\ref{eq:hop}) and the smearing
parameters $\alpha$ and $N$ determine the form and width of the source
that is approximately gaussian for large $N$ on a trivial gauge
field.\footnote{See \cite{Allton:1993wc,Lacock:1994qx,Burch:2004he}
  for different constructions of nonlocal gauge invariant operators.}
In order to suppress gauge noise, one can use smeared links (see
sect.~\ref{sect:fimp}) for the gauge field in (\ref{eq:wuppsmear}). A
variant of this method is the Laplace-Heaviside (LapH) smearing
\cite{Peardon:2009gh} where the smearing kernel is defined as a
Heaviside step function on the lowest eigenmodes of the covariant
Laplacian. This method requires the explicit inversion of the fermion
matrix on a number of lowest eigenmodes of the covariant Laplacian
that grows with the volume of the system. This growth with volume of
the number of required fermion matrix inversions can be countered by
not computing propagators on all eigenmodes exactly but instead using
stochastic techniques, similar to the ones described above, in the
eigenspace of low modes to estimate them
\cite{Morningstar:2010ae,Morningstar:2011ka}.

Replacing the original quark sources in local hadron operators such as
(\ref{eq:simpp},\ref{eq:simpa0},\ref{eq:nuop1},\ref{eq:nuop2}) with
smeared ones usually does improve the overlap with the desired
state. Especially in the case of excited states however, it is often
useful to go further. When trying to find an operator that has maximal
overlap with a certain state, one can use its quantum numbers and
expected wave function to model a lattice operator. The two quantum
numbers of interest are spin and parity. Parity is not broken by the
lattice regularization and therefore one can construct operators that
couple only to states of a given parity exclusively. Note however that
the backward contribution in (\ref{eq:gcf}) will be of opposite parity
than the forward one. For interpolating operators with an odd
  number of quark fields, it is possible to use the relative sign flip
  in the backward amplitude $\mathcal{A}_b$ (\ref{eq:hamp}) between
  periodic and antiperiodic boundary conditions to eliminate
  these. \cite{Sasaki:2001nf,Csikor:2003ng,Leinweber:2004it,Sasaki:2005ug}

Spin on the other hand is the quantum number corresponding to the
continuum rotational symmetry group $SU(2)$ that is broken down to the
symmetry group of cubic lattice rotations. This group, the octahedral
group $\mathcal{O}$ has been studied in \cite{Johnson:1982yq} and it
was found that there are five irreducible representations
corresponding to integer spin and three corresponding to half integer
spin, all of them containing a whole tower of partially overlapping
continuum spin representations. One can construct lattice operators
that transform irreducibly under the octahedral group $\mathcal{O}$
\cite{Basak:2005aq,Basak:2005ir,Basak:2007kj} and are especially
beneficial for the extraction of highly excited baryon resonances.  In
addition to the quark source smearing, these operators generally
involve quark sources that can each be covariantly displaced from a
reference point by one lattice unit in any direction. The general form
of such a baryon operator is an appropriate linear combination of
terms of the form
\begin{equation}
\label{eq:exbarop}
\mathcal{B}=
\epsilon_{abc}
\left(D^\prime_i\Psi\right)^a
\left(D^\prime_j\Psi\right)^b
\left(D^\prime_k\Psi\right)^c
\end{equation}
where $abc$ is color and we have suppressed spin and flavor
indices. The operators $D^\prime_i$ are gauge covariant displacement
operators by one step in direction $i$ (or the unit operator if
$i=0$). For more detailed reviews on the construction of operators for
excited state baryon spectroscopy see
e.g. \cite{Leinweber:2004it,Basak:2006ki,Lang:2007mq}.

A different approach of finding an operator that has optimal overlap
with the ground state was developed\footnote{See \cite{Draper:1993qj}
  for a similar approach for heavy-light systems.} by
\cite{Babich:2005ay,Babich:2007ah}. They considered at the sink side a
general meson operator of the form
\begin{equation}
\label{eq:claudmes}
\mathcal{M}(t,r)=\bar\Psi_1(t,\vec{x})\Gamma\Psi_2(t,\vec{y})\times\delta(|\vec{y}-\vec{x}|-r)
\end{equation}
and studied the profile in $r$ of the resulting correlation function
as a function of $t$. Once a plateau is reached, the profile was found
to settle into an asymptotic form $\phi(r)$ which can then be used to
construct an optimal sink operator
\begin{equation}
\label{eq:claudsink}
\mathcal{O}(t)=\sum_r\phi(r)\mathcal{M}(t,r)
\end{equation}

A special problem occurs when one tries to extract the energy level of
a scattering state or a bound state of hadrons with a non-minimal
number of valence quarks. As repeatedly noted in the literature (see
e.g. \cite{Engel:2010my,Bulava:2010yg}), the coupling of single hadron
operators to multi hadron states is extremely small. A prominent
example is the $\rho$ resonance which has a $\pi-\pi$ scattering state
as a ground state in infinite volume. A closely related phenomenon
occurs in heavy quark physics where string breaking between static
quarks can only be observed when one introduces explicitly a ``broken
string'' operator that consists of two separated static-light mesons
\cite{Knechtli:1998gf,Bali:2005fu}.

Both phenomena suggest that it is very difficult to produce a pair of
appropriate sea quarks from the vacuum that allows for the propagation
of an intermediate multihadron state. Intuitively this is
understandable as the occurrence of large sea quark loops is
suppressed in the path integral (\ref{eq:expolint}). 

For spectroscopy in channels where there is a scattering state below
the relevant resonance or for the spectroscopy of exotic objects that
consist of a non-minimal number of valence quarks - such as tetraquarks
or pentaquarks - one therefore needs to use appropriate interpolating
operators that correctly reflect the valence quark structure of the
desired object.

\begin{figure*}
\includegraphics[width=\textwidth]{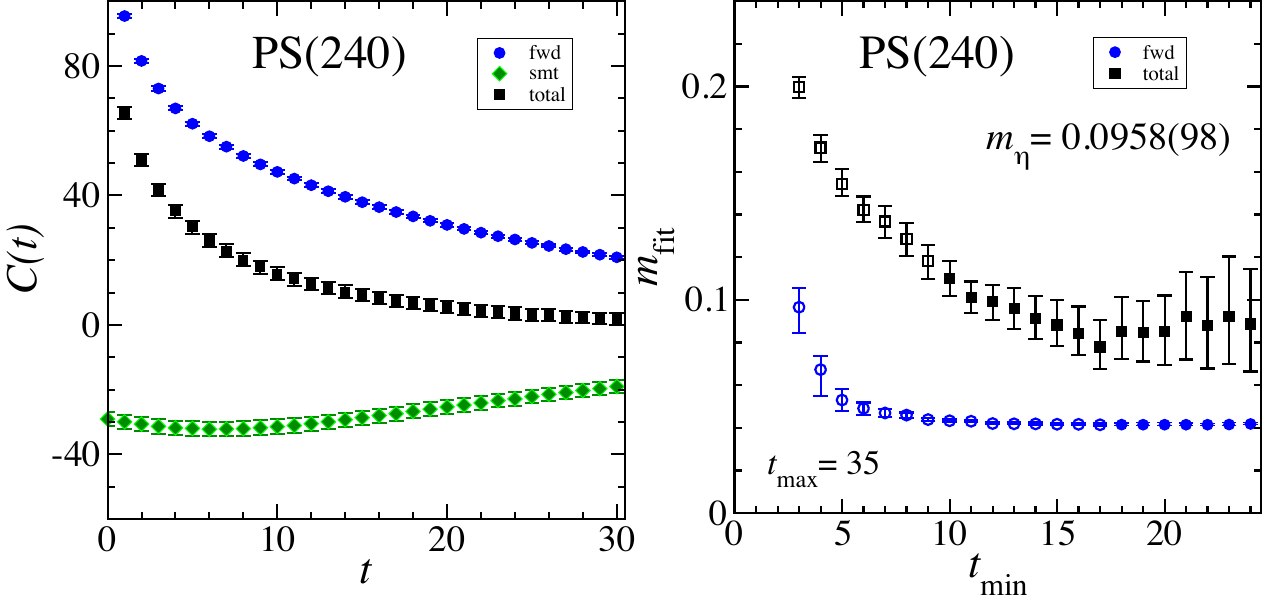}
\caption{
\label{fig:singl}
Correlation function (left panel) and effective mass (right panel) in
the $\eta$ channel using a stochastic LapH source. Connected (fwd) and
in the left panel disconnected (smt) contributions are plotted
additionally.  Plot taken from \cite{Morningstar:2011ka} with friendly
permission of K.~J.~Juge.}
\end{figure*}

When constructing interpolating operators one usually tries to avoid
situations where a quark line may start and terminate at the same time
slice since these operators are known to be much noisier than
operators where all the quark lines run from the source to the sink
timeslice. In some cases however, e.g. for isoscalar mesons or
generally for the multihadron operators mentioned above, quark
propagators that attach to one time slice with both ends are
unavoidable. As an example of how a combination of the above mentioned
techniques can lead to a decent signal in these notoriously difficult
channels fig.~\ref{fig:singl} displays one current determination of
the correlation function and mass plateau in the $\eta$ channel that
was obtained using the stochastic LapH technique described above.
(For similar results with a different mix of techniques see also
\cite{Alexandrou:2010jr}).

Finally, there are also suggestions to completely avoid the problem of
computing disconnected diagrams by finding relations between them and
disconnected diagrams in partially quenched chiral perturbation theory
\cite{DellaMorte:2010aq}.

\subsection{Extracting multiple energy levels}
\label{sect:exst}

In sect.~\ref{sect:lvl} we have seen how to extract the ground state
of a channel by going to asymptotic euclidean time. In order to
extract excited state masses, one can in principle fit the correlation
function to a multiexponential form
\begin{equation}
\label{eq:multex}
G(t,0)=
\sum_n
\mathcal{A}_n
e^{-M_nt}
\end{equation}
where we have ignored backward contributions. A fit of the form
(\ref{eq:multex}) with free parameters $\mathcal{A}_n$ and $M_n$ can
typically not be stabilized numerically for more than two states and
even extracting reliable first excited state masses with this
technique is challenging.

One can overcome these problem by using a variational method
\cite{Michael:1985ne,Luscher:1990ck,Blossier:2009kd}. The basic idea
is to expand the basis to include $N$ initial and final state
operators $\mathcal{O}_{i_k}$ and $\mathcal{O}_{f_k}$ with the same
quantum numbers that ideally couple to different energy states
preferentially and then construct the complete cross-correlator matrix
\begin{equation}
\label{eq:multprop}
G_{lm}(t,0)=\langle 0|\mathcal{O}_{f_l}(t)\mathcal{O}^\dag_{i_m}(0)|0\rangle
\end{equation}
One can then define a matrix $M(t,t_0)$ from
\begin{equation}
\label{eq:multtm}
G(t,0)=M(t,t_0)G(t_0,0)
\end{equation}
and analyze its eigenvalues $\lambda_n(t,t_0)$ and eigenvectors
$v_n(t,t_0)$. At large euclidean times $t$ and $t_0$ the eigenbasis of
the matrix $M$ will then align with the eigenbasis of the
Hamiltonian and the eigenvectors will behave as
\begin{equation}
\label{eq:multen}
\lambda_n=e^{-E_n(t-t_0)}
\end{equation}
from which one can extract an effective mass for each of the energy
levels similar to (\ref{eq:meff}). The energies of a number of lowest
lying states in a given channel can thus be determined provided that
the operator basis chosen has sufficient overlap with all of these
states and the quality of the data is good enough do determine all the
elements of the transfer matrix with sufficient accuracy. 

A robust variant of the variational method was suggested by
\cite{Mahbub:2009nr}.\footnote{See \cite{Draper:1994bi} for a similar
  method for heavy-light systems.} Instead of extracting the energies from the
eigenvalues directly one can use the elements of the eigenbasis to
project out effective single state components from the matrix valued
propagator and analyze them with standard single channel methods. 

\begin{figure}
\includegraphics[width=0.5\textwidth]{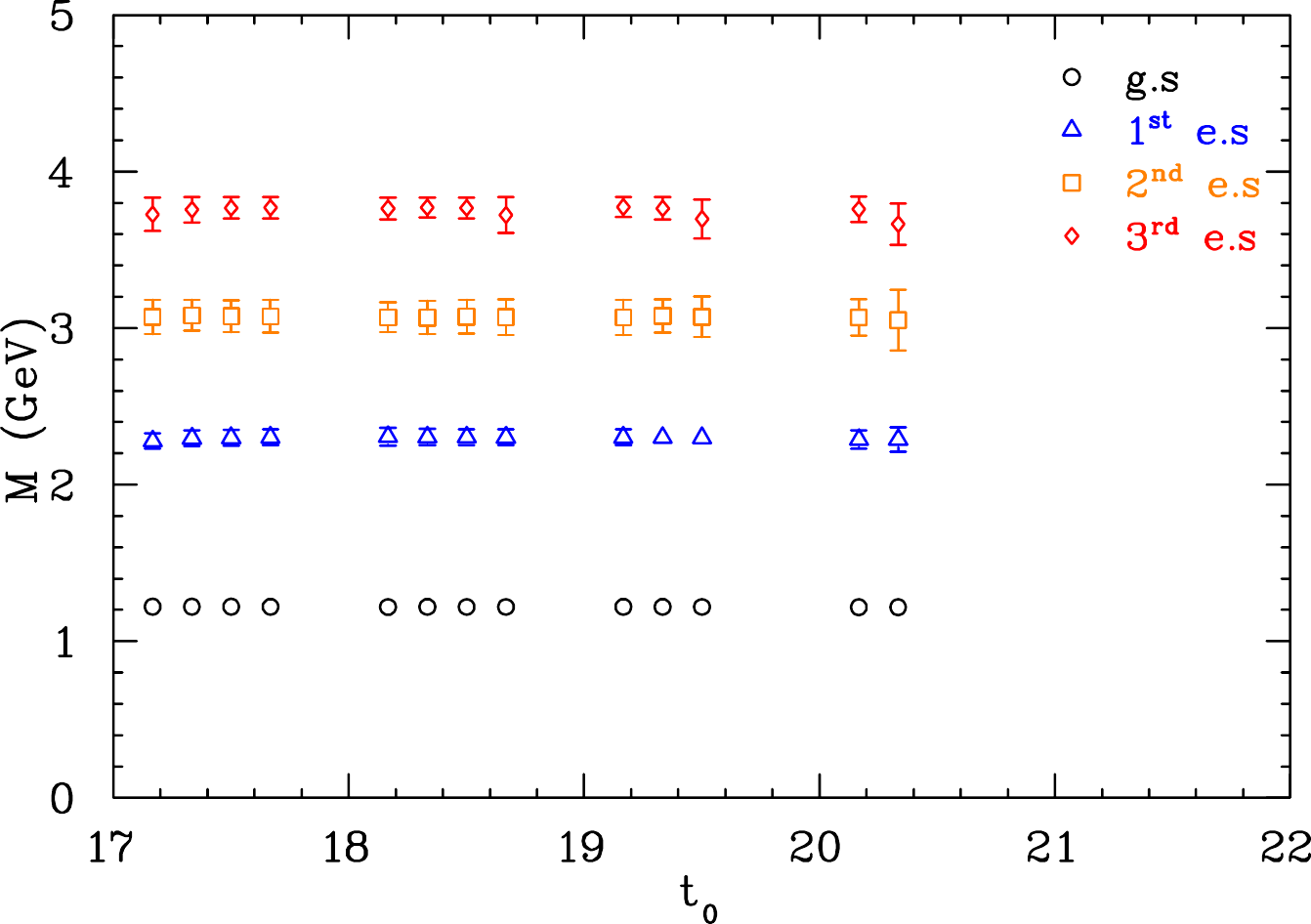}
\caption{
\label{fig:effmaex}
Plot of the masses of the ground and excited states in the nucleon
channel. The extraction has been performed by fitting single state
correlators that were obtained by projection of a $4\times 4$ matrix
correlator onto the elements of the eigenbasis of the transfer matrix
$M(t,t_0)$ for different $t$ and $t_0$.  Plots are taken from
\cite{Mahbub:2010vu} with friendly permission of D.~Leinweber.}
\end{figure}

An example fig.~\ref{fig:effmaex} displays masses of the ground state
and first three excited states of the nucleon channel extracted with
the variational projection method.

\section{Physical predictions}
\label{sect:physpred}

Having extracted hadron masses from nonperturbative lattice QCD
calculations, we now need to turn them into physical predictions. 

QCD, and also lattice QCD, is a theory without any intrinsic scale -
it is formulated in terms of dimensionless quantities entirely. In
order to extract from it dimensionful quantities such as hadron
masses, we need to supply it with a scale. We can do this in general
by picking a scale setting observable, a quantity that is dimensionful
in the continuum theory. In the lattice theory we can then measure the
appropriate dimensionless combination of any target observable we wish
to extract with the scale setting observable and get a dimensionful
prediction for the target observable by fixing the scale setting
observable to its dimensionful input value. One such target observable
would be the lattice spacing $a$ itself.

The scale setting procedure outlined above is not entirely sufficient
for making physical predictions. Once we have supplied lattice QCD
with a scale we still need to fix its remaining parameters - the bare
quark masses. For $N_f$ non-degenerate quark flavors we can generally
do so by fixing $N_f$ linearly independent dimensionless observables
in the lattice theory to their desired input value. What one would
ideally like to do then is to fix the $N_f+1$ dimensionless bare
parameters of the lattice theory, the bare quark masses and the gauge
coupling, such that the $N_f$ dimensionless observables on the lattice
assume their physical values exactly and the lattice spacing $a$ is of
the desired size. One could then measure any observable on the lattice
for a range of lattice spacings $a$ and, with the appropriate
functional form that is given by the discretization effects of the
specific action used, extrapolate them into the continuum $a=0$.

There are a few obstacles towards implementing this ideal procedure in
nonperturbative lattice calculations. The first one being that
  one can not simply go to the physical point by setting the bare
  coupling and input quark masses in a lattice calculation to their
  physically observed values since these physical values can not
  directly be measured in experiment. Quantities that are
accessible experimentally, such as hadron masses or decay widths, have
relations to the bare parameters that need to be determined on the
lattice themselves. One is therefore left with the choice of either tuning the bare
parameters or computing both target observables and those used for
scale setting and parameter fixing at various unphysical points
followed by an interpolation to the physical point.

When trying to implement either of the two procedures, tuning or
interpolating to the physical point, one faces the more technical
problem that it is very difficult to reach the physical point for
light quarks. As discussed in sect.~\ref{sect:doubl} there is a
variety of different fermion discretizations but each one does have a
specific problem when making the quark mass light. In the case of
Wilson fermions one is faced with exceptional
configurations, staggered and twisted mass fermions have the problem
of flavor/taste splitting and chiral fermions are simply expensive in
terms of computer time needed. None of these problems is
insurmountable but they turn out to be sufficiently severe so that an
extrapolation in the light quark mass to the physical point is still
the rule rather than the exception in hadron spectroscopy calculations.

A third problem that is less severe in practice but still has to be
considered is that nature is not QCD alone, even for the light hadron
spectrum. Quarks are electrically charged and there are QED
corrections to hadron masses. Similarly, isospin symmetry is usually
assumed in lattice calculations but broken in nature. Both of these
effects are relatively small as can be seen from the isospin splitting
of the experimentally observed hadron spectrum, but they still need to
be considered.

We start in sect.\ref{sect:physpt} by reviewing the problem of
reaching the physical point. In sect.~\ref{sect:cont} we discuss how
to remove the cutoff i.e. how to reach the continuum limit for the
various lattice discretizations. Finally, in sect.~\ref{sect:fv}, we
consider finite volume corrections that are especially relevant for
resonant states and we conclude with the discussion of subleading
effects from QED and isospin breaking in sect.~\ref{sect:EM}.

\subsection{Reaching the physical point}
\label{sect:physpt}

Having extracted hadron masses from simulations we now need to turn
them into physical predictions. As discussed in the beginning of
sect.~\ref{sect:physpred}, we need to go or extrapolate to the physical
point and we need to set the scale.

On a technical level tuning to the physical or any other target point
is achieved through tuning the bare parameters of the lattice theory -
the coupling $\beta$ and the bare quark masses in the lattice
Lagrangian. Defining the physical or any other target point on the
other hand is generally done by comparing dimensionless combinations
of continuum observables with the corresponding lattice observables
and the scale can be set by comparing any one dimensionful continuum
observable. Provided that all effects beyond QCD with the given number
of flavors are correctly accounted for and provided also that the
chosen lattice discretization of QCD does have the correct continuum
limit, all possible choices of finding the physical point are
equivalent in the continuum limit. If any of the above assumptions is
violated, such as is the case e.g. in quenched QCD, continuum
predictions are not unique and depend on the specific choice of
defining the physical point.

It is usual to treat the two light quark flavors $u$ and $d$ as
degenerate and to include isospin breaking as well as electromagnetic
effects as corrections. One can then define a light quark mass
$\hat{m}=(m_u+m_d)/2$. In order to tune to the correct light quark
mass one combination that is often used is the ratio of the pion mass,
the square of which is proportional to the light quark mass at leading
order, to an observable that depends less on the light quark mass. In
early quenched work on lattice hadron spectroscopy the QCD string
tension was often used for this purpose
\cite{Creutz:1980wj,Pietarinen:1981cd,Bernard:1982hh,Marinari:1981nu,Hamber:1981zn,Weingarten:1981jy,Creutz:1983ev}
as later on was $M_\rho$ \cite{Fucito:1982ip} or $M_\Phi$
\cite{Lipps:1983pi}.  In full QCD these choices are not optimal. The
vector meson $\rho$ is a broad resonance and not the ground state in
its channel, the singlet $\phi$ contains disconnected diagrams and the
string tension is ill defined with dynamical quarks. Even in the
quenched approximation any resonant state mass is not an ideal choice
for a scale setting observable as the connection of the measured
ground state mass to the experimentally measured mass above decay
threshold is not obvious. Quantities that are frequently used today
are either baryon masses that are stable in QCD ($M_N$, $M_\Xi$,
$M_\Omega$)
\cite{Aoki:2008sm,Aoki:2009ix,Durr:2008zz,Alexandrou:2009qu,Brandt:2010ed,Bulava:2010yg,Lin:2008pr,Engel:2010my},
average masses of baryon multiplets \cite{Bietenholz:2010jr} or
distance measures ($r_0$, $r_1$) in the heavy quark potential
\cite{Sommer:1993ce} that are in turn determined from $\Upsilon$
spectroscopy \cite{Bernard:2001av,Davies:2009tsa,Bazavov:2009bb}. The
use of matrix elements, such as the pseudoscalar decay constants,
although common in other areas
\cite{Giusti:2001pk,Blossier:2009bx,Noaki:2009sk,Borsanyi:2010bp} is
less often seen in lattice hadron spectroscopy.

At the physical point the scale can then be set by comparing a
dimensionful continuum observable with the corresponding lattice
observable. At any other non-physical point the scale setting is
conceptually ill defined. It is nonetheless usual to either use the
scale defined at the physical point for all theories with the same
coupling $\beta$ and arbitrary quark masses (mass independent scale
setting) or to obtain the scale by comparing a dimensionful continuum
observable with its lattice counterpart at any nonphysical point
(mass dependent scale setting).

In computations with a dynamical strange quark its mass also has to be
set. In analogy to the light quark case, this is usually done via the
kaon mass, the combination $M_K^2-M_\pi^2/2$, which is proportional to
the strange quark mass to leading order, or the mass of the fictitious
$\eta_s$, the pseudoscalar meson that is composed of two strange
valence quarks.

As for the specific functional form of an extrapolation or
interpolation to the physical point the most straightforward form is
one that is linear in the quark mass. Since $M_\pi^2\propto\hat{m}$
and $\bar{M}_K^2=M_K^2-M_\pi^2/2\propto m_s$ to leading order
\cite{GellMann:1968rz}, one can fit any other hadron mass $M_X$ to a
form
\begin{equation}
\label{eq:loxpt}
M_X=a+b M_\pi^2+c \bar{M}_K^2
\end{equation}

Going beyond this leading order one can perform systematic expansions
around either a point with two or three massless quark flavors or a
general, massive point. For the first case, chiral perturbation theory
($\chi$PT), an effective field theory based upon the chiral symmetry
pattern of QCD has been developed
\cite{Weinberg:1978kz,Gasser:1983yg,Gasser:1984gg}. It provides an
asymptotic expansion around either the two or three flavor massless
point.  As it is built around the assumption of spontaneous chiral
symmetry breaking in the massless theory, it is particularly suited
for treating properties of the pseudo-Goldstone bosons of this
symmetry, i.e. the pions and, to a somewhat lesser extent, the kaons.

For masses other than the pseudoscalars, $\chi$PT generically predicts
a leading nonanalytic term of the form $M_\pi^3$
\cite{Langacker:1974bc}. More formally, one can include baryons in
$\chi$PT \cite{Gasser:1987rb} but the resulting series is only slowly
converging. An alternative formulation with better convergence
properties is heavy baryon $\chi$PT
\cite{Jenkins:1990jv,Bernard:1992qa} which treats baryons as
nonrelativistic particles and currently is most commonly used to fit
lattice baryon data. An extension of heavy baryon $\chi$PT for
staggered fermions was developed by \cite{Bailey:2007iq}. Recently,
the covariant approach \cite{Becher:1999he} that promises better
convergence behavior for heavier pion masses has been revived
\cite{Dorati:2007bk,Durr:2010ni,Durr:2011mp} and used for chiral fits of the
baryon octet.

Partially quenched heavy baryon $\chi$PT is an extension of heavy
baryon $\chi$PT where the sea and valence quark masses may have
different values
\cite{Labrenz:1996jy,Savage:2001dy,Chen:2001yi,Beane:2002vq}. It is
useful to describe lattice data with multiple valence quark masses for
each sea quark mass and for describing ``hybrid'' calculations where
the sea quarks are regularized differently than the valence
quarks. This technique is sometimes employed to keep the computational
costs of dynamical ensembles small by using a fast fermion
discretization like the staggered one while having the advantages of a
more computationally demanding regularization - like domain wall
fermions or overlap fermions - in the valence sector.

An alternative to the chiral expansion for describing the pion mass
dependence of any hadronic observable is a Taylor expansion around a
finite pion mass. In contrast to the chiral expansion, the Taylor
expansion is performed around a nonsingular point and has a finite
radius of convergence. Typically this convergence radius is given by
the distance of the expansion point ${M_\pi}_0$ to the chiral
limit. Usually, the expansion is performed in powers of the
pseudoscalar mass square $M_\pi^2-M_{\pi_0}^2$ resulting in
\begin{equation}
\label{eq:tayl}
M_X=a+b\left(M_\pi^2-M_{\pi_0}^2\right)+c \bar{M}_K^2+d\left(M_\pi^2-M_{\pi_0}^2\right)^2
\end{equation}
but the chiral behavior of baryon masses can also been fairly
successfully described - at least within the statistical accuracy of
current data sets - by a linear expansion in $M_\pi$
\cite{WalkerLoud:2008bp}. Optimal convergence is achieved in principle
by placing the expansion point at the middle of the interval spanned
by all simulation points and the physical point
\cite{Durr:2008zz,Lellouch:2009fg}.  Note however that from a
practical perspective the choice of the expansion point $M_{\pi_0}^2$
does not play a role in the fit itself as a redefinition of
$M_{\pi_0}^2$ may be absorbed by redefining the lower order fit
coefficients $a$ and $b$ of (\ref{eq:tayl}).

One may also try to fit ratios \cite{Durr:2008zz} or differences
\cite{Bietenholz:2010jr} of baryon masses in order to cancel common
contributions and obtain a more regular chiral behavior. Further it is
possible to study SU(3) breaking effects in baryon multiplets in the
$1/N_c$ expansion \cite{Jenkins:2009wv}, which offers an alternative
way of describing the chiral behavior of baryon mass multiplets.

An alternative to extrapolating or interpolating results to the
physical point is tuning the bare parameters of the lattice theory
such that the physical point is directly reached, i.e. that the
dimensionless combinations of continuum variables mentioned above
assume their physical value on the lattice. While recent advances in
lattice discretizations, algorithms and computer technology have made
such an approach possible in principle, the computational overhead
that is associated with the parameter tuning is still large and the
physical point is generally only reached within the precision of the
tuning procedure.

In order to avoid these problems, reweighting techniques have recently
been applied to this problem. In \cite{Aoki:2009ix}, the PACS-CS
collaboration has reweighted one ensemble to the physical point
directly while the RBC-UKQCD collaboration \cite{Aoki:2010dy} followed
a mixed strategy where the ensembles were first reweighted to the
physical strange quark mass and a subsequent extrapolation to the
physical pion mass was performed.

The general idea behind reweighting \cite{Ferrenberg:1988yz} is to
reuse an ensemble produced with a certain set of parameters
$p_0=\{\beta_0,{m_i}_0\}$ to obtain predictions with a different set
$p=\{\beta,m_i\}$. As discussed in sect.~\ref{sect:numerics}, gauge
configurations $U\in\mathcal{U}$ with the original set of parameters $p_0$ in the
action are produced according to the weight
\begin{equation}
\label{eq:w0}
w(p_0;U)\propto\prod_i\det M({m_i}_0;U)\times e^{-S_G(\beta_0;U)}
\end{equation}
such that expectation values of observables (\ref{eq:expolint}) may be
formed by just summing them over gauge configurations
\begin{equation}
\label{eq:expo0}
\langle O\rangle_{p_0}=\frac{\sum_{U\in\mathcal{U}}O(U)}{\sum_{U\in\mathcal{U}}1}
\end{equation}
For a new set of parameters $p$, one may in principle circumvent the
generation of a new ensemble with the weight
\begin{equation}
\label{eq:w}
w(p;U)\propto\prod_i\det M(m_i;U)\times e^{-S_G(\beta;U)}
\end{equation}
by reusing the old ensemble generated with the weight $w(p_0;U)$ from
(\ref{eq:w0}) and putting the ratio of weights into the observable
\begin{equation}
\label{eq:rewe}
\langle
O\rangle_{p}=\frac{\sum_{U\in\mathcal{U}}O(U) w(p;U)/w(p_0;U)}{\sum_{U\in\mathcal{U}} w(p;U)/w(p_0;U)}
\end{equation}

Although (\ref{eq:rewe}) would in principle allow for a combined
reweighting in both the coupling and the masses, a reweighting was
carried out in the quark masses only in
\cite{Aoki:2009ix,Aoki:2010dy}. For this purpose it is necessary to
compute ratios of fermion determinants at different quark masses. As
it is prohibitively expensive to compute them exactly, stochastic
methods were applied.

Of course the reweighting method has its limitations. As one can see
from (\ref{eq:rewe}), reweighting does exponentially enhance or
suppress the weight of individual configurations in an ensemble with
an exponent that is extrinsic, i.e. contains an explicit volume
factor. As it is crucial for any observable to be computed on the
relevant subset of configurations for the specific parameters used,
these exponential factors should not be so large as to allow for one
configuration to dominate the expectation value entirely. This in turn
limits the allowed range in parameter space that one may reach safely
with reweighting depending on the original set of parameters $p_0$ and
the volume. Within this safe range, the relative suppression of some
configurations is that of effectively decreasing the statistics. As
the relative weight factors are explicitly computed
\cite{Csikor:2004ik}, one has a very good handle on these
effects. Note also that one could in principle bypass some of the
negative effects with a multihistogram technique
\cite{Ferrenberg:1989ui}.

\begin{figure*}
\includegraphics[width=\textwidth]{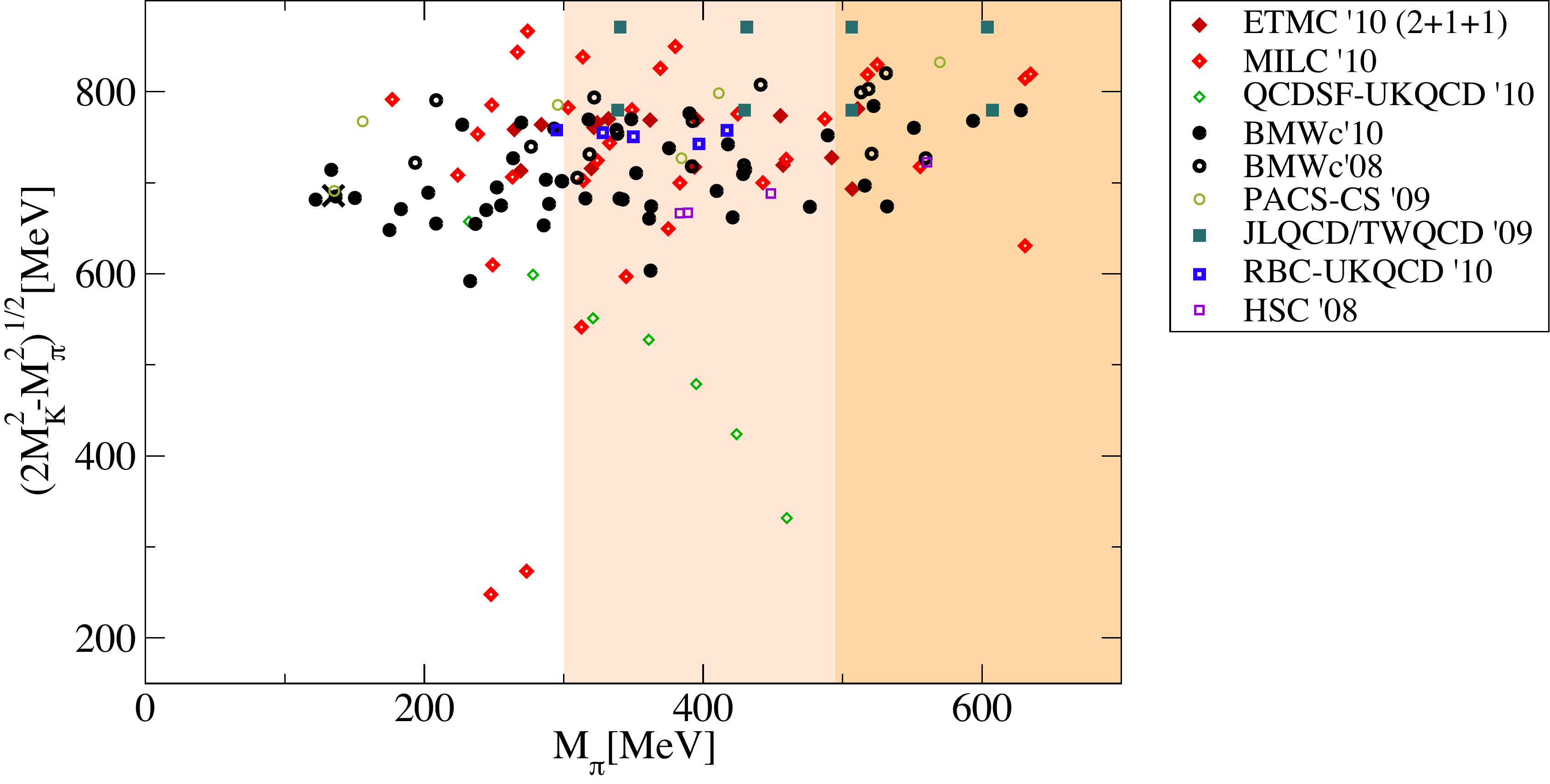}
\caption{\label{fig:l1}
The landscape of recent dynamical fermion simulations projected to the
$\sqrt{2M_K^2-M_\pi^2}$ vs. $M_\pi$ plane. The cross marks the physical
point while shaded areas with increasingly light shade indicate
physically more desirable regions of parameter space.  Data points are
taken from the following references:
ETMC'10(2+1+1) \cite{Baron:2011sf}, 
MILC'10 \cite{Bazavov:2009bb},
QCDSF-UKQCD'10 \cite{Bietenholz:2010si},
BMWc'08 \cite{Durr:2008zz},
BMWc'10 \cite{Durr:2010aw},
PACS-CS'09 \cite{Aoki:2009ix,Aoki:2008sm},
RBC-UKQCD'10 \cite{Aoki:2010dy,Mawhinney:2010xx},
JLQCD/TWQCD'09 \cite{Noaki:2009sk},
HSC'10 \cite{Lin:2008pr} and
all ensembles are from $N_f=2+1$ simulations except explicitly noted
otherwise. For staggered respectively twisted mass ensembles, the Goldstone
respectively charged pion masses are plotted.
}
\end{figure*}

Fig.~\ref{fig:l1} provides an overview of currently used lattice QCD
ensembles with respect to their position in the
$\sqrt{2M_K^2-M_\pi^2}$ vs. $M_\pi$ plane. Note that in leading order
of the chiral expansion the axes are proportional to $\sqrt{\hat{m}}$
and $\sqrt{m_s}$. The location of the physical point also indicated.
As fig.~\ref{fig:l1} indicates, the physical point has already been
reached. 

\subsection{Continuum extrapolation}
\label{sect:cont}

The removal of the cutoff, also known as continuum extrapolation, is
an unavoidable part of any lattice calculation that wants to make a
statement about the underlying fundamental continuum theory. The
severity of the continuum extrapolation however depends very strongly
on both the action used and the combination of scale setting
observable and measured observable.

As discussed in sect.~\ref{sect:improve}, the simplest gluonic action
does already have cutoff effects of $O(a^2)$ that can be improved to
at least $O(g^2 a^2)$ through various techniques.  In contrast, the
scaling behavior of the various fermion actions is typically not as
good. Formally, staggered fermions and twisted mass fermions at
maximal twist as well as exactly chiral fermions show $O(a^2)$
continuum scaling while Wilson type fermions generically start with
$O(a)$ scaling. Only improved staggered fermions such as asqtad have a
leading scaling behavior of $O(g^2 a^2)$. There are several caveats to
this statement however.

For twisted mass fermions, $O(a^2)$ scaling is realized at maximal
renormalized twist \cite{Frezzotti:2003ni,Aoki:2004ta}, which requires
the tuning of one additional parameter. This tuning is routinely done
as part of any twisted mass calculation (see
e.g. \cite{Baron:2011sf}). Since this tuning has a typical accuracy on
the few percent level, it is expected that the $O(a^2)$ terms are
numerically dominant. In addition, twisted mass calculations often
employ a doublet of valence fermions with oposite Wilson parameter to
cancel remnant $O(a)$ effects.

Similarly $O(a^2)$ scaling is only strictly realized for chirally
symmetric fermions if the chiral symmetry is exact. Fermion
formulations that incorporate an inexact chiral symmetry, such as
domain wall fermions at a finite fifth dimension, do formally have a
remaining $O(g^{2n} a)$ scaling behavior. The smallness of the
residual mass and other numerical evidence \cite{Aoki:2010dy} however
suggest that, similar to the twisted mass case, the $O(a^2)$ term is
dominant although it is formally subleading.

Wilson type fermions on the other hand are typically Symanzik improved
by the addition of a Sheikholeslami-Wohlert (clover) term
(\ref{eq:swac}). At tree level ($c_\text{SW}=1$), this results in an
$O(g^2 a)$ scaling of on-shell observables while with a suitable
nonperturbative tuning one can in principle obtain $O(a^2)$
scaling. In addition, there is numerical evidence
\cite{Hoffmann:2007nm,Durr:2008rw,Kurth:2010yk,Durr:2010aw} that the
scaling behavior of clover fermions is substantially improved by gauge
link smearing, which is commonly used today.

Apart from the action used, the continuum scaling is also largely
dependent on the observables considered. As discussed in
sect.~\ref{sect:physpt}, all observables in lattice QCD are
dimensionless quantities and in order to extract dimensionful
quantities such as baryon masses a scale setting observable is needed.
The scaling is of course affected by the choice of scale setting
variable. For baryons and vector mesons good scaling is observed when
choosing a stable light baryon mass as the scale setting observable
\cite{Alexandrou:2009qu,Durr:2008zz}.

Some care has to be taken about the size of the scaling window. While
generally scaling is not expected to set in for lattice spacings
coarser than $a\sim 0.1-0.15\text{~fm}$, it has been observed
\cite{DelDebbio:2002xa,Luscher:2010we,Schaefer:2010hu,Antonio:2008zz,Bazavov:2010xr}
that for fine lattices the autocorrelation time of the topological
charge is rapidly growing. It therefore seems to be prohibitively
expensive with current algorithms to obtain a sufficiently large and
statistically independent ensemble of configurations for lattice
spacings finer than $a\sim 0.05\text{~fm}$.

\begin{figure*}
\includegraphics[width=\textwidth]{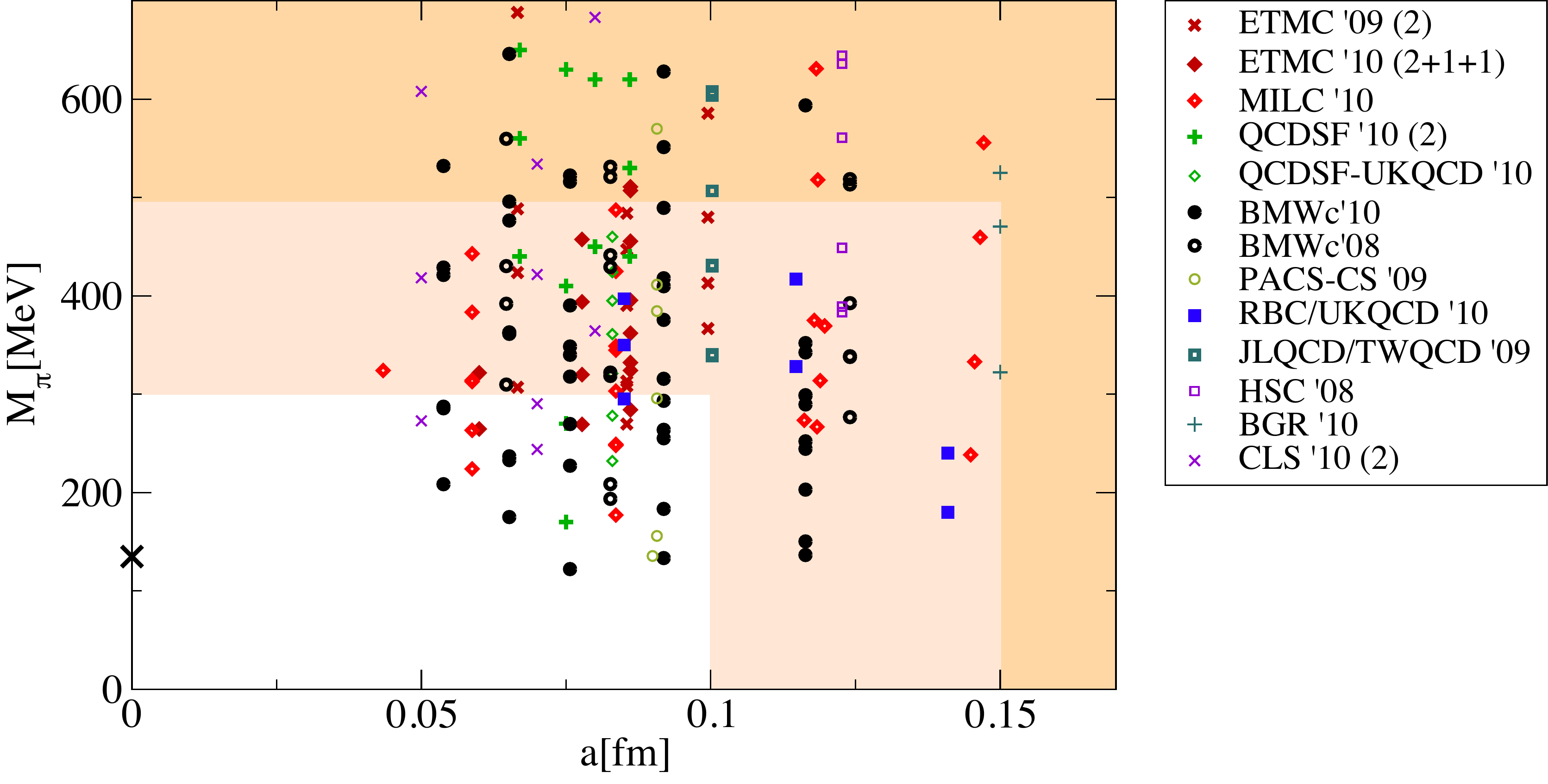}
\caption{\label{fig:l3}
The landscape of recent dynamical fermion simulations projected to the
$M_\pi$ vs. $a$ plane. The cross marks the physical point while shaded
areas with increasingly light shade indicate physically more desirable
regions of parameter space.  Data points are taken from the following
references:
ETMC'09(2) \cite{Blossier:2009bx},
ETMC'10(2+1+1) \cite{Baron:2011sf}, 
MILC'10 \cite{Bazavov:2009bb},
QCDSF'10(2) \cite{Schierholz:2010xx},
QCDSF-UKQCD'10 \cite{Bietenholz:2010si},
BMWc'08 \cite{Durr:2008zz},
BMWc'10 \cite{Durr:2010aw},
PACS-CS'09 \cite{Aoki:2009ix,Aoki:2008sm},
RBC-UKQCD'10 \cite{Aoki:2010dy,Mawhinney:2010xx},
JLQCD/TWQCD'09 \cite{Noaki:2009sk},
HSC'10 \cite{Lin:2008pr},
BGR'10(2) \cite{Engel:2010my} and
CLS'10(2) \cite{Brandt:2010ed}.
All ensembles are from $N_f=2+1$ simulations except explicitly noted
otherwise. For staggered respectively twisted mass ensembles, the Goldstone
respectively charged pion masses are plotted.
}
\end{figure*}

Generally, for the observables considered in this review continuum
scaling is rather mild and not a leading source of systematic error.
Fig.~\ref{fig:l3} gives an overview of the lattice spacing $a$
vs. $M_\pi$ for currently used lattice ensembles. 

\subsection{Finite Volume effects}
\label{sect:fv}

Besides reaching the physical point and removing the cutoff the third
step that generically has to be taken in order to make physical
predictions is the extrapolation to infinite volume. As is the case
for the continuum limit, the infinite volume limit can never be
reached and an extrapolation in the volume is in principle
unavoidable. For most observables however the leading finite volume
corrections are exponentially small in the box size and not
polynomially and can therefore be made sufficiently small in practice
by increasing the volume \cite{Luscher:1985dn}. These finite volume
effects are discussed in sect.~\ref{sect:fvexp}. Resonant states on
the other hand are embedded into a continuum of scattering states at
infinite volume. In finite volume these levels become discrete and
carry a strong volume dependence. Consequently the leading finite
volume effects on resonant states are of a different origin and are
discussed separately in sect.~\ref{sect:fvres}.

Finally we would like to mention that fixing the global topological
charge in QCD is a restriction that becomes irrelevant in the infinite
volume limit, too. For this reason lattice QCD calculations in a fixed
topological sector may be viewed as introducing an additional third
type of finite volume corrections \cite{Brower:2003yx,Aoki:2007ka}.
Since at the time of this writing this technique has not been used in
any work on light hadron spectroscopy we will not discuss it any
further.

\subsubsection{Finite volume effects for stable particles}
\label{sect:fvexp}

In an interacting field theory, the properties of a particle in a
finite box are affected by mirror charge effects. For hadron
spectroscopy this entails that all hadron masses in a finite box
deviate from their infinite volume value with a leading contribution
originating from the pion warping around one spatial lattice
dimension.\footnote{Alternatively in the momentum space view these
  effects may be considered as consequences of the discreteness of the
  momenta in a finite box.}  A generic expectation for the finite
volume correction to any hadron mass $M$ in an $L^3\times T$ box is
therefore\footnote{ For the case of smaller volumes see also
  \cite{Fukugita:1992jj}. They argue that the dominant (polynomial)
  finite size effect is due to the truncation of a hadrons wave
  function.}
\begin{equation}
\label{eq:genfvexp}
1-\frac{M_L}{M_\infty}\propto e^{-M_\pi L}
\end{equation}

As \cite{Luscher:1985dn} demonstrated, there is a relation between the
euclidean finite volume mass correction of a hadron $P$ and the
forward $\pi P$ scattering amplitude in Minkowski space. Concentrating
on the case where a single propagator receives finite volume
corrections, he obtained an explicit expression for the leading term
in an expansion for asymptotically large $L$. Using an alternative
approach, \cite{Gasser:1986vb,Gasser:1987ah,Gasser:1987zq}
incorporated finite volume effects into chiral perturbation
theory. They demonstrated that the finite volume affects only the
propagators and that it can be accounted for by simply replacing the
momentum integration by a summation over the allowed discrete momenta
$p_i=2\pi n_i/L$.

Expanding the relation of \cite{Luscher:1985dn} to include subleading
terms in asymptotic $L$ and using $\chi$PT input for the scattering
amplitudes, \cite{Colangelo:2003hf,Colangelo:2005gd} have combined the
two approaches mentioned above for the case of pseudoscalar mesons. A similar
expansion for baryons has also been pioneered \cite{Colangelo:2010ba}.

\begin{figure*}
  \includegraphics[width=\textwidth]{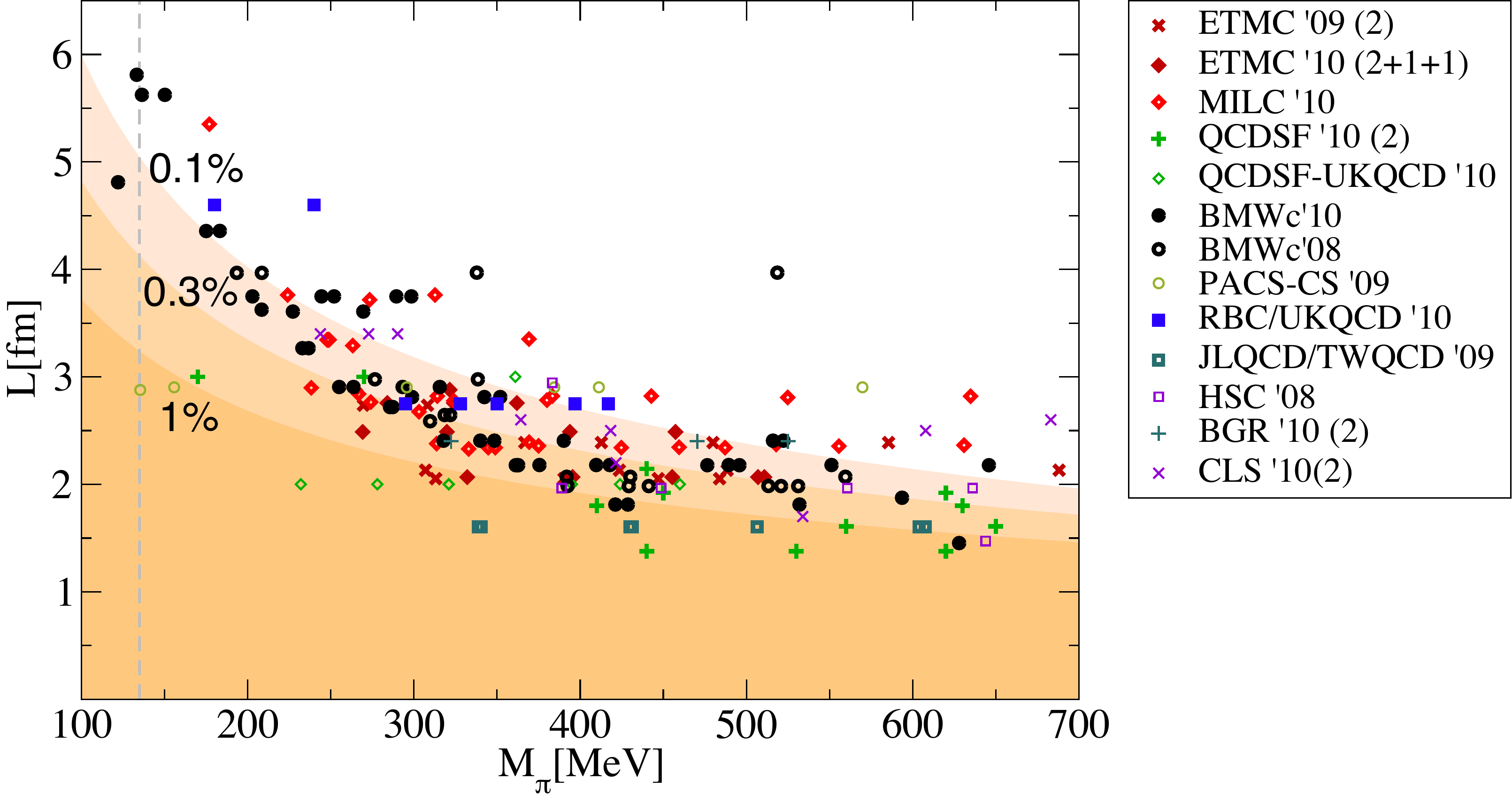} \caption{\label{fig:l2}
    The landscape of recent dynamical fermion simulations projected to
    the $L$ vs. $M_\pi$ plane as given by \cite{Hoelbling:2011kk}.
    The borders of the shaded regions are placed where the expected
    relative error of the pion mass is $1\%$, $0.3\%$ and $0.1\%$
    according to \cite{Colangelo:2005gd}. The vertical dashed line
    indicates the physical pion mass. Data points are taken from the
    following references: ETMC'09(2) \cite{Blossier:2009bx},
    ETMC'10(2+1+1) \cite{Baron:2011sf}, MILC'10 \cite{Bazavov:2009bb},
    QCDSF'10(2) \cite{Schierholz:2010xx}, QCDSF-UKQCD'10
    \cite{Bietenholz:2010si}, 
BMWc'08 \cite{Durr:2008zz},
BMWc'10 \cite{Durr:2010aw},
   PACS-CS'09 \cite{Aoki:2009ix,Aoki:2008sm}, RBC-UKQCD'10
    \cite{Aoki:2010dy,Mawhinney:2010xx}, JLQCD/TWQCD'09
    \cite{Noaki:2009sk}, HSC'10 \cite{Lin:2008pr}, BGR'10(2)
    \cite{Engel:2010my} and CLS'10(2) \cite{Brandt:2010ed}.  All
    ensembles are from $N_f=2+1$ simulations except explicitly noted
    otherwise. For staggered respectively twisted mass ensembles, the Goldstone
respectively charged pion masses are plotted.}
\end{figure*}

From a practical point of view these results imply that there is a
safe asymptotic region of relatively large lattice volumes where these
finite size effects are exponentially small and in addition can be
systematically corrected for. As a rule of thumb for lattice
computations with pion masses above $\sim 300\text{~MeV}$, lattices
with $M_\pi L>4$ are considered safe while those with $m_\pi L<3$ are
widely affected by finite volume corrections. For a more quantitative
statement, fig.~\ref{fig:l2} shows a plot of box size $L$ vs. pion
mass $M_\pi$ where regions are identified that according to
\cite{Colangelo:2005gd} imply the finite volume effect on the pion
mass to be $<1\%$, $<0.3\%$ and $<0.1\%$ respectively. On top of these
regions parameters of current or recent lattice computations are
superimposed.  As one can see current lattices are typically large
enough to have percent level or smaller finite volume corrections on
the pion mass. Note however, that corrections to baryon masses can be
substantially larger \cite{Colangelo:2010ba}.

\subsubsection{Finite volume effects for unstable particles}
\label{sect:fvres}

Finite volume corrections are not always just exponentially small at
large $L$ as discussed in sect.~\ref{sect:fvexp}. In case where one is
interested in extracting the mass of a resonant state that in infinite
volume is embedded into a continuous spectrum of scattering states
finite volume effects are more complicated.  For illustration we
start by considering the hypothetical case where there is no coupling
between the resonance (which we will refer to as ``heavy state'' in
this paragraph) and the scattering states.  In a finite box of size
$L$, the spectrum in the center of mass frame consists of two particle
states with energy
\begin{equation}
\label{eq:epa}
\sqrt{M_1^2+{\vec{k}}^2}+\sqrt{M_2^2+{\vec{k}}^2}
\end{equation}
where $k_i=2n_i\pi/L$ and $M_1$, $M_2$ are the finite volume masses of
the lighter particles (cf. sect.~\ref{sect:fvexp}) and, in addition,
of the state of the heavy particle with finite volume mass $M_X$.  As
we increase $L$, the energy of any one of the two particle states
decreases and eventually becomes smaller than the energy $M_X$ of
$X$. An analogous phenomenon can occur when we fix $L$ but reduce the
quark mass since the energy of the two light particles changes more
than $M_X$.  In the presence of interactions, this level crossing
disappears and, due to the mixing of the heavy state and the
scattering state, an avoided level crossing phenomenon is
observed. Such mass shifts due to avoided level crossing can distort
the chiral extrapolation of hadron masses to the physical pion mass.

\begin{figure}
\includegraphics[width=0.5\textwidth]{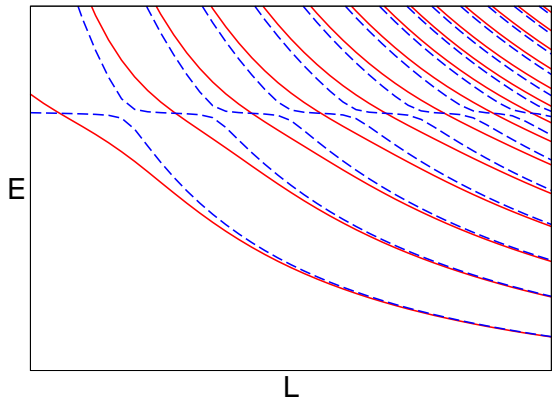}
\caption{
\label{fig:rescom}
Plot of the finite volume energy levels vs. box size $L$ according to
\cite{Luscher:1990ux}. The full lines roughly correspond to the
$\rho-\pi\pi$ system, the dashed lines show the behavior in the case
of a much smaller coupling $g_\rho$.
}
\end{figure}

The literature
\cite{Luscher:1986pf,Luscher:1990ux,Luscher:1991cf,Rummukainen:1995vs,Durr:2008zz}
provides a conceptually satisfactory basis to study resonances in
lattice QCD: each measured energy corresponds to a momentum, $|{\bf
  k}|$, which is a solution of a complicated non-linear equation. We
follow \cite{Luscher:1991cf} where the $\rho$-resonance was taken as
an example and it was pointed out that other resonances can be treated
in the same way without additional difficulties. The $\rho$-resonance
decays almost exclusively into two pions. The absolute value of the
pion momentum is denoted by $k=|{\bf k}|$.  The total energy of the
scattered particles is
\begin{equation}
\label{eq:etot}
W=2\sqrt{M_\pi^2+k^2}
\end{equation}
in the center of mass frame. The $\pi\pi$ scattering phase
$\delta_{11}(k)$ in the isospin $I=1$, spin $J=1$ channel passes
through $\pi/2$ at the resonance energy, which correspond to a pion
momentum $k$ equal to
\begin{equation}
\label{eq:ppiq}
k_\rho=\sqrt{\frac{M_\rho^2}{4}-M_\pi^2}
\end{equation}
In the effective range formula
\begin{equation}
\label{eq:effrang}
(k^3/W)\cdot\cot \delta_{11}=a+bk^2
\end{equation}
this behavior implies
\begin{equation}
\label{eq:adef}
a=-bk_\rho^2=\frac{4k_\rho^5}{M_\rho^2\Gamma_\rho}
\end{equation}
where $\Gamma_\rho$ is the decay width the resonance (which can be
parametrized by an effective coupling between the pions and the
$\rho$). The basic result of \cite{Luscher:1990ux} is that the
finite-volume energy spectrum is still given by (\ref{eq:etot}) but
with $k$ being a solution of a complicated non-linear equation, which
involves the $\pi\pi$ scattering phase $\delta_{11}(k)$ in the isospin
$I=1$, spin $J=1$ channel and reads
\begin{equation}
\label{eq:kdef}
n\pi-\delta_{11}(k)=\phi(q)
\end{equation}
Here $k$ is in the range $0<k<\sqrt{3}M_\pi$, $n$ is an integer,
$q=kL/(2\pi)$ and $\phi(q)$ is a known kinematical function which can
be evaluated numerically. In the limit of small $q$, $\phi(q)\propto
q^3$ and $\phi(q)\approx \pi q^2$ for $q\ge 0.1$ to a good
approximation.  Solving the above equation leads to energy levels for
different volumes and pion masses. Fig.~\ref{fig:rescom} illustrates
these solutions as a function of the box size.

Thus, the spectrum is determined by the box length $L$, the infinite
volume masses of the resonance $M_X$ and the two decay products $M_1$
and $M_2$ and one parameter, $g_X$, which describes the effective
coupling of the resonance to the two decay products and is thus
directly related to the width of the resonance. 

In infinite volume the resonance manifests itself as an increased
state density in a continuous spectrum. Identifying the infinite
volume resonance on a lattice at a given finite volume is therefore
not straightforward - generically it is not possible to identify the
resonance with a single energy level. Scanning over different system
lengths it is however possible to identify energies with an increased
probability of finding a state.\footnote{Note that depending on the
  specific volume this might be either the ground state or any of the
  excited states (cf. fig.~\ref{fig:rescom}).} as seen in
fig.~\ref{fig:rescom}.  This property may in principle be used to
identify a resonance in a lattice calculation
\cite{Bernard:2008ax,Giudice:2010ch,Bernard:2010fp}.  An alternative
method based on finite time correlators has also recently been
suggested by \cite{Meissner:2010rq}.

Although conceptually clear, the treatment of resonant states in a
region where they are not the ground state faces the huge challenge of
reliably extracting the ground state as well as a number of excited
states. One therefore often uses the assumption that an operator which
does not mirror the valence quark structure of a scattering state will
almost exclusively couple to the resonance for extracting directly the
desired resonance level. Recent studies \cite{Lin:2008pr,Engel:2010my}
provide some evidence for the validity of this assumption.

\subsection{Electromagnetic effects and isospin breaking}
\label{sect:EM}

Including dynamical fermions into a lattice QCD calculation is
numerically expensive due to the occurrence of the fermion determinant
in (\ref{eq:expolint}). For this reason, early lattice calculations
have been performed ignoring the effect of dynamical fermions all
together (quenched approximation). With improved algorithmic
understanding and the increase in available computer power the
inclusion of some dynamical fermion effects has become possible. The
first step is usually the inclusion of a degenerate pair of light
quarks followed by the inclusion of a single strange quark.

Although the physical $u$ and $d$ quarks are far from degenerate
$m_u/m_d\sim0.56$ \cite{Nakamura:2010zzi}, both masses are much
smaller than the QCD scale $\Lambda_\text{QCD}$ and therefore can be
treated as a pair of degenerate light quarks with mass
$\hat{m}=(m_u+m_d)/2$ to a very good approximation. In addition,
quarks are electrically charged and a full understanding of the
experimentally observed hadron spectrum therefore necessarily includes
QED effects, too. For most observables related to hadron
  spectroscopy, these are however subdominant as can be readily seen
by the smallness of the coupling constant $\alpha_\text{EM}$ relative
to the QCD coupling constant $\alpha$. Effects of other interactions
or of heavier quarks are negligible for light hadron spectroscopy
within the currently attainable precision.

The relatively largest electromagnetic/isospin breaking effects can be
observed in the pions and also for kaons the effect is still at the
percent level. Since both pion and kaon masses are often used to
define the physical point, it is necessary to define a properly
isospin averaged pion and kaon mass as an input to lattice QCD
calculations with a degenerate pair of light quarks.  We denote the
physical mass splittings in the pion/kaon sector as
$\Delta_\pi=M_{\pi^\pm}^2-M_{\pi^0}^2$ and
$\Delta_K=M_{K^\pm}^2-M_{K^0}^2$.  Calling the pion and kaon mass in
pure, isospin averaged QCD $M_\pi$ and $M_K$ respectively, we can
write the pion and kaon masses in full QCD+QED as
\begin{equation}
\label{eq:mpidef}
M_\pi^2+
\begin{array}{l}
I_{\pi^+}+\Gamma_{\pi^+}=M_{\pi^+}^2\\
I_{\pi^0}+\Gamma_{\pi^0}=M_{\pi^0}^2
\end{array}
\end{equation}
and
\begin{equation}
\label{eq:mkdef}
M_K^2+
\begin{array}{l}
I_{K^+}+\Gamma_{K^+}=M_{K^+}^2\\
I_{K^0}+\Gamma_{K^0}=M_{K^0}^2
\end{array}
\end{equation}
where the $I_x$ are the isospin and the $\Gamma_x$ are the QED
corrections to the squared mass of the particle $x$.  In order to
obtain these, we start by noting that upon interchanging $u$ and $d$
in the pion sector $\pi^+$ and $\pi^-$ are interchanged while for
kaons a $K^+$ goes over into a $K^0$ and vice versa. Disregarding QED
effects, we therefore see that both $M_{\pi^+}^2$ and
$\left(M_{K^+}^2+M_{K^0}^2\right)/2$ may only contain even powers when
expanding in the isospin breaking parameter $(m_u-m_d)$ and are
therefore $I_{\pi^+}\propto (m_u-m_d)^2$ and $I_{K^0}+I_{K^+}\propto
(m_u-m_d)^2$ are free of leading (linear) isospin breaking
effects. Moreover, as \cite{Gasser:1983yg} have demonstrated,
$I_{\pi^+}\propto (m_u-m_d)^2(m_u+m_d)$ leading to an even stronger
suppression. Within currently attainable precision, isospin breaking
corrections to both $M_{\pi^+}^2$ and
$\left(M_{K^+}^2+M_{K^0}^2\right)/2$ are therefore negligible and we
can assume $I_{\pi^+}=I_{K^0}+I_{K^+}=0$.

Regarding the isospin correction to the charged pion,
$I_{\pi^0}-I_{\pi^+}\propto (m_u-m_d)^2$ itself and
\cite{Gasser:1984gg} have found a parameter free expression at NLO in
terms of physical meson masses that yields
\begin{equation}
\epsilon_m\equiv\frac{I_{\pi^+}-I_{\pi^0}}{\Delta_\pi}\simeq0.04
\end{equation}
indicating that the bulk of the pion mass splitting is due to
electromagnetic effects.

Regarding electromagnetic effects, according to \cite{Dashen:1969eg}
the picture in the leading order of the $SU(3)$ chiral expansion is
that while neutral pseudo-Goldstone masses stay unaffected by
electromagnetic corrections, i.e. $\Gamma_{\pi^0}=\Gamma_{K^0}=0$, the
difference of the square of the charged masses receive the same
correction $\Gamma_{\pi^+}=\Gamma_{K^+}$. The absence of
electromagnetic corrections in the $\pi^0$ and to a lesser extent the
$K^0$ mass is further justified by \cite{Das:1967it} who demonstrated
that these corrections vanish in the massless limit $m_u=m_d=0$
respectively $m_u=m_d=m_s=0$.  On the other hand, a range of model
calculations \cite{Bijnens:1996kk,Donoghue:1996zn} that are partly
based on the inclusion of QED effects into chiral perturbation theory
\cite{Urech:1994hd} and dispersive calculations based on the $\eta\rightarrow 3\pi$ decay
\cite{Kambor:1995yc,Anisovich:1996tx,Ditsche:2008cq,Colangelo:2009db,Bijnens:2007pr,Leutwyler:1996qg}
suggest that there are noticeable corrections to the other parts of
Dashen's theorem.  A recent world average of these corrections was
given in \cite{Colangelo:2010et} as
\begin{equation}
\label{eq:dashvi}
\epsilon\equiv\frac{\Gamma_{\pi^0}-\Gamma_{\pi^+}-\left(\Gamma_{K^0}-\Gamma_{K^+}\right)}{\Delta_\pi}=0.7(5)
\end{equation}
Assuming Dashen's theorem to hold we would thus get for the QED
corrected, isospin averaged pion and kaon masses $M_\pi\simeq
134.8\text{~MeV}$ and $M_K\simeq 495\text{~MeV}$. Including the
correction (\ref{eq:dashvi}) while keeping the reasonable assumption
$\Gamma_{\pi^0}=\Gamma_{K^0}=0$, the kaon mass is shifted by less than
$1\%$ to $M_K\simeq 494.6\text{~MeV}$ while the pion mass is
unaffected.

In order to go beyond these results, we need to treat QED and isospin
breaking effects on the lattice. Since it has been difficult to reach
the physical point even in the isospin limit with two degenerate light
quarks and since at least for the pion QED effects are substantially
larger, it is these QED effects that have received most attention in
the literature to date.

Regularizing QED on the lattice poses a very different set of problems
than regularizing QCD. There is a straightforward way of including the
leading QED corrections into QCD calculations that was first employed
by \cite{Duncan:1996xy} to obtain an estimate for the pion mass
splitting. Since QED is an abelian gauge theory, the electromagnetic
gauge field is trivial when ignoring the electrical charge of the sea
quarks. In this partially quenched approximation one can therefore
sample free QED $U(1)$ fields independently of the QCD $SU(3)$
fields. This is most conveniently achieved by not sampling the
parallel transports $U_\mu(x)=e^{ieA_\mu(x)a}$ directly but by instead
fixing the gauge and sampling the underlying gauge fields $A_\mu(x)$
in which the Maxwell equations are linear. The corresponding action
\begin{equation}
\label{eq:lagem}
S_\text{QED}=\frac{1}{4e^2}\sum_x\left(
\partial_\mu A_\nu(x)-\partial_\nu A_\mu(x)
\right)^2
\end{equation}
with the forward difference operator $\partial_\mu$ decouples in
Fourier space and gauge field configuration with the proper weight
$e^{-S_\text{QED}}$ may thus be produced by simply producing each
Fourier component with a proper random weight. The only further
restriction is the vanishing of the $p=0$ component due to the compact
space provided by the finite lattice.

In a calculation with dynamical
sea quarks, the electromagnetic corrections to the light quark masses
that is introduced in the valence sector leads to a mismatch of sea
and valence quark masses. In order to minimize these unitarity
violating effects, one can retune the valence light quark masses such
that after the inclusion of quenched QED effects both $m_u$ and $m_d$
have the same value as they had in pure QCD \cite{Portelli:2010yn}.

In the quenched approximation calculation of \cite{Duncan:1996xy} a
large violation of Dashen's theorem was found corresponding to
$\epsilon\sim 0.5$. In later work with $N_f=2$ dynamical flavors of
domain wall fermions, \cite{Blum:2007cy} have reported a somewhat
larger value while in the most recent update $N_f=2+1$
\cite{Blum:2010ym} found a value compatible with the original estimate
of \cite{Duncan:1996xy}.  Preliminary results are also available from
the MILC collaboration using $N_f=2+1$ flavors of staggered fermions
\cite{Basak:2008na} and the Budapest-Marseille-Wuppertal collaboration
with $N_f=2+1$ Wilson fermions \cite{Portelli:2010yn}.

The general picture that emerges is that the corrections to Dashen's
theorem that are parameterized in $\epsilon$ are in agreement with the
phenomenological determinations. Taking lattice determinations into
account, \cite{Colangelo:2010et} have combined recent results on QED
and isospin splitting effects into a world average of the QED
corrected, isospin averaged pion and kaon masses. For the individual
symmetry breaking parameters they find
\begin{equation}
\begin{aligned}
\epsilon&=0.7(5)
&
\epsilon_m&=0.04(2)\\
\Gamma_{\pi^0}&=0.07(7)\Delta_\pi
&
\Gamma_{K^0}&=0.3(3)\Delta_\pi
\end{aligned}
\end{equation}
and consequently $M_\pi=134.8(3)\text{~MeV}$
and $M_K=494.2(5)\text{~MeV}$, which agree within error with the
values quoted above that were obtained under the assumption that
Dashen's theorem holds.

Isospin splitting effects in the baryon spectrum are less dramatic
than for the pseudoscalar mesons. As an example, the mass splitting in
the nucleon system is $M_n-M_p=1.2933321(4)\text{~MeV}$
\cite{Nakamura:2010zzi} and an effective theory estimate of the
electromagnetic contribution turns out to be negative
$\left(M_n-M_p\right)_\text{QED}=-0.76(30)\text{~MeV}$
\cite{Gasser:1982ap}. If one is interested in isospin averaged baryon
masses only, a straight isospin average is therefore sufficient for
the accuracies that are presently obtainable in lattice calculations.

A first dedicated lattice study of the nucleon mass splitting was
carried out by \cite{Beane:2006fk}. They used a hybrid setup with
domain wall valence on $N_f=2+1$ staggered sea quarks with pion masses
in the range of $\sim 290-350\text{~MeV}$. A single lattice spacing
$a\sim 0.125\text{~fm}$ was used and an extrapolation to the physical
point was carried out in the framework of NLO partially quenched heavy
baryon $\chi$PT. Using the experimental value of $M_\Delta-M_N$ as an
input they obtain for the isospin part of the mass splitting
$\left(M_n-M_p\right)_\text{QCD}=2.26(57)(43)\text{~MeV}$.

In addition to the isospin part of the nucleon mass diference,
\cite{Blum:2010ym} have also calculated the QED part. They use
$N_f=2+1$ partially quenched domain wall fermions with pion masses in
the range of $\sim 250-400\text{~MeV}$ at a single lattice spacing
$a\sim 0.11\text{~fm}$. Two volumes were used to estimate finite size
effects in the final result and an extrapolation to the physical point
was performed using NLO partially quenched heavy baryon $\chi$PT. They
quote the final results
$\left(M_n-M_p\right)_\text{QCD}=2.24(12)\text{~MeV}$,
$\left(M_n-M_p\right)_\text{QED}=-0.383(68)\text{~MeV}$ and
$M_n-M_p=1.86(14)(47)\text{~MeV}$ where the first error is statistical
and the second part of the systematic error.

\section{Lattice results}
\label{sect:results}

In this section we discuss lattice results on the light hadron
spectrum. Historically, the first results were from the quenched
approximation that is discussed in sect.~\ref{sect:quenched}. The
inclusion of dynamical fermions was pioneered with heavy, degenerate
quarks (sect.~\ref{sect:nf2}) before it developed into the study of
theories with non-degenerate light and strange quarks that we review
in sect.~\ref{sect:nf21}. While sect.~\ref{sect:quenched} and
\ref{sect:nf2} are now of mainly historical interest, the three-flavor
(and coming four-flavor) dynamical calculations are the definitive
modern calculations.

Our review of lattice results does not include glueballs. For recent
reviews on this topic see
e.g. \cite{Teper:1998kw,Klempt:2007cp,McNeile:2008sr,Mathieu:2008me}.

\subsection{Results in the Quenched approximation}
\label{sect:quenched}

Although the quantitative understanding of the light hadron spectrum
is an obvious and essential check for any candidate theory of the
strong interaction, it took several years from the original proposal
of \cite{Wilson:1974sk} that contained the lattice discretization of
gauge theories and the strong coupling picture of quark confinement to
the first numerical studies of the hadron spectrum
\cite{Marinari:1981nu,Hamber:1981zn,Weingarten:1982qe,Hasenfratz:1982bw,Hamber:1982mz,Martinelli:1982eu,Fukugita:1982ap,Fucito:1982ip}.
Due to the lack of viable dynamical fermion algorithms and computer
power, these pioneering studies were carried out in the quenched
approximation sometimes with an $SU(2)$ gauge group and even further
discrete truncations. Lattices had a typical size of $6^3\times 12$
and $O(10)$ gauge configurations were generated with the Wilson gauge
action. Naive or plain Wilson fermion actions were typically used to
extract hadron masses and physical point predictions were obtained by
linear extrapolation of either squares of meson masses or baryon
masses. A first world average of these pioneering results was given by
\cite{Creutz:1983ev}
\begin{equation}
\begin{split}
m_\rho&=800(100)\text{~MeV}\\
m_{a_0}&=950(150)\text{~MeV}\\
 m_{a_1}&=1100(150)\text{~MeV}\\
 m_p&=1000(150)\text{~MeV}\\
 m_\Delta&=1300(150)\text{~MeV}\\
\end{split}
\end{equation}
From a modern perspective, these results should be viewed with some
caution as these calculations were clearly exploratory and
pioneering. The computer power of the times was not sufficient to
properly clarify many systematic effects. As an example, the inverse
lattice spacing of $SU(3)$ gauge theory with the Wilson gauge action
at $\beta=6.0$ used by \cite{Marinari:1981nu} was
$a^{-1}=1.12\text{~GeV}$ whereas modern determinations from various
observables agree that it is $a^{-1}\simeq 2.1-2.3\text{~GeV}$
\cite{Lipps:1983pi,Gutbrod:1983yh,Otto:1984qr,Aoki:1999yr,Giusti:2001pk,Necco:2001xg,Durr:2006ky}.

It was quickly realized
\cite{Gupta:1982hu,Bowler:1982uq,Politzer:1983qm,Hasenfratz:1982ft,Bernard:1982hh}
that physical volumes were not big enough and that one should use
larger time extents in order to safely extract a ground state
signal. In the following years, quenched calculations with unimproved
Wilson and staggered fermions on Wilson gauge action were pushed to
larger lattice volumes and higher statistics
\cite{Lipps:1983pi,Billoire:1984jm,Bowler:1984dh,Kunszt:1984zu,Konig:1984rz,Gilchrist:1984jb,Billoire:1984xj,Bowler:1985hr,Itoh:1985vv,Itoh:1986gy}
where lattices were often doubled in time direction in order to obtain
a clean signal. With gauge couplings typically $\beta\sim 5.7-6$ and
spatial lattice extents typically $10-16$ lattice units and time
extents typically twice as much, a qualitatively consistent picture of
the hadron masses started to emerge although large systematic effects
were present that could not clearly be identified yet. (For reviews of
this generation of results see
\cite{Montvay:1985wh,Hasenfratz:1985pd}). In particular, the ratio of
the nucleon mass to the $\rho$ mass, which experimentally is
$M_N/M_\rho\simeq 1.21$ turned out to be consistently too high
$M_N/M_\rho>1.6$, which is even larger than the static quark limit
$M_N/M_\rho=1.5$. Another stumbling block for these early calculations
was the absence of sufficient splitting between the masses of the
nucleon and the $\Delta$. 

During the following years, the focus shifted slightly towards
inclusion of sea quark effects with steady progress in quenched
spectroscopy
\cite{Fukugita:1987wm,Gupta:1987zc,Bacilieri:1988fh,Bacilieri:1988zq}
until the first precision calculations of the quenched light hadron
spectrum emerged in the early 1990's
\cite{Bacilieri:1989dn,Gupta:1990mr,Cabasino:1990zu,Cabasino:1991mr,Allton:1992sg,Allton:1993ps,Butler:1992ki,Guagnelli:1992zq,Bitar:1992dk,Daniel:1992ek,Kim:1993gc,Butler:1994em}.

Among these, the first landmark precision calculation of the quenched
light hadron spectrum was carried out by the GF11 collaboration
\cite{Butler:1992ki,Butler:1994em}. Wilson fermions were used on a
Wilson gauge action at three different values of the lattice spacing
in the range $a\sim 0.07-0.14\text{~fm}$. Propagators were extracted
using point and gaussian smeared sources at different quark masses
corresponding to $M_\pi/M_\rho>0.5$ i.e. with pion masses
$M_\pi\gtrsim 400\text{~MeV}$. Lattice volumes in the range
$16^3\times 32$ to $30^2\times 32\times 40$ were used corresponding to
a spatial lattice extent of $\sim 2.3\text{fm}$ at all three lattice
spacings. At the coarsest lattice spacing, dedicated runs at larger
and smaller volumes were performed in order to extract the finite
volume dependence of the result. They were used in the end to correct
the physical predictions to infinite volume.\footnote{See
  \cite{Gottlieb:1996hy} for a detailed discussion of the finite
  volume effects.} Considering degenerate quarks only, a linear
relation was established between the degenerate quark mass $m_q$ and
$M_\pi^2$ while for all other hadrons a fit of the form $M=a+b m_q$
described the data. Assuming that these linear relations extend to the
non-degenerate case with two quarks of mass $m_1$ and $m_2$,
i.e. $M_\pi^2\propto (m_1+m_2)^2$ and $M=a+b_1 m_1+b_2 m_2$ the
physical point was found using $M_\pi$, $M_K$ and $M_\rho$ input with
the later used as scale setting observable. A continuum extrapolation
linear in $a$ was performed that turned out to be rather mild.

\begin{table}
  \includegraphics[width=0.5\textwidth]{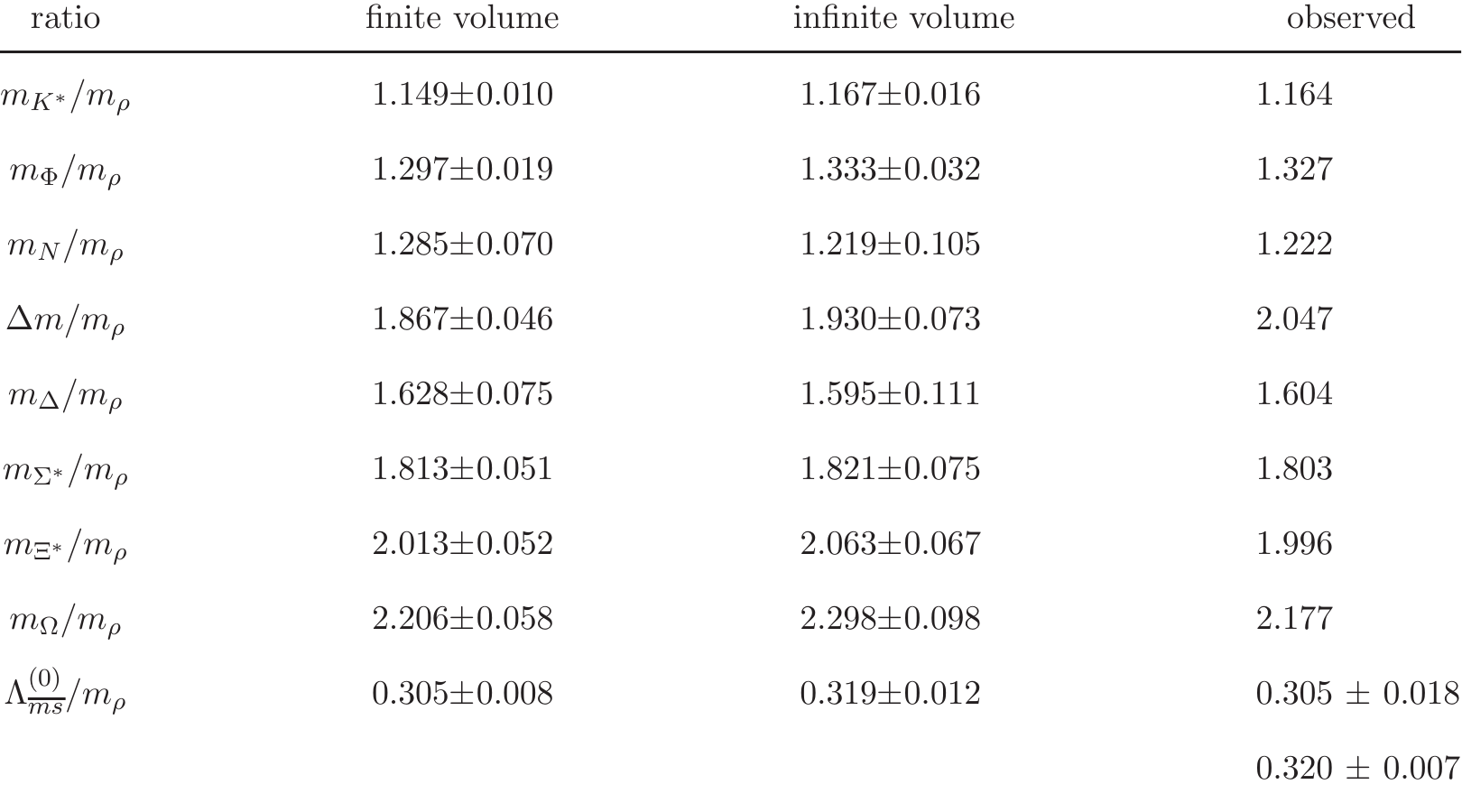}
  \caption{\label{tab:gf11} Quenched lattice QCD prediction of the
    light hadron spectrum according to
    \cite{Butler:1992ki,Butler:1994em}. The table gives the ratio of
    various hadron masses and the QCD scale
    $\Lambda_{\overline{\text{MS}}}^{(0)}$ to the
    mass of the $\rho$ that was used to set the scale. The label
    $\Delta m$ refers to the combination $\Delta
    m=m_\Xi+m_\Sigma-m_N$. Observed values are experimental results
    from \cite{Hikasa:1992je} except for the case of the QCD scale
    $\Lambda_{\overline{\text{MS}}}^{(0)}$ where they refer to two
    previous results from the literature
    \cite{ElKhadra:1992vn,Bali:1992xm,Bali:1992ru}.
    Note that some experimental values, notably the mass of the $\rho$, have been
    updated since \cite{Nakamura:2010zzi}.}
\end{table}

Table~\ref{tab:gf11} shows the resulting spectrum obtained by
\cite{Butler:1992ki,Butler:1994em}. Despite the many approximations
used, the overall agreement with experiment is rather remarkable and
at the $<10\%$ level.

Similarly sophisticated quenched analyses were soon after performed
for the $\eta-\eta^\prime$ system \cite{Kuramashi:1994aj} and for
excited state mesons \cite{Lacock:1995tq,Lacock:1996vy}. These
calculations and detailed studies of systematic effects such as finite
size \cite{Aoki:1993gi}, excited state contaminations
\cite{Iwasaki:1995cm} or quenched chiral logarithms and $SU(3)$
splittings \cite{Kim:1995tg,Bhattacharya:1995fz} revealed potential
inconsistencies of the quenched approximation of up to $20\%$.  On the
other hand, results with improved Wilson actions
\cite{Collins:1996zc,Gockeler:1996jx,Edwards:1997nh,Gockeler:1997fn}
and for staggered fermions that reached finer lattice spacings and
smaller quark masses \cite{Bernard:1998db,Kim:1999ur} indicated
quenching effects that were less dramatic at $O(5\%)$. In the case of
the latter two results it was especially noted that a simple linear
extrapolation in the light quark mass was no more sufficient. Several
$\chi$PT motivated fit forms were found to describe the chiral
behavior of the $\rho$ and nucleon masses but the coefficients were
not found to be in agreement with quenched $\chi$PT expectations at
all.

\begin{figure}
  \includegraphics[width=0.5\textwidth]{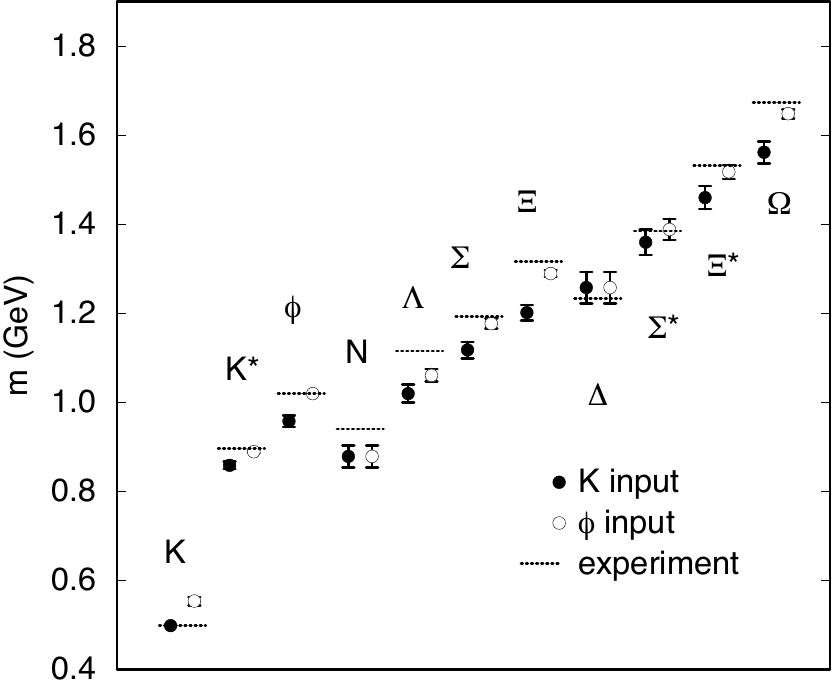}
  \caption{\label{fig:cppacs} Quenched lattice QCD prediction of the
    light hadron spectrum according to
    \cite{Aoki:1999yr,Aoki:2002fd}. For both data sets plotted
    $M_\rho$ and $M_\pi$ were used to set the scale and the light
    quark mass. In order to set the strange mass, either $M_K$ (filled
    circles) or $M_\phi$ (open circles) are used. Plot reproduced with
    friendly permission of the CP-PACS collaboration.}
\end{figure}

The accuracy of the quenched approximation was addressed in the large
scale calculation by the CP-PACS collaboration
\cite{Aoki:1999yr,Aoki:2002fd}. They used lattices of $\sim
3\text{~fm}$ spatial extent at four values of the lattice spacing in
the range $a\sim 0.05-0.1\text{~fm}$ with quark masses corresponding
to $M_\pi/M_\rho\sim 0.4-0.75$. Both fermion and gauge action used
were plain Wilson and non-degenerate quark masses were used to
investigate splittings in the $SU(3)$ multiplets. While pseudoscalar
meson masses were found to have a chiral behavior compatible with the
quenched $\chi$PT expectations, the discrepancy in the vector meson
and baryon sector found in the staggered results of
\cite{Bernard:1998db,Kim:1999ur} was confirmed. For these masses,
$\chi$PT motivated fits were used to extrapolate to the physical
point. The physical point in the light quark mass was defined using
$M_\pi$ and $M_\rho$ and either $M_K$ or $M_\phi$ were used to define
the physical strange mass.\footnote{Note that although the $\phi$ is a
  singlet meson, its disconnected part is usually disregarded in
  lattice studies.} The final result that has a precision of $\sim
1-3\%$ is displayed in fig.~\ref{fig:cppacs}.

A statistically significant deviation from the experimentally observed
spectrum was noted with discrepancies up to $\sim 10\%$. This
discrepancies however are particularly pronounced due to the choice of
$M_\rho$ as a scale setting observable. Since the $\rho$ does not
decay in the quenched approximation and therefore represents the
ground state in the vector channel, it is in principle a viable scale
setting variable from a pure lattice perspective. Nonetheless, the
identification of the stable quenched ground state energy with the
mass of a resonance with $\sim 150\text{~MeV}$ experimental width is
not optimal. On top of that, an accurate determination of $\rho$ meson
properties is a challenging experimental task. This is highlighted by
the fact that the experimental value of $M_\rho$ itself has moved by
$\sim 1\%$ or more than ten standard deviations over the last two
decades \cite{Hikasa:1992je,Nakamura:2010zzi}.  

As \cite{Garden:1999fg} noted, one can derive from the CP-PACS results
predictions for hadron masses with the scale set by the nucleon mass
instead of $M_\rho$. In this case, the maximum deviation from
experiment turns out to be significantly lower at $\sim 4\%$ indicating
that indeed the quenched approximation is substantially worse for
resonance masses than for masses of hadrons that are stable within
QCD. A confirmation of these results with somewhat lower statistical
accuracy was reported by \cite{Bowler:1999ae}.

The CP-PACS calculation was one of the last large scale calculation
aimed at a precision determination of the quenched ground state light
hadron spectrum. As quenching effects had become statistically
significant, the focus of efforts to get a quantitative confirmation
of QCD reproducing the experimentally observed ground state light
hadron spectrum moved towards inclusion of dynamical fermion
effects. Nonetheless, due to its relatively low numerical cost, the
quenched approximation continues to be used to this day as a testbed
for new numerical approaches and as a first step in studying
computationally demanding observables. In the following years, the
quenched ground state hadron spectrum was used to check various
variants of chirally symmetric fermion actions
\cite{Gattringer:2003qx,Babich:2005ay,Galletly:2006hq} or to develop
improved \cite{Melnitchouk:2002eg} or anisotropic \cite{Nemoto:2003ft}
actions that were intended for studying excited hadrons.

Following the CP-PACS determination of the ground state quenched
hadron spectrum the attention in quenched hadron spectroscopy turned
towards resonant and singlet states
\cite{Lee:1999kv,McNeile:2000xx,Gockeler:2001db,Sasaki:2001nf,Melnitchouk:2002eg,Mathur:2003zf,Gattringer:2003qx,Nemoto:2003ft,Brommel:2003jm,Sasaki:2005ap,Mathur:2006bs,Burch:2006dg,Burch:2006cc,Lasscock:2007ce,Wada:2007cp,Basak:2007kj,Fleming:2009wb,Engel:2010my}. In
particular, many groups reported on the splitting between the nucleon
and its lightest negative parity partner, the $N^*(1535)$. In all
cases, a clear signal of the mass splitting between the nucleon ground
state and the $N^*(1535)$ could be seen on the lattice. The splitting
is increasing as one lowers the light quark masses towards the
physical point and in all cases is roughly consistent with the
experimentally observed mass splitting. It is an interesting
peculiarity however, that the lightest experimentally observed nucleon
excited state is not the nucleon parity partner $N^*(1535)$ but in
fact the $N^*(1440)$, the so-called Roper resonance which carries
positive parity and $J=1/2$ as the nucleon.  With the exception of
\cite{Mathur:2003zf,Sasaki:2005ap} who employed Bayesian techniques to
extract excited state information from a single channel\footnote{See
  \cite{Sasaki:2005ug} for a discussion of possibly large finite size
  effects for this technique} however, the first positive parity
excitation of the nucleon turned out to lie above the first negative
nucleon state in all lattice calculations. A possible solution
to this discrepancy has recently been proposed by
\cite{Mahbub:2009aa,Mahbub:2010jz} who demonstrated that a mix of
excited states enters typical interpolating operators. By using a
large operator basis they could explicitly disentangle up to eight
states and demonstrate a level crossing between the negative parity
ground state and the first positive parity excitation for light quark
masses that is consistent with the finding of
\cite{Mathur:2003zf,Sasaki:2005ap} and the experimentally observed
level ordering between the $N^*(1535)$ and $N^*(1440)$ states.

Another interesting case in excited state baryon spectroscopy is the
$\Lambda(1405)$. In the quark model picture, it is the lightest
negative parity partner of the $\Lambda$ with a valence quark
structure $uds$. It is however the lightest negative octet baryon -
more than $100\text{~MeV}$ lighter than the lightest negative parity
nucleon, the $N^*(1520)$ - even though it does contain a strange
quark. This is the most striking of many peculiar features that have
given rise to a number of suggestions for a nontrivial structure of
the $\Lambda(1405)$ such as that of a $N\bar{K}$ hadronic molecule or
a pentaquark state (for a recent review see \cite{Klempt:2009pi}).
\cite{Melnitchouk:2002eg,Nemoto:2003ft} have studied the negative
parity $\Lambda$ states in the quenched approximation with standard
interpolating operators and found it impossible to reproduce the
$\Lambda(1405)$ which they interpret as an indication for a nontrivial
structure of the $\Lambda(1405)$ that might not be properly reflected
in the quenched approximation. In contrast, \cite{Burch:2006cc} found
the $\Lambda(1405)$ to be consistent with the negative parity octet
state.

With the exception of the $N^*(1440)$ and the $\Lambda(1405)$ that
were discussed above, no qualitative tension between experiment and
the above mentioned quenched excited baryon studies was found. For a
more detailed review see \cite{Leinweber:2004it}.

\subsection{Results with degenerate dynamical quarks}
\label{sect:nf2}

The spectrum calculation of the CP-PACS collaboration
\cite{Aoki:1999yr,Aoki:2002fd} reached a numerical precision such that
quenching effects could clearly be seen. In order to obtain a
quantitative understanding of the ground state light hadron spectrum
on the few percent level it is therefore necessary to include
dynamical fermion effects into the lattice calculation. From the six
fermion flavors in nature, the charm, bottom and top each have masses
much larger than the QCD scale $\Lambda_\text{QCD}$. Their
contribution to light hadron masses through quark loop effects is
therefore believed to be negligible. Among the remaining three
flavors, $m_{u/d}\ll\Lambda_\text{QCD}$ while
$m_s\sim\Lambda_\text{QCD}$. Consequently and because an even number
of quark flavors is usually easier to implement
(cf. sect.~\ref{sect:numerics}), the first attempts at unquenching
lattice QCD calculations were performed with two degenerate quark
flavors. In the case of staggered fermions calculations with four
dynamical quark flavors are even easier due to the remnant doubling
(cf. sect.~\ref{sect:stag}). In this section we review results
obtained with (a usually even number of) degenerate\footnote{We use
  the term non-degenerate here in the sense that no explicit term was
  added that does break the flavor symmetry. In the case of staggered
  quarks the unavoidable taste splitting is of course present.}
dynamical fermions. While this still represents an approximation that
is necessitated by the lack of proper computational resources, it is
still a very significant step forward from the quenched approximation.

Pioneering work on lattice hadron spectroscopy with dynamical fermions
was done by \cite{Langguth:1984ue} where dynamical fermions were
implemented using a strong coupling expansion of the determinant
ratio. During the following years unquenching via the inclusion of
$N_f=2$ and $4$ dynamical Wilson and $N_f=2,3$ and $4$ staggered
fermions was investigated by several groups
\cite{Fukugita:1986wf,Fucito:1986ui,Billoire:1986tg,Fukugita:1986tg,Hamber:1987mq,Gottlieb:1987cw,Fukugita:1987qb,Campostrini:1987ht,DeTar:1987xb,DeTar:1987ar,Grady:1987yd}. These
early works demonstrated that the main effect of including dynamical
fermions was a change in dependence of the lattice spacing on bare
coupling constant $\beta$. Apart form this effect, no clear sign of
unquenching could be observed. In particular, the $M_\rho/M_N$ ratio
tended to stay constant or even increase. In studies with staggered
fermions, the taste breaking effects were also observed to be rather
severe. For a comprehensive review of these early studies see
\cite{Fukugita:1987ca}.

During the following years it became clear that with staggered
fermions one could go to substantially lighter quark masses than with
Wilson fermions
\cite{Patel:1989pv,Bitar:1990cb,Bitar:1990wk,Gupta:1991sn,Brown:1991qw,Altmeyer:1992dd,Fukugita:1992hr,Bitar:1992dk}
in the sense that the $M_\pi/M_\rho$ ratio attainable with Wilson
fermions was limited to $M_\pi/M_\rho\gtrsim 0.7$ while if taking the
lightest pion one could go down to about half this number in the
staggered case. Since reducing the mass of the valence quarks only was
substantially easier, some of these studies started exploring
partially quenched setups, where the valence quark masses are varied
independently of the sea quark masses, and even hybrid calculations
with valence Wilson quarks on a dynamical staggered sea. None of these
calculations however gave a clear signal for a $M_\rho/M_N$ ratio that
was substantially better than the ones obtained in contemporary
quenched calculations. Although there was steady progress over the
following few years \cite{Bitar:1993xd,Allton:1998gi,Eicker:1998sy},
the focus of large scale calculations shifted more towards precision
computations in the quenched approximation. This was in large part
due to the tremendous computational effort that was needed for
dynamical fermion computations which exceeded the computer capabilities
of that time. A first unquenched study of the $\eta-\eta^\prime$
mixing was performed by \cite{McNeile:2000hf} which found a mixing
angle of $\theta\sim -10^\circ$ albeit without continuum and chiral
extrapolation.\footnote{See also \cite{Lesk:2002gd}.}

These efforts culminated in the first large scale project to compute
the light hadron spectrum in $N_f=2$ QCD by the CP-PACS collaboration
\cite{Ali_Khan:2001tx}. They used two degenerate flavors of mean field
improved clover fermions on an Iwasaki gauge action. The strange
valence quark was included in a quenched setup. Three relatively coarse
lattice spacings in the range $a\sim 0.11-0.22\text{~fm}$ were used
with an approximately constant physical volume $L\sim 2.5\text{~fm}$
and $T=2L$. Four sea and valence quark masses in a range corresponding
to $M_\rho/M_N\sim 0.6-0.8$ and an additional valence quark mass at
$M_\rho/M_N\sim 0.5$ were investigated. Point sources and
exponentially smeared quark sources on a gauge fixed background were
chosen for optimal plateau onset. Chiral extrapolation was performed
by a combined fit to all partially quenched masses for each channel on
a given sea quark mass. Vector meson and baryon masses were
extrapolated to the physical point using quadratic functions in the
valence and sea $M_\pi^2$ with certain restrictions on the quadratic
terms. In the case of vector mesons, $\chi$PT motivated nonanalytic
$M_\pi^3$ type terms were also used instead of $M_\pi^4$ type terms to
estimate the systematic error. Following the example of the quenched
CP-PACS calculation discussed in sect.~\ref{sect:quenched}, the
physical light quark masses and the scale are defined via $M_\pi$ and
$M_\rho$ while two options, $M_K$ or $M_\phi$, were used to set the
strange quark mass. The continuum limit is obtained by linear
extrapolation in $a$.

\begin{figure*}
  \includegraphics[width=0.49\textwidth]{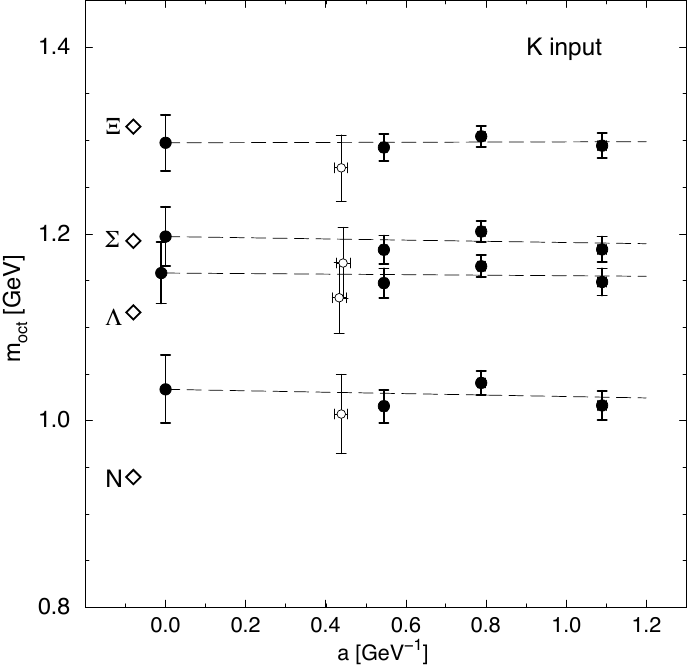}
  \includegraphics[width=0.49\textwidth]{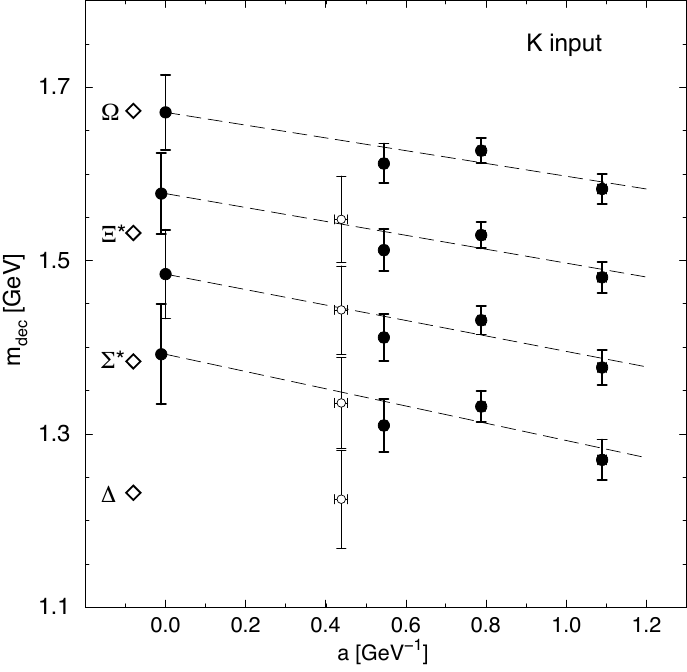}
  \caption{\label{fig:cppacsuq} Dynamical $N_f=2$ lattice QCD
    prediction of the light baryon spectrum according to
    \cite{Ali_Khan:2001tx}. The two figures display the continuum
    extrapolation of the ground state light octet and decuplet baryon
    masses in the case where the strange mass was set via $M_K$. The
    continuum extrapolation was performed using three lattice spacings
    that are displayed with solid circles. Open circles represent a
    fourth, finer lattice which was not included in the analysis due
    to a too small volume.  Plot reproduced with friendly permission
    of the CP-PACS collaboration.}
\end{figure*}

The resulting light hadron spectrum is plotted in
fig.~\ref{fig:cppacsuq}. Clearly the heavier baryon states are in good
agreement with experiment while the lighter ones, especially the
nucleon and the $\Delta$, seem to be systematically too high. This
does not come as a big surprise though since the extrapolation to the
physical point is substantially more severe for the baryons containing
more light valence quarks.

Similar efforts to that of the CP-PACS collaboration were reported by
the UKQCD and JLQCD collaboration in \cite{Allton:2001sk,Aoki:2002uc}.
The UKQCD collaboration worked at a single lattice spacing $a\sim 0.1
\text{fm}$ that was set with $r_0$. The range of sea quark masses was
chosen such that $M_\pi/M_\rho\sim 0.55-0.9$ and the spatial lattice
extent was $L\sim1.7\text{~fm}$.  The JLQCD collaboration worked at
one single lattice spacing $a\sim 0.09\text{~fm}$ at a spatial extent
$L\sim1.8\text{~fm}$ using clover fermions on a Wilson gauge action
and the same range of sea masses $M_\pi/M_\rho\sim 0.6-0.8$ than
CP-PACS. The findings of both collaborations on the light hadron
spectrum are in good agreement with the continuum CP-PACS results.

The conclusion from these two large scale projects regarding the
feasibility of computations with light dynamical quarks was summarized
in a plot that became to be known as the ``Berlin wall plot'' by
\cite{Ukawa:2002pc}. He conjectured that the cost of dynamical fermion
simulations would rise towards the chiral limit essentially as
$\propto M_\pi^6$ effectively rendering any dynamical calculations
with Wilson type fermions near the physical point impossible in any
foreseeable future without substantial algorithmic improvements. Due
to their lower computational cost staggered fermions were able to push
further towards the chiral limit and did already do so with also a
dynamical strange quark (see sect.~\ref{sect:nf21}), but extracting
especially baryonic states is less straightforward in this formulation
(cf. sect.~\ref{sect:lvl}). For a more comprehensive review of these
results see \cite{McNeile:2003dy}.

Because the inclusion of a non-degenerate sea quark incurs little
extra expense, from this point on, the main development in light
hadron spectroscopy continued with the inclusion of a strange quark in
addition to two degenerate light quarks.  This is usually referred to
as $N_f=2+1$ and is discussed in sect.~\ref{sect:nf21}.  The question
of how much two-flavor calculations differ from experiment has not
been answered as definitively as it has been for the quenched
calculations discussed earlier. A number of $N_f=2$ calculations of
light hadron masses were still being performed however for a number of
reasons such as algorithmic tests
\cite{DelDebbio:2006cn,DelDebbio:2007pz} or as a first step for
formulations that allow even number of quark flavors only
\cite{Alexandrou:2009qu}. Similarly, flavor singlet spectroscopy that
typically requires substantially more statistical precision has still
been investigated in $N_f=2$ QCD
\cite{McNeile:2001cr,Kunihiro:2003yj,Allton:2004qq,Prelovsek:2004jp,Hart:2006ps,McNeile:2006nv,McNeile:2009mx}. For
a recent comprehensive review of these results see
\cite{McNeile:2007fu}.

\begin{figure}
\centerline{\includegraphics[width=0.5\textwidth]{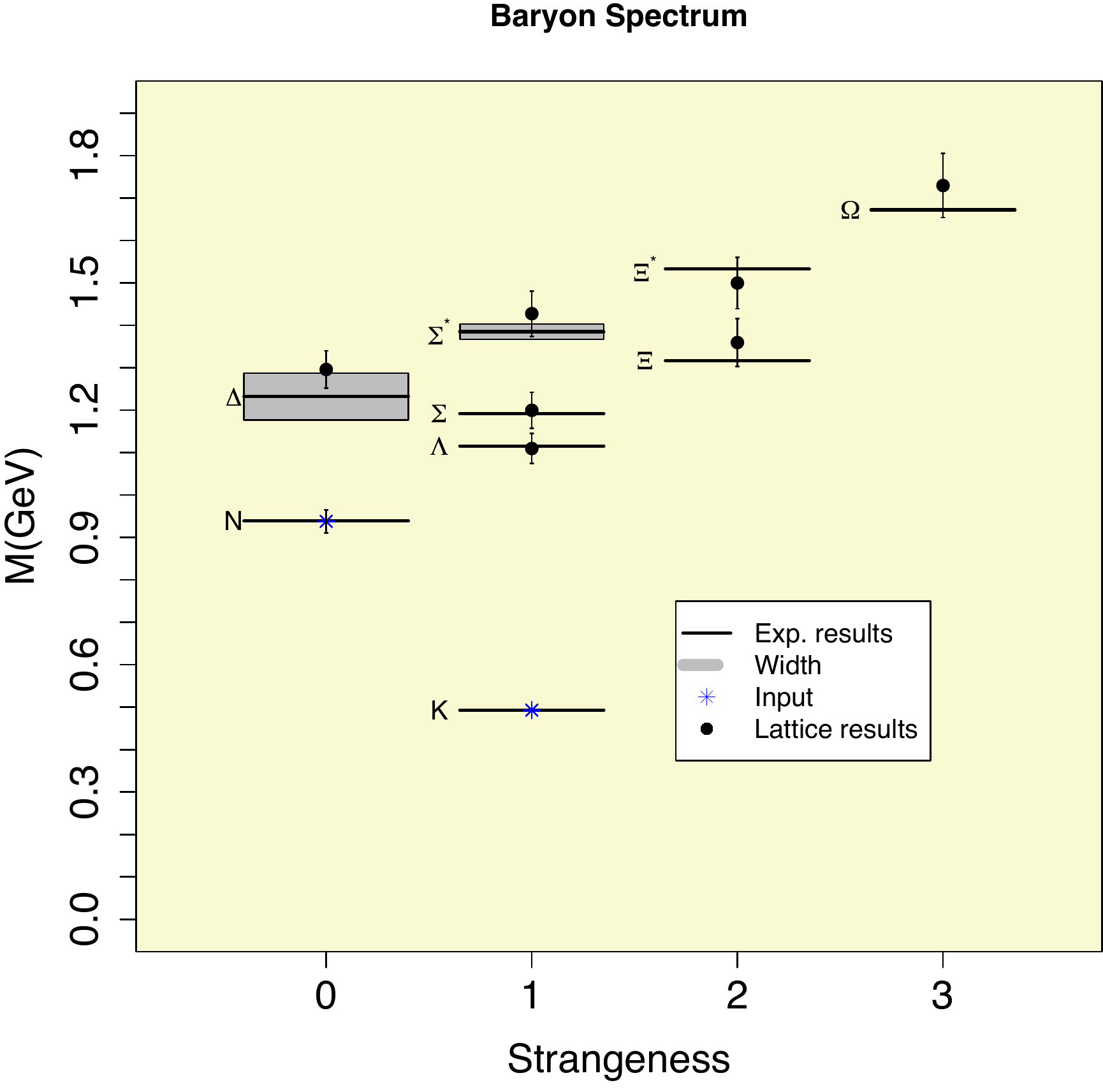}}
\caption{\label{fig:etmc}
Baryon spectrum obtained by the ETM
collaboration with $N_f=2$ twisted mass fermions. The plot is reproduced
from \cite{Alexandrou:2009qu} with friendly permission of the ETM
collaboration.
}
\end{figure}

The ETM collaboration \cite{Alexandrou:2009qu} has published results
for the ground state light baryon spectrum with $N_f=2$ twisted mass
fermions.\footnote{Ses also \cite{Alexandrou:2008tn} for some results
  with more lattice spacings and different scale setting.} They used
$N_f=2$ twisted mass fermions at maximal renormalized twist on a tree
level Symanzik improved gauge action. Two lattice spacings ($a\sim
0.07\text{~fm}$ and $a\sim 0.09\text{~fm}$) were used with charged
pion masses in the range $270$ to $500\text{~MeV}$.\footnote{The
  isospin splitting of the pions is ${M^\pm_\pi}^2-{M^0_\pi}^2\sim
  (150-220\text{~MeV})^2$ \cite{Baron:2009wt}} The lattice spacing was
set via the nucleon mass and chiral extrapolations were performed with
a variety of different ans\"atze. The valence strange quark mass is
set by tuning the kaon mass to its physical value. The final result
employs two different heavy barion $\chi$PT ans{\"a}tze ($O(p^3)$
respectively NLO $SU(2)$) for the extrapolation of baryons without respectively with
valence strange quarks to the physical mass point. The continuum
extrapolation was performed using a constant which was demonstrated to
be sufficient at the given level of accuracy.  Exponential finite
volume corrections were taken into account in the final fit
form. Resonant state finite volume corrections were not performed but
are believed to be irrelevant in the region of parameter space covered
by the simulations. Effects of the twisted mass isospin breaking were
observed to be negligible except in the case of the $\Xi$ where they
amounted to a $6\%$ correction. Their final result is displayed in
fig.~\ref{fig:etmc} shows good agreement with experiment at the level
of precision of the calculation which is $\sim 5\%$. The ETM
collaboration also investigated the $\rho$-$\omega$ mass splitting and
mixing \cite{McNeile:2009mx} excluding electromagnetic effects. While
a clear signal and qualitatively correct behavior was found, the
quantitative understanding of the experimentally observed splitting
remains a challenging task.

\begin{figure}
\centerline{\includegraphics[width=0.5\textwidth]{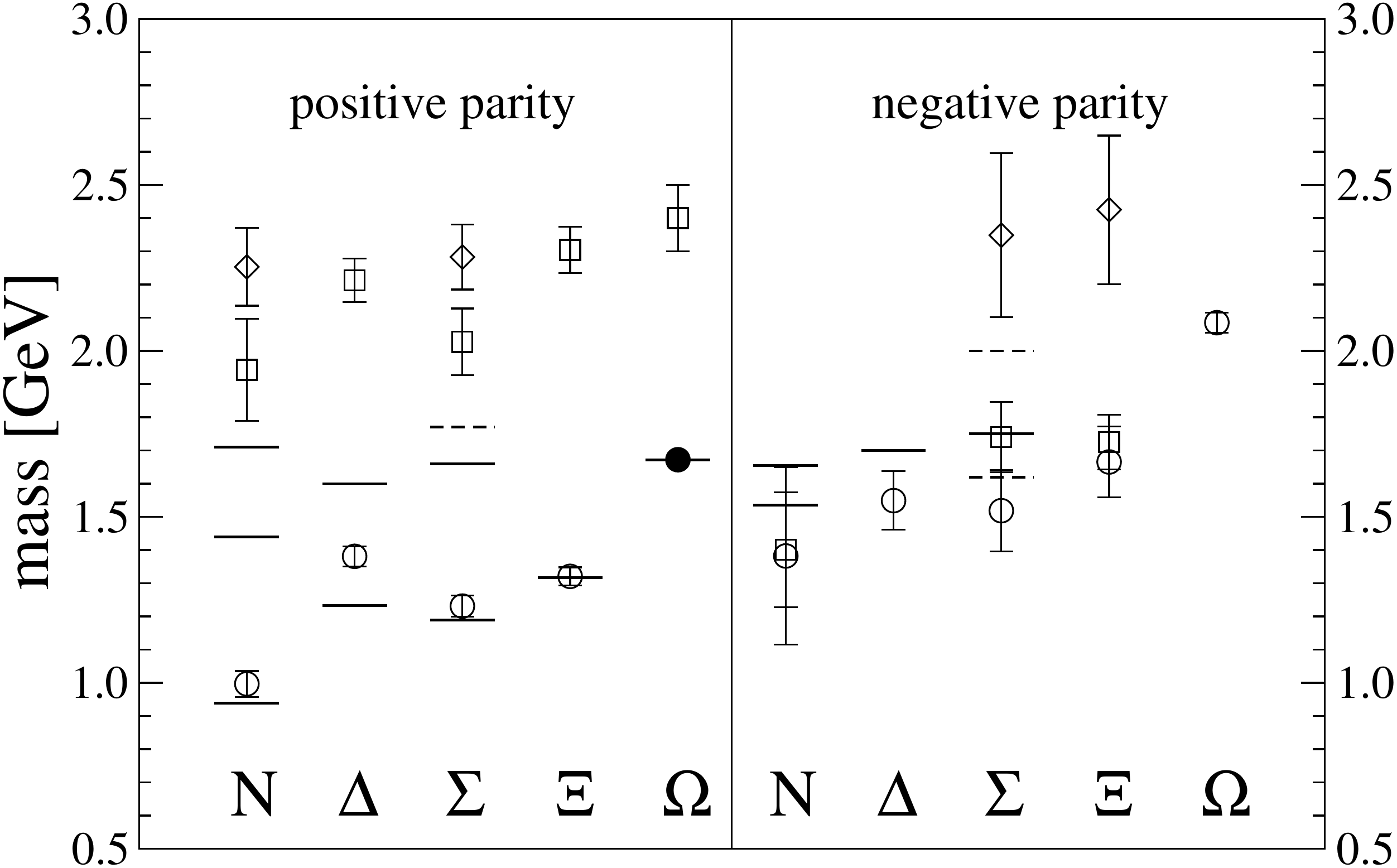}}
\caption{\label{fig:bgr}
Positive and negative baryon spectrum obtained by the BGR
collaboration with $N_f=2$ chirally improved fermions. The plot is reproduced
from \cite{Engel:2010my} with friendly permission of the BGR
collaboration.
}
\end{figure}

Turning to excited states, the BGR collaboration has computed ground
and excited state hadron spectra using $N_f=2$ single step stout
smeared chirally improved fermions on a tadpole improved
L\"uscher-Weisz gauge action at a singe lattice spacing $a\sim
0.15\text{~fm}$ \cite{Prelovsek:2010kg,Engel:2010my}. Three pion
masses in the range $320-530\text{~MeV}$ were used and the scale was
set with $r_0$ . Gaussian smeared quark sources were used in
combination with a variational method based on three interpolating
operators to extract the energy levels. A chiral extrapolation linear
in $M_\pi$ was performed and the strange quark was introduced in a
partially quenched setup. The results for positive and negative baryon
states are plotted in fig.~\ref{fig:bgr}. A good signal for the ground state was found
but excited and scattering state signals were generally weak. Some
evidence was also presented that the $\sigma$ and $\kappa$ resonances
contain a sizable exotic admixture of a tetraquark
($\bar{q}\bar{q}qq$) state.

In an extension of the work of \cite{Melnitchouk:2002eg,Nemoto:2003ft}
in the quenched approximation, \cite{Takahashi:2009bu} have studied
the $\Lambda^*(1405)$ on $N_f=2$ CP-PACS lattices and essentially
reach the same conclusion as \cite{Melnitchouk:2002eg,Nemoto:2003ft}
that the $\Lambda^*(1405)$ can not be reproduced using standard baryon
octet and singlet interpolating operators.

\subsection{Results with dynamical light and strange quarks}
\label{sect:nf21}

\begin{figure}
\centerline{\includegraphics[width=0.5\textwidth]{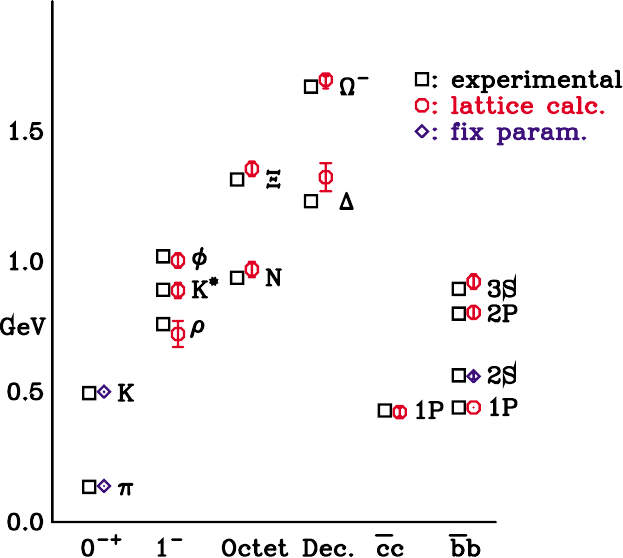}}
\caption{\label{fig:milc}Comparison of the $N_f=2+1$ light hadron
  spectrum results from the MILC collaboration \cite{Bazavov:2009bb}
  with experiment.  Diamonds are input quantities while circles are
  predictions. Experimental masses of hadrons from
  \cite{Amsler:2008zzb} are indicated by squares. Note that the plot
  also includes charmonium and bottomonium masses with some of the
  later ones used to set the scale.  Plot reproduced with friendly
  permission of the MILC collaboration.}
\end{figure}

The first large scale computation of the light hadron spectrum with a
pair of light and one strange sea quark was performed by the MILC
collaboration \cite{Bernard:2001av,Aubin:2004fs}\footnote{See also
  \cite{Davies:2003ik} where the effects of unquenching are discussed
  for observables beyond the light hadron spectrum.}. With asqtad
fermions on a one-loop Symanzik improved gauge action, they reached
Goldstone (i.e. taste pseudoscalar) pion masses down to $M_\pi\sim
260\text{~MeV}$ on lattices of spatial size $L\sim 2.4\text{~fm}$ and
$L\sim 3.4\text{~fm}$ at two values of the lattice spacing $a\sim
0.09\text{~fm}$ and $a\sim 0.12\text{~fm}$. Finite volume effects were
explicitly checked for and found to be under control. The fermion
update algorithm used was the R-algorithm and explicit checks for the
absence of step size dependent effects were performed. The scale was
set via b-meson spectroscopy, in particular the $\Upsilon$ 1P-1S mass
splitting, and physical light and strange quark masses were defined by
$M_\pi$ and $M_K$. Ground state meson and some baryon masses were
computed as well as the radially excited pseudoscalar meson state. The
extrapolation to physical pion masses was performed using various
heavy baryon $\chi$PT motivated fit functions and a continuum
extrapolation was done using $g^2a^2$ terms were possible. An update of
these results including data from finer lattices as well as a
comprehensive review is available in \cite{Bazavov:2009bb}. The
resulting light hadron spectrum is displayed in
fig.~\ref{fig:milc}. Note that due to the particular difficulties in
extracting baryon masses in the staggered formulation
(cf. sect.~\ref{sect:extr}) there are only predictions for a subset of
the ground state baryons. Again, the numbers turn out to be in good
agreement with experiment.

A subset of the MILC ensembles with $a\sim 0.12\text{~fm}$ and with a
smallest pion mass of $\sim 290 \text{~MeV}$ has been studied in
\cite{WalkerLoud:2008bp} in a mixed action setup with domain wall
valence quarks. Comparing different chiral fit forms for the nucleon
mass it was demonstrated that a simple linear fit in $M_\pi$ gives a
good description of the given data set and extrapolates to the correct
value at the physical point. In the same paper, this feature has also
been found in other collaborations data.

\begin{figure}
\centerline{\includegraphics[width=0.5\textwidth]{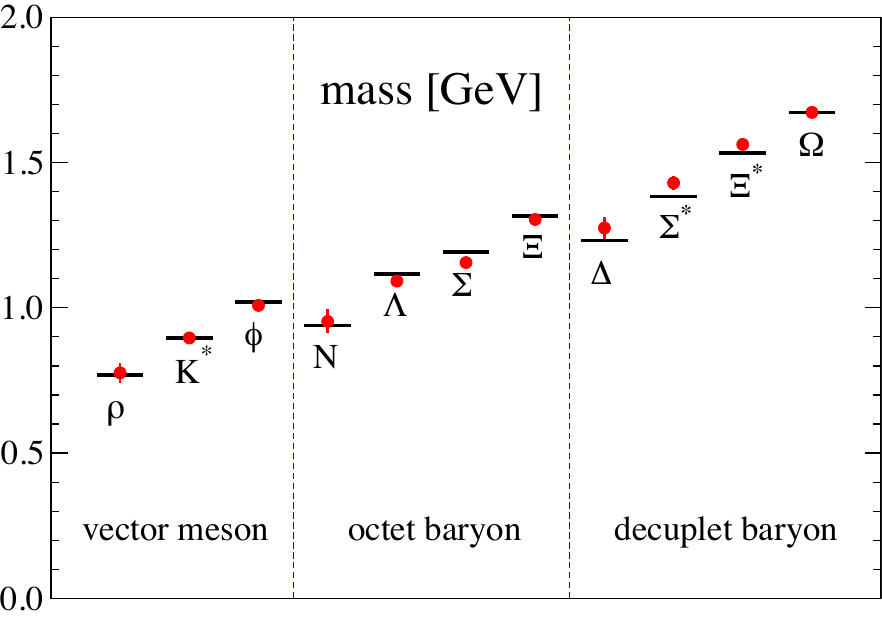}}
\caption{\label{fig:pacscs}The extrapolated $N_f=2+1$ light hadron
  spectrum results from the PACS-CS collaboration. Experimental data
  are from \cite{Amsler:2008zzb}. The plot is reproduced from
  \cite{Aoki:2008sm} with friendly permission of the PACS-CS
  collaboration.}
\end{figure}

The PACS-CS collaboration has published results for the light hadron
spectrum using both a chiral extrapolation \cite{Aoki:2008sm} and a
direct reweighting to the physical point \cite{Aoki:2009ix}. In both
cases $N_f=2+1$ nonperturbatively $O(a)$ improved cover fermions on an
Iwasaki gauge action were used at a single lattice spacing $a\sim
0.09\text{~fm}$ and a spatial lattice extent of $L\sim
2.9\text{~fm}$. Pion masses down to $\sim 150\text{~MeV}$ were
directly simulated and a reweighting to the physical point was carried
out with the lightest ensemble. In the extrapolated ensemble finite
size effects on the pseudoscalar masses were corrected using SU(2)
$\chi$PT at NLO. The tiny chiral extrapolation was performed linearly
in the light quark mass and $M_\Omega$ was used to set the scale. More
involved chiral forms were subsequently investigated in
\cite{Ishikawa:2009vc}.  Similarly in the reweighted ensemble the
masses of the $\pi$, $K$ and $\Omega$ were used to tune to the
physical point. The final result from the extrapolation method is
plotted in fig.~\ref{fig:pacscs}. Very similar results have been found
with the reweighting method as detailed in \cite{Aoki:2009ix}.

\begin{figure}
\centerline{\includegraphics[width=0.5\textwidth]{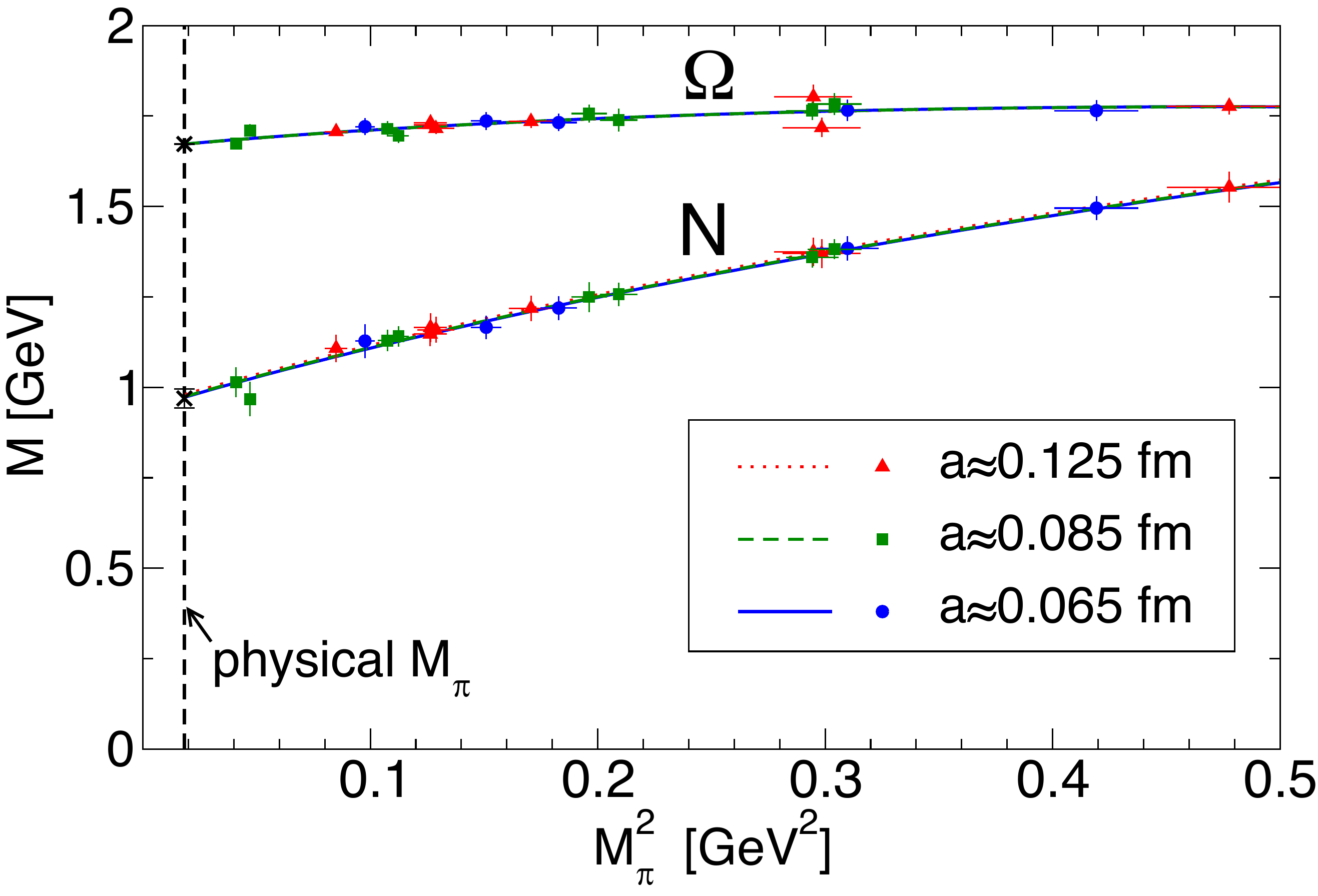}}
\caption{\label{fig:bmwscal} 
Sample chiral and continuum extrapolation
  of the lattice hadron masses of \cite{Durr:2008zz} at
  physical $M_K^2-M_\pi^2/2$ in physical units. The scale
setting variable $M_\Omega$ and the nucleon mass are plotted vs. the
square of the pion mass together with a fit of the data at every
lattice spacing. The vertical dashed line represents the physical pion
mass.}
\end{figure}

\begin{figure}
  \includegraphics[width=0.5\textwidth]{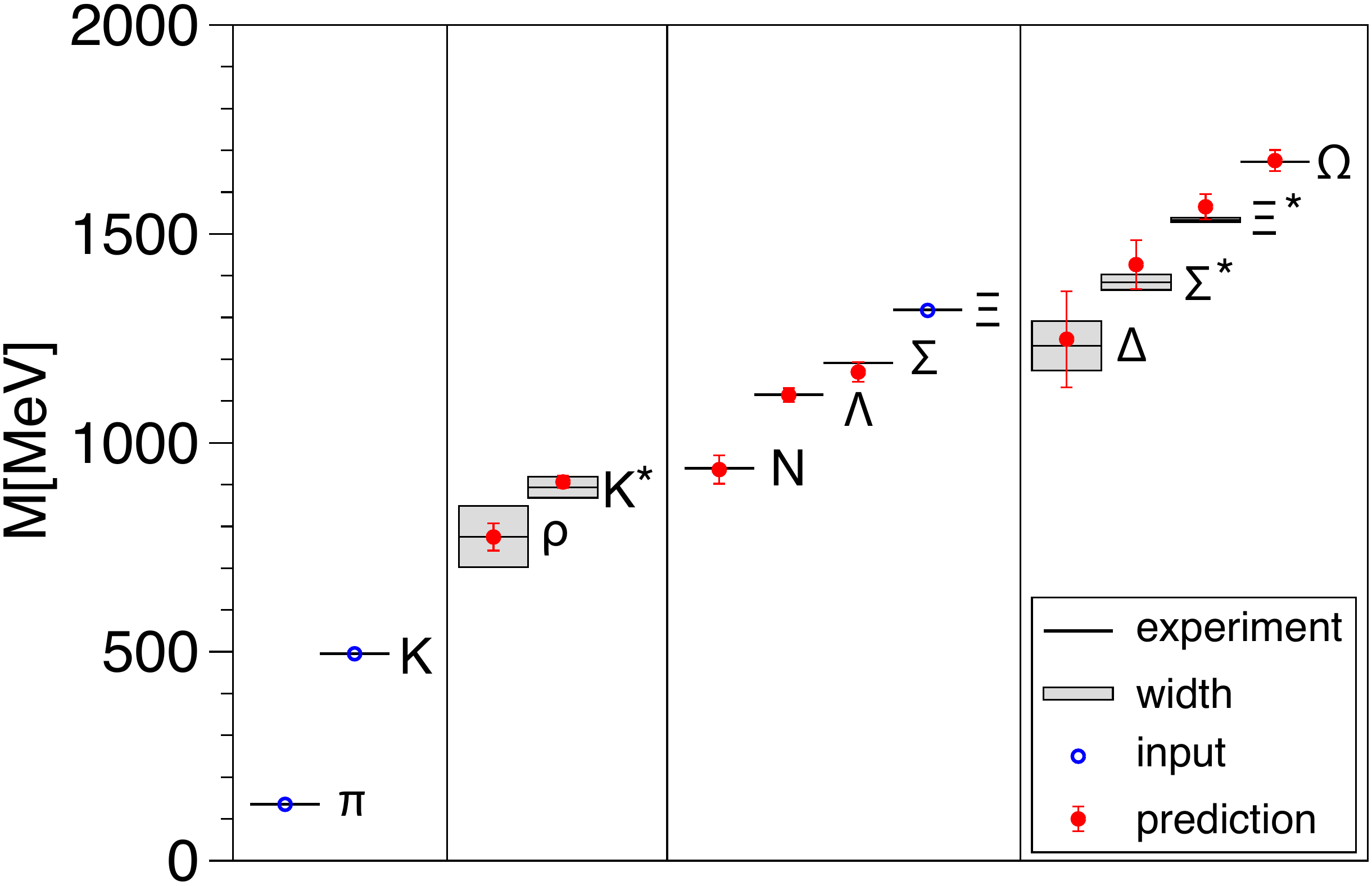}
  \caption{\label{fig:bmwc} Prediction of the light hadron spectrum
    in full $N_f=2+1$ QCD according to \cite{Durr:2008zz}. Open
    circles are input quantities while filled circles are
    predictions. Experimental masses of hadrons that are stable in QCD
    are given with a vertical bar while for resonant states the box
    indicates the decay width. Experimental numbers are from
    \cite{Amsler:2008zzb}.}
\end{figure}

Full control over all systematic uncertainties at the few percent
level was achieved in the light hadron spectrum calculation of the
Budapest-Marseille-Wuppertal collaboration \cite{Durr:2008zz}. They
used tree level improved 6-step stout smeared $N_f=2+1$ clover
fermions on a tree level Symanzik improved gauge action on lattices of
spatial extent of $L\sim 2.0-4.1\text{~fm}$. Both the gauge and the
fermion action are known to be in the correct universality classes and
the updating algorithm is exact and free of possible ergodicity
problems. Pion masses down to $190\text{~MeV}$ and three lattice
spacings $a\sim0.065\text{~fm}$, $a\sim0.85\text{~fm}$ and
$a\sim0.125\text{~fm}$ were used which allowed for a fully controlled
extrapolation to the continuum and the physical point with various
ans\"atze for both. Possible contamination of the propagators from
excited states were accounted for by varying the fit range. Finite
volume corrections were applied including energy shifts for resonant
states (as described in sect.~\ref{sect:fvres}) that allowed for a
detailed treatment of resonant states, too. The continuum
extrapolation was performed with a term linear in $a$ or $a^2$ and
chiral fits were done with both Taylor and NLO heavy baryon $\chi$PT
with a free coefficient (see fig.~\ref{fig:bmwscal} for an example
extrapolation to the physical point and continuum limit). The above
procedure allowed for a fully controlled calculation of the systematic
uncertainty via the spread of the results of all analyses weighted by
the fit quality. The ground state light hadron spectrum was reproduced
at the percent level (cf. fig.~\ref{fig:bmwc}).

\begin{figure}
\centerline{\includegraphics[width=0.5\textwidth]{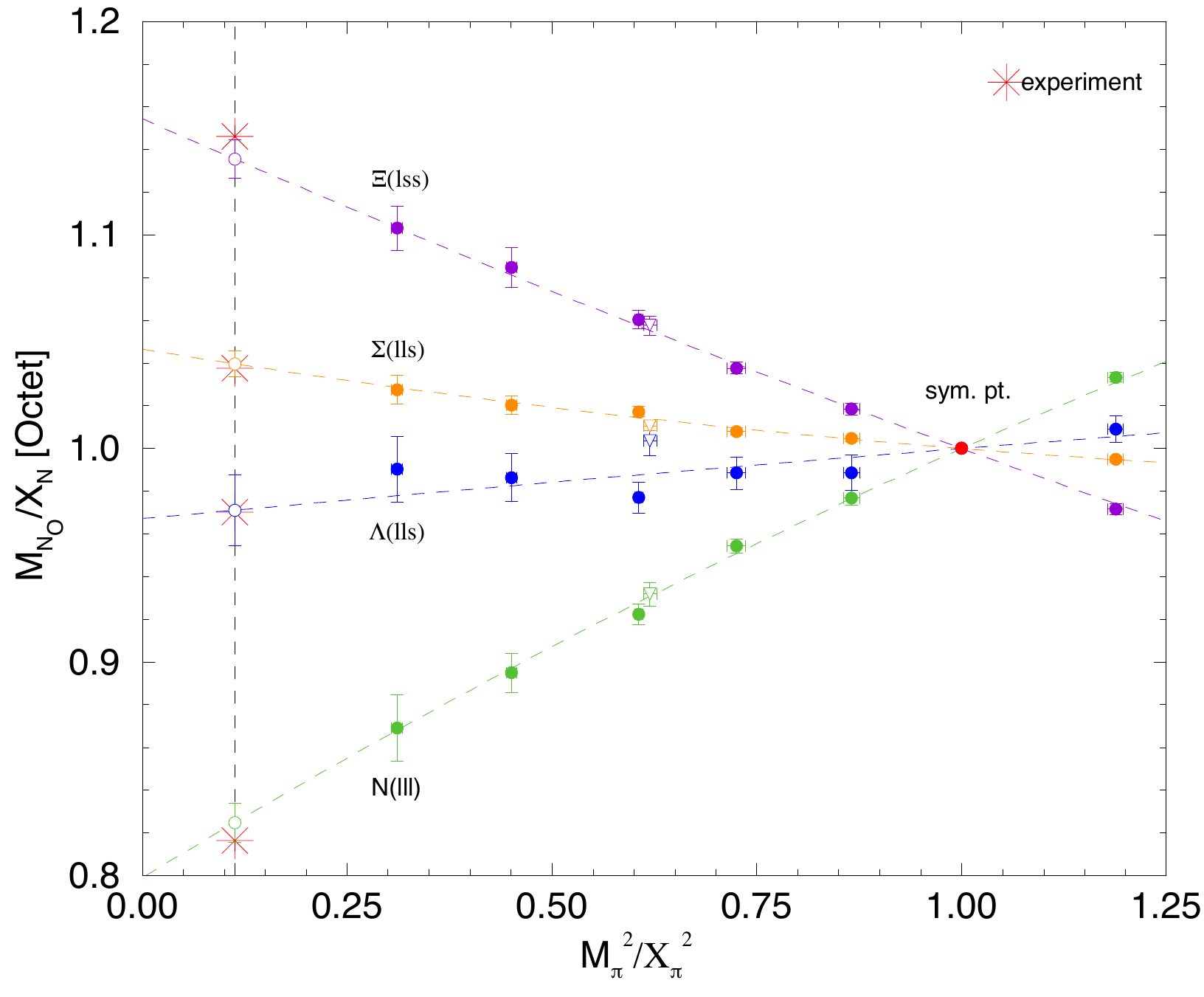}}
\caption{\label{fig:qcdsf}
Chiral behavior of the ratio of individual octet masses over the average
octet mass $X_N=\frac{1}{3}\left(M_N+M_\Sigma+M_\Xi\right)$ vs. the
ratio of the square of the pion mass over the square average of
pseudoscalar meson masses
 $X_\pi^2=\frac{1}{3}\left(2M^2_K+M^2_\pi\right)$ as obtained by the
 QCDSF-UKQCD collaboration. The plot is reproduced from
 \cite{Bietenholz:2010si} with friendly permission of the QCDSF-UKQCD collaboration.}
\end{figure}

The QCDSF-UKQCD collaboration has recently proposed a different
approach to the physical point starting from an SU(3) symmetric theory
and systematically expanding in the SU(3) breaking parameter while
keeping $2M_K^2+M_\pi^2$ constant
\cite{Bietenholz:2010jr,Bietenholz:2010si,Bietenholz:2011qq}.
Preliminary results at a single lattice spacing $a\sim
  0.076\text{~fm}$ and a spatial lattice extent of $L\sim
  1.2-2.5\text{~fm}$ are displayed in fig.~\ref{fig:qcdsf}. They show
a linear dependence of the octet and decuplet masses considered and a
good agreement with the experimentally observed hadron spectrum. An
$N_f=2+1$ nonperturbatively improved single step stout smeared clover
action on a tree level Symanzik improved gauge action was used for
this study. Finite size corrections are not yet applied.

There is also an ongoing effort to compute ground state baryons with
twisted mass fermions including a dynamical strange quark. As the
twisted mass formalism necessitates an even number of fermion flavors
(cf. sect.~\ref{sect:twist}), these calculations also include a charm
quark ($N_f=2+1+1$). First preliminary results of this effort are
reported in \cite{Drach:2010hy}.

The RBC-UKQCD collaboration has recently performed a pioneering
calculation of the $\eta$ and $\eta^\prime$ masses using $N_f=2+1$
flavor domain wall ensembles on an Iwasaki gauge action
\cite{Christ:2010dd}. Three pion masses in the range
$400-700\text{~MeV}$ with a single lattice spacing $a\sim
0.11\text{~fm}$ on latices with a spatial extent of $L\sim
1.8\text{~fm}$ were used. A two operator basis with gauge fixed wall
sources was used to extract the correlation functions. A mixing angle
of $\Theta=-9.2(4.7)^\circ$ and masses $M_\eta=583(15)\text{~MeV}$ and
$M_{\eta^\prime}=853(123)\text{~MeV}$ were found.

\begin{figure*}
\includegraphics[width=0.32\textwidth]{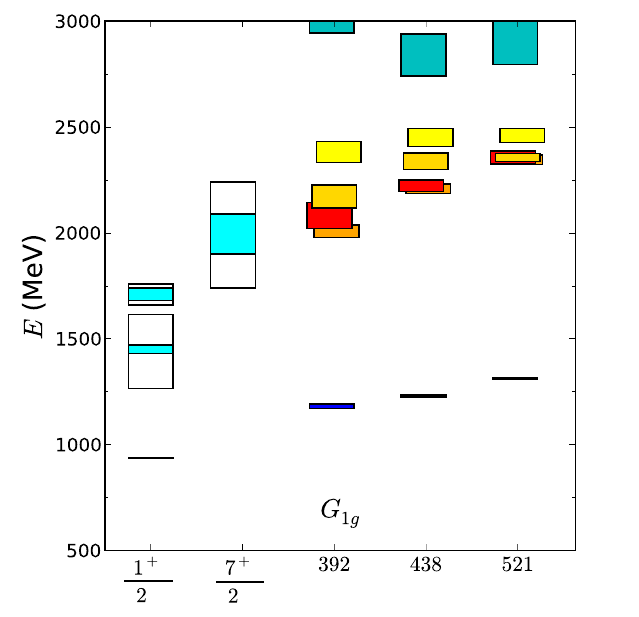}
\includegraphics[width=0.32\textwidth]{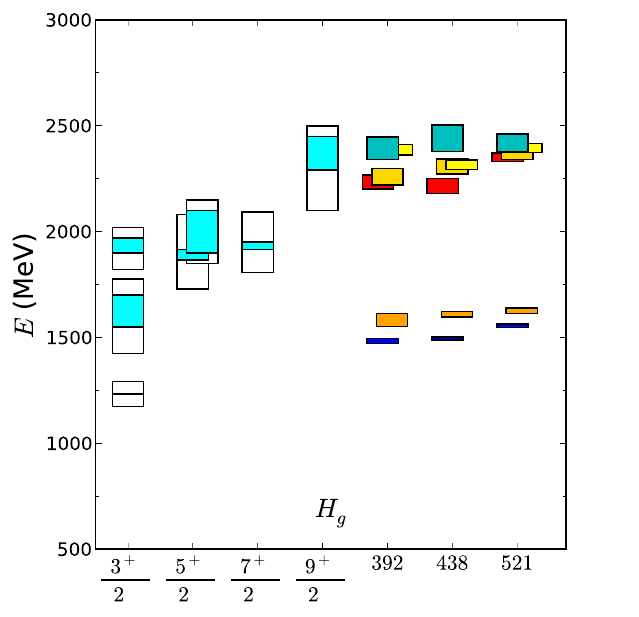}
\includegraphics[width=0.32\textwidth]{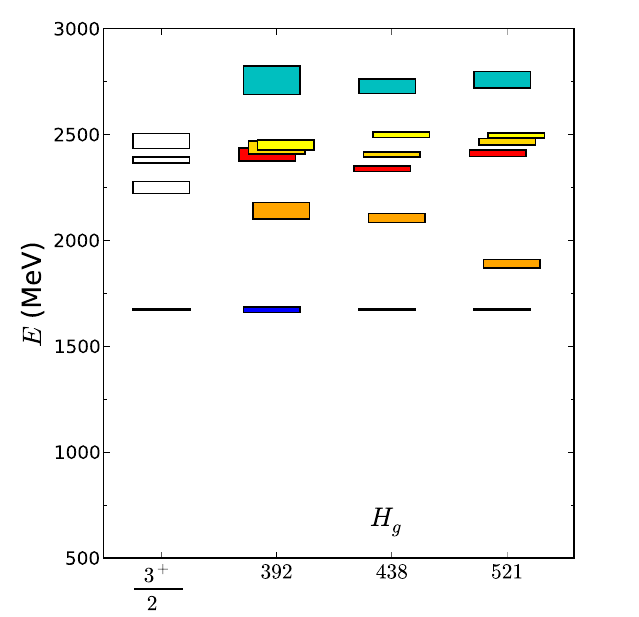}
\caption{\label{fig:hsc}
Comparison of part of the excited state spectrum of nucleon (left)
$\Delta$ (middle) and $\Omega$ (right) type baryons as computed by the hadron
spectrum collaboration at three different pion masses with
experiment. More details can be found in the original paper.  The plot
is reproduced from \cite{Bulava:2010yg} with friendly permission of the
hadron spectrum collaboration.
}
\end{figure*}

The Hadron Spectrum Collaboration is using anisotropic lattices in
order to obtain a fine time resolution of the propagators. These
ensembles are mainly used to extract the highly excited baryon
spectrum. The lattice spacing in time direction is tuned to be smaller
by a factor of $\xi\sim 3.5$ than the lattice spacing in the spatial
directions \cite{Edwards:2008ja}. In their excited state spectroscopy
studies \cite{Lin:2008pr,Bulava:2010yg,Dudek:2011tt} they employ
$N_f=2+1$ anisotropic clover fermions on a tree level tadpole improved
Symanzik gauge action. A single spatial lattice spacing $a_s\sim
0.12\text{~fm}$ and three pion masses in the range
$390-530\text{~MeV}$ are used. The scale is set with $M_\Omega$. A
variational method based on a large number (6-10) of specifically
tailored interpolating operators are used to extract the tower of
excited states in the different channels. Results are reported at
three different pion masses and show a nice overall qualitative
agreement with the experimentally observed excited hadron spectrum
(see fig.~\ref{fig:hsc}). The authors emphasize the need for multi
hadron interpolating operators in order to reliably identify
scattering states. More recently, also the spins of nucleon and
$\Delta$ excitations up to spin $7/2$ have been identified by
\cite{Edwards:2011jj}.

Ground and excited state meson spectra are also being studied with
overlap valence on dynamical domain wall fermions. Some preliminary
results can be found in \cite{Mathur:2010ed}

The quenched studies of
\cite{Mahbub:2009aa,Mahbub:2010jz,Mahbub:2010vu} on the excited baryon
spectrum, especially the excited states of the nucleon, were recently
extended to $N_f=2+1$ dynamical configurations by
\cite{Mahbub:2010me,Mahbub:2010rm}. Fat link irrelevant clover (FLIC)
valence fermions were used on the PACS-CS dynamical ensembles
discussed above. Large operator bases of up to 8 were used and signals
for up to 3 excited states were identified. The chiral behavior of
both positive and negative nucleon excitations was studied and some
evidence was found for the correct ordering of the negative parity
ground state and the Roper resonance as one approaches physical pion
masses.

\section{Concluding remarks}

Although it has taken over 30 years from the formulation of QCD as the
theory of the strong force and Wilsons lattice regularization, it is
fair to say that today we have a firm, quantitative understanding of
the most relevant part of its particle content. It has taken so long
to reach this level of understanding because low energy QCD is a very
rich and nonperturbative theory. The mechanism of permanent quark
confinement and the subsequent emergence of a particle spectrum that
does not at all reflect the fundamental degrees of freedom required
the development of an entirely new set of techniques that have now
matured to a point where the experimentally observed spectrum of
ground state, light non-singlet hadrons can be reproduced to an
accuracy of a few percent.

This quantitative understanding was gained in a process that spanned
several decades. Although the fundamental theory and the general
strategy towards its nonperturbative first-principles solution was
clear from the beginning, it required a substantial amount of
conceptual development and physical insight.

It is however still not a trivial task today to obtain a precise
prediction with fully controlled uncertainties from QCD in the regime
where it is a strongly coupled gauge theory. One needs to be careful 
of optimizing all aspects of the calculation to such a degree
that no single one of them does fully dominate the total error while
at the same time keeping the formalism simple and transparent enough
that computations are manageable in a reasonable amount of time. While
ground state non-singlet hadron masses can be computed to a few
percent accuracy today, reaching the same level of precision for
excited states or singlet hadrons is still a challenging task. There
has been substantial progress regarding the extraction of excited
states and disconnected diagram contributions and the current
understanding is approaching the precision level. A detailed treatment
of resonant finite volume effects, the continuum extrapolation and
even reaching the physical point is work currently in progress.

Lattice calculations of the ground state, non-singlet hadron masses
are currently trying to enter the sub-percent level precision
region. In order to reach this goal, the next challenges involve a
first principles treatment of electromagnetic and isospin breaking
effects as well as an improved treatment of finite volume effects in
the case of resonant states.

In spite of these many open questions and future challenges, we do
believe however that the percent level understanding of relevant parts
of the light hadron spectrum with fully controlled systematic uncertainties
that has been achieved by lattice QCD is a milestone that marks the
overall maturity of the method. While a lot of interesting problems
such as excited state spectroscopy still require substantial work,
lattice QCD today represents a reliable tool of extracting from first
principles properties of a strongly coupled quantum field theory.

\begin{acknowledgments}
  We would like to thank Stephan D\"urr, Stefan Krieg, Thorsten Kurth,
  Laurent Lellouch, Alberto Ramos, Kalman Szab{\'o}, Balint Toth and
  especially Craig Mc~Neile for discussions and comments on the
  manuscript. This work was in part funded by the ``Deutsche
  Forschungsgemeinschaft'' under the grant SFB-TR55.
\end{acknowledgments}
\bibliography{references}{}
\bibliographystyle{apsrmp}

\end{document}